\documentclass[12pt]{article}
\usepackage{latexsym,amssymb}
\usepackage[vsentermath]{youngtab}
\newcommand{\ft}[2]{{\textstyle\frac{#1}{#2}}}
\def\tilde{\widetilde}

\newsavebox{\uuunit}
\sbox{\uuunit}
                      {\setlength{\unitlength}{0.825em}
                             \begin{picture}(0.6,0.7)
                                                  \thinlines
                                                  \put(0,0){\line(1,0){0.5}}
                                                  \put(0.15,0){\line(0,1){0.7}}
                                                  \put(0.35,0){\line(0,1){0.8}}
                                                 \multiput(0.3,0.8)(-0.04,-0.02){10}{\rule{0.5pt}{0.5pt}}
                             \end {picture}}

\makeatletter \@addtoreset{equation}{section} \makeatother

\def\bfone{\relax{\rm 1\kern-.35em 1}}

\def\bfone{\relax{\rm 1\kern-.35em 1}}
 
\begin{document}
\begin{titlepage}
\vskip 0.5cm
\begin{center}
{\LARGE \bf  The general pattern of  Ka\v c Moody extensions \\}
\vskip 0.3cm
{\LARGE \bf in  supergravity  } \\
\vskip 0.3cm
{\LARGE \bf and the issue of cosmic billiards$^\dagger$}\\ 
\vskip 2cm
{\large
Pietro Fr\'e$^\star$, Floriana Gargiulo$^\#$ \\
\vskip 0.2cm
Ksenya Rulik$^\star$  and Mario Trigiante$^\#$ } \\
\vskip 0.5cm {\small
$^\star$ Dipartimento di Fisica Teorica, Universit\'a di Torino, \\
$\&$ INFN -
Sezione di Torino\\
via P. Giuria 1, I-10125 Torino, Italy\\
$^\#$ Dipartimento di Fisica Politecnico di Torino,\\
 C.so Duca degli Abruzzi, 24, I-10129 Torino, Italy
 }
\end{center}
\begin{abstract}
{In this paper we study the systematics of the affine extension of
supergravity duality algebras when one steps down from $D=4$ to
$D=2$ which is instrumental for the study of cosmic billiards.  For all $D=4$
supergravities (with $N\ge 3$) there is a universal field theoretical
mechanism promoting the extension, which relies on the coexistence
of two non locally related lagrangian descriptions of the
corresponding $D=2$ degrees of freedom: the Ehlers lagrangian and
the Matzner--Misner one. This and the existence of a
generalized Kramer--Neugebauer non local transformation relating
the two models, provide a Chevalley-Serre presentation of the
affine Ka\v c--Moody algebra which follows a universal pattern for
 all supergravities. This  is an extension of the mechanism considered by Nicolai for pure N=1 supergravity,
 but has general distinctive features in extended theories (N$\geq$3) related to the presence of vector fields and to their symplectic
 description. Moreover the novelty is that in the general case
the Matzner-Misner lagrangian is structurally different from the
Ehlers one, since half of the scalars are replaced by gauge
$0$--forms subject to $\mathrm{SO(2n,2n)}$ electric--magnetic duality rotations
representing in $D=2$ the $\mathrm{Sp(2n,\mathbb{R})}$  rotations  of $D=4$.
 The role played by the symplectic bundle of vectors in this context, suggests that the mechanism
 of the affine extension can be studied also for $N=2$ supergravity, where one deals with geometries rather than
 with algebras, the scalar manifold being not necessarily a homogeneous manifold $\mathrm{U/H}$. We also show that the mechanism
 of the affine extension commutes with the Tits Satake projection of the relevant duality algebras. This is very important for the issue of cosmic billiards, as we show in a separate paper. Finally we also comment on the general field theoretical mechanism of the further hyperbolic extension obtained in $D=1$, which we plan to analyze in detail in a forthcoming paper. The possible uses of our results and their relation to outstanding problems are pointed out.}
\end{abstract}
\vspace{2mm} \vfill \hrule width 3.cm {\footnotesize $^ \dagger $
This work is supported in part by the European Union network contract
MRTN-CT-2004-005104}

\end{titlepage}
\section{Introduction}
The discovery that  General Relativity hides dynamical
symmetries associated with infinite Lie algebras of the Ka\v c--Moody
type is an old one and dates back to the work of Geroch
\cite{Geroch}. Considering metrics that admit two commuting, space-like Killing
vectors, Geroch effectively reduced Einstein Gravity to $D=2$ and
showed that the space of such solutions is mapped into itself by an
infinite group of symmetries, whose Lie algebra, in modern parlance,
is  $A_1^\wedge$, namely the affine Ka\v c--Moody extension of $A_1$,
the Lie algebra of the three-parameter group
$\mathrm{SL(2,\mathbb{R})}$. These quite intriguing symmetries were
extensively analyzed at the end of the seventies and in the beginning
of the eighties by several groups (for a comprehensive review see
\cite{Maison&Breiten}) and a formalism was developed to cope with
them in a mathematical rigorous and effective way that is centered on the notion of
\textit{linear system}. 
To summarize a complicated story in a few words we just recall that
Ka\v c--Moody algebras \cite{GO} are the Lie algebras of
centrally extended loop groups, namely the groups of analytic maps
from a circle into an ordinary simple Lie group $G$:
\begin{equation}
 G^{\infty} \, \ni \,  \gamma \,  : \, \mathbb{S}^1 \, \mapsto \, G
\label{looppo}
\end{equation}
Calling $z=\exp\left[ {\rm i} \, \theta \right] $ a local coordinate
on $\mathbb{S}^1 $, a group element $\hat{g} \, \in \, G^\wedge$ is
represented by a pair
\begin{equation}
\hat{g} = \left(g(z),e^\alpha\right)
\end{equation}
where $g(z) \in G^{\infty}$ is a group element of the simple Lie
group which depends analytically on $z$ and $e^\alpha$ is an
exponent of the central element. In the set up of linear systems all
possible solutions of Einstein gravity reduced to $D=2$ are
associated with all possible maps from the $D=2$ space-time
$\mathcal{M}_{D=2}$ to the infinite-dimensional coset
$\mathrm{SL(2,\mathbb{R})}^\wedge/H^\wedge$, where
$\mathrm{SL(2,\mathbb{R})}^\wedge$ is the centrally extended loop
group of $\mathrm{SL(2,\mathbb{R})}$ and $H^\wedge$ is its
\textit{maximal compact subgroup}. In other words, reduced to $D=2$
Einstein Gravity becomes a sort of $\sigma$--model:
\begin{equation}
  \mathcal{M}_{D=2} \mapsto \frac{\mathrm{SL(2,\mathbb{R}})^\wedge}{H^\wedge}
\label{sortofsigma}
\end{equation}
where the target space space is the infinite dimensional affine
generalization of the maximally split, non compact coset
$\mathrm{SL(2,\mathbb{R})}/\mathrm{O(2)}$. As we are going to stress
in the present paper by recalling and generalizing results which
were obtained times ago by Nicolai \cite{NicolaiEHMM}, naive
dimensional reduction of the Einstein lagrangian produces a
$2D$--gravity coupled $\sigma$-model, where the target manifold is
indeed $\mathrm{SL(2,\mathbb{R})}/\mathrm{O(2)}$:
\begin{equation}
  \mathcal{M}_{D=2} \mapsto
  \frac{\mathrm{SL(2,\mathbb{R})}}{\mathrm{O(2)}}
\label{normsigma}
\end{equation}
The replacement of equation (\ref{normsigma}) with
eq.(\ref{sortofsigma}) occurs because there are two different ways
of obtaining (\ref{normsigma}) via dimensional reductions which are
non locally related to each other. Covering the two pictures at the
same time is equivalent to the affine extension, namely to
eq.(\ref{sortofsigma}). The main point of the present article is
that this mechanism of generation of the infinite symmetries is
actually completely general and follows a regular algebraic pattern
for all supergravity theories which we plan to describe hereby,
emphasizing its role in the discussion of cosmic billiard dynamics.
This, however, is just an anticipation of our subsequent
discussions. For the time being it suffices to note that an explicit
map of the type advocated in eq. (\ref{sortofsigma}) is given by a
coset representative
\begin{equation}\label{affinecosrep}
\hat{\mathbb{L}} = \left(\mathbb{L}\left(z(x^\pm), x^\pm
\right),e^{-d(x)}\right)
\end{equation}
where $\mathbb{L}\left ( z(x^\pm), x^\pm \right)$ is an element of
$\mathrm{SL(2,\mathbb{R})}$ depending on the $2D$ coordinates $x^\pm
$ both explicitly and implicitly through the dependence of $z$ on
$x^\pm $, and $d(x)$ is a logarithm of the conformal factor of
two--dimensional metric. This is just the mathematical transcription
of the formal map defined in eq.(\ref{sortofsigma}). In the context
of such a discussion, the coordinate $z$ on the circle is named the
\textit{spectral parameter}. Once the algorithm that  to each coset
representative $\hat{\mathbb{L}}$ associates a solution of gravity
and viceversa to each solution associates a coset representative has
been established:
\begin{equation}
\mathcal{A}   \, : \, \hat{\mathbb{L}} \, \Leftrightarrow \,
\mbox{solution of Einstein gravity} \label{Aalgo}
\end{equation}
then the action of the affine group on the space of solutions is
easily defined by the left action of $\hat{g}(z)$ on $\hat{\mathbb{L}}$.

\par
After this clarification occurred in the context of pure gravity the
interest in infinite symmetries was stimulated by the advent of
supergravity. By dimensionally reducing $D=10$ or $D=11$
supergravity to lower dimensions one discovers the so called hidden
symmetries, that act as isometries of the kinetic  metric on the scalars and
as generalized electric/magnetic duality rotations on the vector
fields or $p$--forms. Hidden symmetries were discovered specially
through the work of Cremmer and Julia  who found $E_{7(7)}$ \cite{N=8CremmerJulia} in
$\mathcal{N}=8,D=4$ supergravity and later clarified that
$\mathrm{E_{11-D,(11-D)}}$ is the duality symmetry for maximal supergravity in
$D$-dimensions, obtained by compactification of M--theory on a
$\mathrm{T^{11-D}}$ torus. The continuation of the series to $D<3$
leads to algebras that are no longer finite dimensional, rather they are
Ka\v c--Moody algebras. Indeed it was Julia \cite{oldJulia} who already long time ago
noted this phenomenon, pointing out that $\mathrm{E_{9(9)}}$ is just
the affine extension of $\mathrm{E_{8(8)}}$, while $\mathrm{E_{10(10)}}$ is its double
hyperbolic extension. Later, extensive work on $\mathrm{E_{10}}$ and in general
on the infinite symmetries of lower dimensional supergravities was
performed by Nicolai and collaborators in a large set of papers \cite{Nicolaistream}.
Notwithstanding the interest  and the potential relevance of all these
results it must be said that for a long time they did not find a
convincing arena of applications in the context of superstring
and brane theory, remaining an unexploited truth. This is even more
surprising if one considers that precisely the finite dimensional
relatives of these symmetries, namely the hidden symmetries in $D \ge
4$ played a fundamental role in the so named second string
revolution, namely in establishing the non perturbative dualities
among the various string models. Indeed, as it is well known, those dualities that unify
the various perturbative quantum strings into a unique M-theory are
elements of a unified group $\mathrm{U}(\mathbb{Z})$ which is the suitable restriction
to integers of a corresponding Lie group $\mathrm{U}(\mathbb{R})$ encoded in
compactified supergravity and given, for toroidal compactifications, by the earlier mentioned
$\mathrm{E_{11-D,(11-D)}}$ series.
\par
This situation changed significantly, after 2001, with the renewed
interest in \textit{Ka\v c--Moody symmetries} generated by the
discovery of the \textit{cosmic billiard phenomenon}. This
phenomenon encodes a profound link between the features of time
evolution of the cosmological scale factors and the  structure of
the hidden symmetry algebra $\mathrm{U}(\mathbb{R})$. Let us name
$N_Q$ the number of supersymmetry charges. For $N_Q > 8$ the
scalar manifold is always a homogeneous space $\mathrm{U/H}$ and
what actually happens is that the cosmological scale factors
$a_i(t)$ associated with the various  dimensions of supergravity
can be interpreted as exponentials of those scalar fields $h_i(t)$
which lie in the Cartan subalgebra of $\mathbb{U}$, while the
other scalar fields in $\mathrm{U/H}$ correspond to positive roots
$\alpha >0$ of the Lie algebra $\mathbb{U}$. In this way the
cosmological evolution is described by the motion of a
\textit{fictitious ball} in the CSA of $\mathbb{U}$. This space is
actually a billiard table whose walls are the hyperplanes
orthogonal to the various roots. The fictitious ball bounces on
the billiard walls and this means that there are inversions in the
time evolution of the scale factors. Certain dimension that were
previously expanding almost suddenly begin to contract and others
do the reverse. Such a scenario was introduced by Damour,
Henneaux, Julia and Nicolai in \cite{cosmicbilliardliterature1},
and in a series of papers with collaborators
\cite{cosmicbilliardliterature2},
\cite{cosmicbilliardliterature3},\cite{cosmicbilliardliterature4},
which generalize classical results obtained in the context of pure
General Relativity
\cite{Kassner},\cite{cosmicbilliardliterature5}.
\par
In their approach the quoted authors analyzed the cosmic billiard
phenomenon  as an asymptotic regime in the neighborhood of
space-like singularities and the billiard walls were seen as delta
function potentials provided by the various $p$--forms of
supergravity localized at sharp instants of time. The main focus of
attention was centered on establishing whether and under which
conditions there may be a \textit{chaotic behaviour} in the evolution
of the scale factors and the same authors established that this may
occur only when the billiard table, identified with the Weyl chamber
of the duality algebra $\mathbb{U}$ is hyperbolic, namely when the
double Ka\v c--Moody extension of the hidden symmetries is taken in proper
account.
\par
With a different standpoint it was observed in \cite{noiconsasha}
that the fundamental mathematical setup underlying the appearance
of the billiard phenomenon is the so named \textit{Solvable Lie
algebra parametrization} of supergravity scalar manifolds,
pioneered in \cite{primisolvi} and later applied to the solution
of a large variety of superstring/supergravity problems
(\cite{primisolvi},\cite{noie7blackholes},\cite{otherBHpape},\cite{gaugedsugrapot}
).  In particular we showed that one can implement the following
programme:
\begin{enumerate}
  \item Reduce the original supergravity in higher  dimensions $D \ge 4$ (for instance $D=10,11$) to
  a gravity-coupled $\sigma$--model in $D \le 3$ where gravity is
  non--dynamical and where the original higher dimensional bosonic field
  equations reduce to geodesic equations for a solvable
  group-manifold, metrically equivalent to a non compact
  coset manifold $\exp \left[ Solv\left(\mathrm{U/H} \right)  \right]
  \cong \mathrm{U/H}$.
  \item Utilize the algebraic structure of the solvable Lie algebra
  $Solv\left(\mathrm{U/H} \right) $ associated with the pair of the algebra $\mathbb{U}$ and its maximal compact
  subalgebra $\mathbb{H}$ in order to integrate analytically the geodesic
  equations.
  \item Dimensionally oxide the solutions obtained in this way to
  exact  time dependent solutions of  $D \ge 4$ supergravity.
\end{enumerate}
Within this approach it was proved in \cite{noiconsasha} that the
\textit{cosmic billiard phenomenon} is indeed a general feature of exact time
dependent solutions of supergravity and has \textit{smooth
realizations}. Calling $\mathbf{h}(t)$ the $r$--component
vector of Cartan fields (where $r$ is the split rank of $\mathbb{U}$) and
$\mathbf{h}_\alpha(t) \equiv \mathbf{\alpha} \, \cdot \,
 {\mathbf{h}}(t)$ its projection along any positive root $\alpha$,
a \textit{bounce} occurs at those instant of times $t_i$ such that:
\begin{equation}
\exists \,  \mathbf{\alpha} \, \in \, \Delta_+ \, \quad  \backslash
\quad \quad
 \dot{\mathbf{h}}_\alpha (t)\mid_{t=t_i} \, = \, 0
\label{bouncino}
\end{equation}
namely when the Cartan field in the direction of some root $\alpha$ inverts
its behaviour and begins to shrink if it was growing or viceversa
begins to grow if it was shrinking. Since all higher dimensional
bosonic fields (off-diagonal components of the metric $g_{\mu\nu}$ or
$p$--forms $A^{[p]}$) are, via the solvable parametrization of $\mathrm{U/H}$, in one-to-one correspondence with roots
$ \phi_\alpha \, \Leftrightarrow \, \alpha$, it follows that
the bounce on a \textit{wall} (hyperplane orthogonal to the root $\alpha$) is
caused by the sudden growing of that particular field $\phi_\alpha$.
Indeed  in the exact smooth solutions
which we  obtained in \cite{noiconsasha},  each bounce is associated with
a typical bell-shaped behaviour of the root field $\phi_\alpha$ and
the whole process can be interpreted as a temporary localization
of the Universe energy density in a \textit{lump} on a spatial brane associated with the
field $\phi_\alpha$.
\par
Although very much encouraging, the analysis of \cite{noiconsasha}
was limited in various respects. One limitation, whose removal is
the main motivation of the present paper,  consists of the
following. The dimensional reduction process which is responsible
for making manifest the duality algebra $ \mathbb{U}$ and hence
for creating the whole algebraic machinery utilized in deriving
the \textit{smooth cosmic billiard solutions} was stopped at $D=3$
namely at the first point where all the bosonic degrees of freedom
can be represented by scalars. In $D=3$, $\mathbb{U}$ is still a
finite dimensional Lie algebra and the whole richness of the
underlying algebraic structure is not yet displayed. As implied by
the results of \cite{cosmicbilliardliterature4}, in order to
investigate the most challenging aspects of  billiard dynamics, in
particular chaos, within the framework of exact solutions, we
should derive these latter better in a  $D=2$ or $D=1$ context
where Ka\v c--Moody symmetries become manifest. Although the
appearance of Ka\v c--Moody extensions is algebraically well
established, their exploitation in deriving solutions is not as
clear as the exploitation of ordinary symmetries. Indeed the main
issue to clarify is the field theoretical realization of the Ka\v
c--Moody extensions which is the prerequisite for their
utilization in deriving billiard dynamics.
\par
The present paper aims at providing a clearcut step-forward in
this direction. In particular we want to show that there is a
general mechanism underlying the affine Ka\v c--Moody extension of
the D=3 algebra $\mathbb{U}_{D=3}$ when we step down to $D=2$ and
that this mechanism follows a general algebraic pattern for all
supergravity theories, independently of the number of supercharges
$N_Q$. As already mentioned above this mechanism relies on the
existence of two different reduction schemes from $D=4$ to $D=2$,
respectively named the Ehlers reduction and the Matzner--Misner
reduction, which are non locally related to each other. Nicolai
observed this phenomenon time ago  in the case of pure gravity (or
better of N=1 pure supergravity)\cite{NicolaiEHMM} and showed that
one obtains two identical lagrangians, each displaying an
$\mathrm{SL(2,\mathbb{R})}$ symmetry. The fields appearing in one
lagrangian have a non local relation to those of the other
lagrangian and one can put together both
$\mathrm{SL(2,\mathbb{R})}$ algebras. One algebra generates local
transformations on one set of fields the other algebra generates
non local ones. Together the six generators of the two
$\mathrm{SL(2,\mathbb{R})}$ provide a Chevalley basis for the Ka\v
c--Moody extension $\mathrm{SL(2,\mathbb{R})}^\wedge$ namely for
$A_1^\wedge$. Our analysis will be an extension of the argument by
Nicolai. For a generic supergravity theory, the two reduction
schemes Ehlers and Matzner--Misner lead to two different
lagrangians with different local symmetries. The first is a normal
\textit{$\sigma$--model} the second is a \textit{twisted
$\sigma$--model}. We shall discuss in detail the symmetries of
both theories.  Just as in Nicolai case we can put together the
symmetries of both lagrangians and in this way we obtain a
Chevalley basis for the Ka\v c--Moody algebra. In this way we can
write down a precise field theoretic realization of the affine
symmetries setting the basis to exploit them in billiard dynamics.
We shall then comment on the further hyperbolic extension
occurring in $D=1$ and on the nature of the billiard chamber.
\section{$D=4$ supergravity and its duality symmetries}
\label{d4sugrasym}
Rather than starting from $D=10$ supergravity or $11$--dimensional M-theory we begin
our analysis from $D=4$. How we stepped down from $D=10,11$ to $D=4$
is not necessary to specify at this level. It is implicitly encoded
in the number of residual supersymmetries that we consider. If $N_Q
=32$ is maximal it means that we used toroidal compactification.
Lower values of $N_Q$ correspond to compactifications on manifolds of restricted
holonomy, Calabi Yau three-folds, for instance, or orbifolds.
The relevant point is that for  $D=4$ \textit{ungauged supergravity} the bosonic
lagrangian admits a general form which we presently discuss and exploit in our
argument. We have:
\begin{eqnarray}
\mathcal{L}^{(4)} &=& \sqrt{\mbox{det}\, g}\left[-2R[g] - \frac{1}{6}
\partial_{\hat{\mu}}\phi^a\partial^{\hat{\mu}}\phi^b h_{ab}(\phi) \,
+ \,
\mbox{Im}\mathcal{N}_{\Lambda\Sigma}\, F_{\hat{\mu}\hat{\nu}}^\Lambda
F^{\Sigma|\hat{\mu}\hat{\nu}}\right] \nonumber\\
&&+
\frac{1}{2}\mbox{Re}\mathcal{N}_{\Lambda\Sigma}\, F_{\hat{\mu}\hat{\nu}}^\Lambda
F^{\Sigma}_{\hat{\rho}\hat{\sigma}}\epsilon^{\hat{\mu}\hat{\nu}\hat{\rho}\hat{\sigma}}
\label{d4generlag}
\end{eqnarray}
In eq.(\ref{d4generlag}) $\phi^a$ denotes the whole set of $n_S$ scalar fields
parametrizing the scalar manifold $ \mathcal{M}_{scalar}^{D=4}$
which, for $N_Q > 8$, is necessarily a coset manifold:
\begin{equation}
  \mathcal{M}_{scalar}^{D=4} \, =
  \,\frac{\mathrm{U_{D=4}}}{\mathrm{H}}
\label{cosettoquando}
\end{equation}
For $N_Q \le 8$ eq.(\ref{cosettoquando}) is not obligatory but it is
possible. Particularly in the $\mathcal{N}=2$ case, i.e. for $N_Q =8$, a large
variety of homogeneous special K\"ahler
\cite{specHomgeo} fall into the set up of the present general discussion.
The fields $\phi^a$ have $\sigma$--model interactions
dictated by the metric $h_{ab}(\phi)$ of $\mathcal{M}_{scalar}^{D=4}$. 
\par
The theory includes also $n$ vector
fields $A_{\hat{\mu}}^\Lambda$ for which
\begin{equation}
  \mathcal{F}^{\pm| \Lambda}_{\hat{\mu}\hat{\nu}} \equiv \ft 12
  \left[{F}^{\Lambda}_{\hat{\mu}\hat{\nu}} \mp \, {\rm i} \, \frac{\sqrt{\mbox{det}\, g}}{2}
  \epsilon_{\hat{\mu}\hat{\nu}\hat{\rho}\hat{\sigma}} \, F^{\hat{\rho}\hat{\sigma}} \right]
\label{Fpiumeno}
\end{equation}
denote the self-dual (respectively antiself-dual) parts of the field-strengths. As displayed in
eq.(\ref{d4generlag}) they are non minimally coupled to the scalars via the symmetric complex matrix
\begin{equation}
  \mathcal{N}_{\Lambda\Sigma}(\phi)
  ={\rm i}\, \mbox{Im}\mathcal{N}_{\Lambda\Sigma}+ \mbox{Re}\mathcal{N}_{\Lambda\Sigma}
\label{scriptaenna}
\end{equation}
which transforms projectively under $\mathrm{U_{D=4}}$. Indeed the field strengths ${F}^{
\Lambda}_{\mu\nu}$ plus their magnetic duals fill up a $2\,
n$--dimensional symplectic representation of $\mathrm{\mathbb{U}_{D=4}}$
which we call by the name of $\mathbf{W}$.
\par
Following the notations and the conventions of \cite{myparis}, we rephrase the above statements
by asserting that there is always a symplectic embedding of the duality group
$\mathrm{U}_{D=4}$,
\begin{equation}
  \mathrm{U}_{D=4} \mapsto \mathrm{Sp(2n, \mathbb{R})} \quad ; \quad
  n = n_V \, \equiv \, \mbox{$\#$ of vector fields}
\label{sympembed}
\end{equation}
so that for each element $\xi \in \mathrm{U}_{D=4}$ we have its
representation by means of a  suitable real symplectic matrix:
\begin{equation}
  \xi \mapsto \Lambda_\xi \equiv \left( \begin{array}{cc}
     A_\xi & B_\xi \\
     C_\xi & D_\xi \
  \end{array} \right)
\label{embeddusmatra}
\end{equation}
satisfying the defining relation:
\begin{equation}
  \Lambda_\xi ^T \, \left( \begin{array}{cc}
     \mathbf{0}_{n \times n}  & { \mathbf{1}}_{n \times n} \\
     -{ \mathbf{1}}_{n \times n}  & \mathbf{0}_{n \times n}  \
  \end{array} \right) \, \Lambda_\xi = \left( \begin{array}{cc}
     \mathbf{0}_{n \times n}  & { \mathbf{1}}_{n \times n} \\
     -{ \mathbf{1}}_{n \times n}  & \mathbf{0}_{n \times n}  \
  \end{array} \right)
\label{definingsympe}
\end{equation}
which implies the following relations on the $n \times n$ blocks:
\begin{eqnarray}
A^T \, C - C^T \, A & = & 0 \nonumber\\
A^T \, D - C^T \, B & = & \mathbf{1}\nonumber\\
B^T \, C - D^T \, A& = & - \mathbf{1}\nonumber\\
B^T \, D - D^T \, B & =  & 0
\label{symplerele}
\end{eqnarray}
Under an element of the duality groups the field strengths transform
as follows:
\begin{equation}
  \left(\begin{array}{c}
     \mathcal{F}^+ \\
     \mathcal{G}^+ \
  \end{array} \right)  ^\prime \, = \,\left( \begin{array}{cc}
     A_\xi & B_\xi \\
     C_\xi & D_\xi \
  \end{array} \right) \,  \left(\begin{array}{c}
     \mathcal{F}^+ \\
     \mathcal{G}^+ \
  \end{array} \right) \quad ; \quad \left(\begin{array}{c}
     \mathcal{F}^- \\
     \mathcal{G}^- \
  \end{array} \right)  ^\prime \, = \,\left( \begin{array}{cc}
     A_\xi & B_\xi \\
     C_\xi & D_\xi \
  \end{array} \right) \,  \left(\begin{array}{c}
     \mathcal{F}^- \\
     \mathcal{G}^- \
  \end{array} \right)
\label{lucoidale1}
\end{equation}
where, by their own definitions:
\begin{equation}
    \mathcal{G}^+ = \mathcal{N} \, \mathcal{F}^+ \quad ; \quad \mathcal{G}^- = \overline{\mathcal{N}} \,
    \mathcal{F}^-
\label{lucoidale2}
\end{equation}
and the complex symmetric matrix $\mathcal{N}$ transforms as follows:
\begin{equation}
  \mathcal{N}^\prime = \left(  C + D \, \mathcal{N}\right) \, \left( A + B \,\mathcal{N}\right)
  ^{-1}
\label{Ntransfa}
\end{equation}
Eq.(\ref{d4generlag}) is the lagrangian that we are supposed to dimensionally
reduce according to the two available schemes, the Ehlers reduction
and the Matzner-Misner reduction respectively. We will perform such
reductions in later sections. Prior to that we dwell on an algebraic
interlude anticipating the result of the Ehlers reduction and
analyzing the algebraic structure of the $\mathbb{U}_{D=3}$ algebra.
Such an analysis is very important in order to establish the
properties of its affine extension and single out a basis of
candidate Chevalley-Serre generators.
\section{Structure of the duality algebras in D=3\\ from the Ehlers reduction
and their affine extensions}
\label{structadual}
Upon toroidal dimensional reduction from $D=4$ to $D=3$ and then full--dualization of the vector fields, which
is the Ehlers scheme to be described in detail later,
we obtain supergravity theories admitting a \textbf{duality Lie algebra} $\mathbb{U}_{D=3}$
whose structure is universal in the following sense. It  always
contains, as subalgebra, the duality algebra $\mathbb{U}_{D=4}$ of the parent
supergravity theory in $D=4$ times an $\mathrm{SL(2,\mathbb{R})_E}$ algebra
which is produced by the dimensional reduction of pure gravity (see
section \ref{ehlersgrav} for the details).
Furthermore, with respect to this subalgebra $\mathbb{U}_{D=3}$
admits the following universal decomposition, holding for all
$\mathcal{N}$-extended supergravities having semisimple duality algebras:
\begin{equation}
\mbox{adj}(\mathbb{U}_{D=3}) =
\mbox{adj}(\mathbb{U}_{D=4})\oplus\mbox{adj}(\mathrm{SL(2,\mathbb{R})_E})\oplus
W_{(2,\mathbf{W})}
\label{gendecompo}
\end{equation}
where $\mathbf{W}$ is the {\bf symplectic} representation of
$\mathbb{U}_{D=4}$ discussed in the previous section. Indeed the
scalar fields associated with the generators of
$W_{(2,\mathbf{W})}$ are just those coming from the vectors in
$D=4$. Denoting the generators of $\mathbb{U}_{D=4}$ by $T^A$, the
generators of $\mathrm{SL(2,\mathbb{R})_E}$ by $\mathrm{L^x}$ and
denoting by $W^{i\alpha}$ the generators in $W_{(2,\mathbf{W})}$,
the commutation relations that correspond to the decomposition
(\ref{gendecompo}) have the following general form:
\begin{eqnarray}
\nonumber && [T^A,T^B] = f^{AB}_{\phantom{AB}C} \, T^C  \\
\nonumber && [L^x,L^y] = f^{xy}_{\phantom{xy}z} \, L^z , \\ &&\nonumber
[T^A,W^{i\alpha}] = (\Lambda^A)^\alpha_{\,\,\,\beta} \, W^{i\beta},
\\ && [L^x, W^{i\alpha}] = (\lambda^x)^i_{\,\, j}\, W^{j\alpha}, \\
\nonumber &&[W^{i\alpha},W^{j\beta}] =
\omega^{ij}\, (K_A)^{\alpha\beta}\, T^A + \, \Omega^{\alpha\beta}\, k_x^{ij}\, L^x
\label{genGD3pre}
\end{eqnarray}
where the matrices $(\lambda^x)^i_j$, which are $2 \times 2$ are the canonical generators of $\mathrm{SL(2,\mathbb{R})}$
in the fundamental, defining representation:
\begin{equation}
  \lambda_3 = \left(\begin{array}{cc}
     \ft 12 & 0 \\
     0 & -\ft 12 \
  \end{array} \right) \quad ; \quad \lambda_1 = \left(\begin{array}{cc}
     0 & \ft 12  \\
     \ft 12 & 0\
  \end{array} \right) \quad ; \quad \lambda_2 = \left(\begin{array}{cc}
     0 & \ft 12  \\
     -\ft 12 & 0\
  \end{array} \right)
\label{lambdax}
\end{equation}
while $\Lambda^A$ are the generators
of $\mathbb{U}_{D=4}$ in the symplectic representation $\mathbf{W}$. By
\begin{equation}
  \Omega^{\alpha\beta} \equiv \left( \begin{array}{c|c}
     \mathbf{0}_{n\times n} & \mathbf{1}_{n\times n} \\
     \hline
     -\mathbf{1}_{n\times n} & \mathbf{0}_{n\times n} \
  \end{array}\right)
\label{omegamatra}
\end{equation}
we denote the antisymmetric symplectic metric in $2n$ dimensions,
$n=n_V$ being the number of vector fields in $D=4$, as we have
already stressed. The symplectic character of the representation
$\mathbf{W}$ is asserted by the identity:
\begin{equation}
  \Lambda^A\, \Omega + \Omega\, \left( \Lambda^A \right )^T = 0
\label{Lamsymp}
\end{equation}
The fundamental doublet representation of $\mathrm{SL(2,\mathbb{R})}$
is also symplectic and we have denoted by $\omega^{ij}= \left( \begin{array}{cc}
  0 & 1 \\
  -1 & 0
\end{array}\right) $ the
$2$-dimensional symplectic metric, so that:
\begin{equation}
    \lambda^x\, \omega + \omega\, \left( \lambda^x \right )^T = 0,
\label{lamsymp}
\end{equation}
The matrices
$\left(K_A\right)^{\alpha\beta}=\left(K_A\right)^{\beta\alpha}$
and $\left(k_x\right)^{ij}=\left(k_y\right)^{ji}$ are just
symmetric matrices in one-to-one correspondence with the
generators of $\mathbb{U}_{D=4}$ and $\mathrm{SL(2,\mathbb{R})}$,
respectively. Implementing Jacobi identities, however, we find the
following relations:
\begin{eqnarray}
  && \nonumber K_A\Lambda^C +
\Lambda^C K_A = f^{BC}_{\phantom{BC}A}K_B, \quad k_x\lambda^y + \lambda^y k_x
= f^{yz}_{\phantom{yz}x}k_z,
\label{jacobrele}
\end{eqnarray}
which admit the unique solution:
\begin{equation}
    K_A = \alpha \, \mathbf{g}_{AB} \,\Lambda^B\Omega, \quad ; \quad k_x
= \beta \, \mathbf{g}_{xy} \, \lambda^y \omega
\label{uniquesolutK&k}
\end{equation}
where $\mathbf{g}_{AB}$, $\mathbf{g}_{xy}$ are the Cartan-Killing
metrics on the algebras $\mathbb{U}_{D=4}$ and
$\mathrm{SL(2,\mathbb{R})}$, respectively, and  $\alpha$ and
$\beta$ are two arbitrary constants. These latter can always be
reabsorbed into the normalization of the generators $W^{i\alpha}$
and correspondingly set to one. Hence the algebra
(\ref{genGD3pre}) can always be put into the following elegant
form:
\begin{eqnarray}
\label{genGD3}
\nonumber && [T^A,T^B] = f^{AB}_{\phantom{AB}C} \, T^C  \\
\nonumber && [L^x,L^y] = f^{xy}_{\phantom{xy}z} \, L^z , \\ &&\nonumber
[T^A,W^{i\alpha}] = (\Lambda^A)^\alpha_{\,\,\,\beta} \, W^{i\beta},
\\ && [L^x, W^{i\alpha}] = (\lambda^x)^i_{\,\, j}\, W^{j\alpha}, \\
\nonumber &&[W^{i\alpha},W^{j\beta}] = \omega^{ij}\,
(\Lambda_A)^{\alpha\beta}\, T^A + \, \Omega^{\alpha\beta}\,
\lambda_x^{ij}\, L^x
\end{eqnarray}
where we have used the convention that symplectic indices ($\alpha, i$) are raised
and lowered with the symplectic metrics ($\Omega$ and $\omega$), while adjoint representation
indices ($A, x$) are raised and lowered with the eucledian Cartan-Killing metrics.
\subsection{Affine Ka\v c--Moody extension of $\mathbb{U}_{D=3}$ from an
algebraic viewpoint}
Any simple Lie algebra admits an affine Ka\v c--Moody extension. From
a purely algebraic point of view let us discuss the affine extension
of the $\mathbb{U}_{D=3}$ algebra in the presentation provided by
its decomposition with respect to its $\mathbb{U}_{D=4}$--subalgebra, namely
in the presentation given by eq.s (\ref{genGD3}) which are well adapted to
the Ehlers reduction, as we have already stressed. This will be a
preparatory study for the argument we shall develop after presenting
the two dimensional reductions.
In full generality we can write the ansatz:
\begin{eqnarray}
\label{KacMoodyD2}
\nonumber && [T_n^A,T_m^B] = f^{AB}_CT^C_{n+m} +
c_1\delta^{AB}n\delta_{m+n,0}, \\
\nonumber && [L_n^x,L_m^y] = f^{xy}_zL^z_{m+n} +
c_2\delta^{xy}n\delta_{m+n,0}, \\ &&\nonumber [T^A_m,W_n^{i\alpha}]
= (\Lambda^A)^\alpha_\beta W^{i\beta}_{n+m}, \\ && [L_m^x,
W^{i\alpha}_n] = (\lambda^x)^i_jW^{j\alpha}_{m+n}, \\ \nonumber
&&[W^{i\alpha}_n,W^{j\beta}_m] =
\omega^{ij}(\Lambda_A)^{\alpha\beta}T^A_{m+n} +
\Omega^{\alpha\beta}\lambda_x^{ij}L^x_{n+m} +
c_3\omega^{ij}\Omega^{\alpha\beta}n\delta_{n+m,0}
\end{eqnarray}
where $c_{1,2,3}$ are three apparently different central charges.
Implementation of the Jacobi identities immediately shows that these
charges are actually related and there is just one independent charge
$c$.
Explicitly we find the relations:
\begin{equation}
  c_1 \, =\,c_2 \, =\,   c_3 \, =\,  c
\label{relationscharge}
\end{equation}
\par
In eq.(\ref{KacMoodyD2} ) we wrote the affine extension of the
$\mathbb{{U}}_{D=3}$ duality algebra in a compact notation that
emphasizes the role of those generators which are associated with
the dimensional reduction of vector fields. We shall shortly from
now see the relevance of this presentation in order to discuss the
merging of Ehlers symmetries with those of the Matzner--Misner
reduction. For other purposes it is now convenient to fix our
notations for affine Ka\v c--Moody algebras in the Weyl-Dynkin
basis. In this case the generators will be denoted as
$\{\mathcal{H}^i_n,E^\alpha_n\}$ and the relevant commutation
relations are:
\begin{eqnarray}
&& [\mathcal{H}^i_n,\mathcal{H}^j_m] = \frac{1}{2}k\cdot c_g(R)\delta^{ij}\cdot n \delta_{n+m,0}, \nonumber \\
&& [\mathcal{H}^i_n,E^\alpha_m] = \alpha^i\cdot E^\alpha_{n+m} , \nonumber \\
&& [E^\alpha_n,E^\beta_m] = N_{\alpha\beta}E^{\alpha+\beta}_{n+m}, \nonumber \\
&& [E^\alpha_n,E^{-\alpha}_m] = \alpha^i\cdot \mathcal{H}^i_{n+m}
+ \frac{1}{2}k\cdot c_g(R)n\delta_{n+m,0}
\label{KacMWD}
\end{eqnarray}
where $c_g(R)$ is the value of the 1$s$t Casimir operator in the adjoint representation.
\par
It is also useful to introduce the extra generator $d$, which
measures the level
\begin{equation}
\left [d \, ,\, \mathcal{H}^i_n \right] = n \, \mathcal{H}^i_n \quad ; \quad \left [ d \, , \, E^\alpha_n \right ] =
n \, E^\alpha_n
\label{dgradus}
\end{equation}
The Cartan subalgebra of the affine Ka\v c--Moody algebra ${\hat \mathbb{G}}$ consists of the following generators $\hat{C} =
\left \{\mathcal{H}^i_0,k,d \right\}$, where $k$ is the central element. In this way the roots are now vectors
with $r+2$ components, $r$ being the rank of the simple Lie algebra that we extend and they form an infinite set $\Delta$.
The set of positive roots $\Delta^+$ is composed of three type of roots
\begin{equation}
0 \, < \, \hat{\alpha} =\cases{ (\alpha, 0,0) \, \quad \alpha > 0 \mbox{ as a root of $\mathbb{G}$} \cr
 (\alpha,0,n) \, \quad n>0 \mbox{ for both $\alpha >0$ and $\alpha < 0$ as roots of $\mathbb{G}$
 }\cr
(0,0,n) \, \quad n>0\cr}
\label{positiveroots}
\end{equation}
which can be expressed as integer non negative linear combinations of a set of $r+1$ simple positive roots. For these latter we take
\begin{equation}
\hat{\alpha}^i = \left (\alpha^i, 0,0 \right ) \, ; \, \quad \hat{\alpha}^0 = \left (-\psi,0,1\right )
\label{simplerutte}
\end{equation}
where $\alpha^i$ are the simple roots of $\mathbb{G}$ and $\psi$
denotes the highest root also of $\mathbb{G}$.
The invariant bilinear form  on the CSA and hence on the root space
$\hat{\alpha}=\left (\alpha, n , m \right)$ has a
Lorentzian signature and it is given by
\begin{equation}
<\hat{\alpha_1},\hat{\alpha_2}> = <\alpha_1,\alpha_2> + n_1  m_2  +
m_1 n_2
\label{scalarproduct}
\end{equation}
Considering now eq.s (\ref{KacMoodyD2}) we see that $\mathbb{U}_{D=3}^\wedge$, namely the affine extension
of $\mathbb{U}_{D=3}$ contains, as a subalgebra, $A^\wedge_1$, namely
the affine extension of $\mathrm{SL(2,\mathbb{R})_E}$. This is
evident from the second of eq.s (\ref{KacMoodyD2}) and plays an
important role in our argument. To this effect let us focus on an
algebra $A^\wedge_1$ and write it in the Weyl-Dynkin basis as in eq.s
(\ref{KacMWD}). It takes the form:
\begin{eqnarray}
&& [\mathcal{H}_n,\mathcal{H}_m] =\ft 12 c_{A_1} \cdot k\cdot n \, \delta_{n+m,0}\nonumber \\
&& [\mathcal{H}_n,E^\pm_m] =\pm \sqrt{2} \,  E^\pm_{n+m} , \nonumber \\
&& [E^+_n,E^-_m] = \sqrt{2} \, \mathcal{H}_{n+m}
+ \, \ft 12 c_{A_1} \cdot k\cdot n \, \delta_{n+m,0}\nonumber\\
&& [E^\pm_n,E^\pm_m] = 0
\label{hatA1}
\end{eqnarray}
where $c_{A_1}$ is the quadratic Casimir of the Lie algebra $A_1$ in
the adjoint representation.
Next let us observe that the infinite dimensional $A_1^\wedge$
algebra contains not just one but several $A_1\equiv \mathrm{SL(2,\mathbb{R})}$ subalgebras whose
standard commutation relation have originated the KM-extension
(\ref{hatA1}) and are:
\begin{eqnarray}
&& [L_0,L_\pm] =\pm \sqrt{2} \,  L_\pm  , \nonumber \\
&& [L_+,L_-] = \sqrt{2} \, L_0
\label{A1}
\end{eqnarray}
One $A_1$ subalgebra is the obvious one obtained by taking all level zero generators, namely by setting:
\begin{equation}
  L_0 = \mathcal{H}_0 \quad ; \quad L_\pm = E^\pm_0
\label{levelzeroA1}
\end{equation}
yet it must be realized that eq.(\ref{levelzeroA1}) is just one instance in an infinite family of $A_1$
subalgebras obtained by setting:
\begin{equation}
  L_0^{[m]} = \mathcal{H}_0 -m \, \frac{1}{2\sqrt{2}} \, c_{A_1} \,k
  \quad ; \quad L_\pm^{[m]} =E^\pm_{\mp m}
\label{infinitefam}
\end{equation}
Secondly let us observe that by using two distinct elements in this
infinite family of $A_1$ subalgebras we can list six generators that
provide a standard Chevalley-Serre presentation of the entire affine Ka\v c--Moody algebra
$A_1^\wedge$. Let us recall the concept of Chevalley-Serre
presentation. This is the analogue for Lie algebras of the
presentation of discrete groups through generators and relations.
Given a simple Lie algebra of rank $r$ defined by its Cartan matrix
$C_{ij}$, a Chevalley-Serre basis is given by $r$-triplets of generators:
\begin{equation}
  \left( h_i \, , \, e_i \, , \, f_i \right)  \quad ; \quad i=1,\dots \, r
\label{ChSerretriplets}
\end{equation}
such that the following commutation relations are satisfied:
\begin{eqnarray}
\left[ h_i \, , \, h_j\right]  & = & 0 \nonumber\\
\left[ h_i \, , \, e_j\right] & = & C_{ij} \, e_j \nonumber\\
\left[ h_i \, , \, f_j\right] & = & -C_{ij} \, f_j  \nonumber\\
\left[ e_i \, , \, f_j\right] & = & \delta_{ij} \, h_i \nonumber\\
 \mbox{adj $[e_i]$}^{(C_{ji}+1)}\left( e_j\right)   & = & 0 \nonumber\\
 \mbox{adj $[f_i]$}^{(C_{ji}+1)}\left( f_j\right)   & = & 0
\label{RelazieCS}
\end{eqnarray}
When such $r$--triplets are given the entire algebra is defined.
Indeed all the other generators are constructed by commuting these ones
modulo the relations (\ref{RelazieCS}). For simply-laced finite
simple Lie algebras a Chevalley basis is easily constructed in terms
of simple roots. Let $\alpha_i$ denote the simple roots, then it
suffices to set:
\begin{equation}
    \left( h_i \, , \, e_i \, , \, f_i \right) = \left( H_{\alpha_i} \, , \, E^{\alpha_i} \, , \, E^{-\alpha_i} \right)
\label{Csexemplo}
\end{equation}
where $H_{\alpha_i} \equiv \alpha_i \cdot H $ are the Cartan generator
associated with the simple roots and $E^{\pm \alpha_i}$ are the step
operators respectively associated with the simple roots and their
negative.
\par
The Cartan matrix of the affine algebra $A^\wedge_1$ is:
\begin{equation}
  C^{A^\wedge_1}_{ij} = \left(\begin{array}{cc}
     2 & -2 \\
     -2 & 2 \
  \end{array}  \right)
\label{A1WCarta}
\end{equation}
and, as noted by Nicolai \cite{NicolaiEHMM} time ago a
Chevalley-Serre basis for this algebra is provided by setting:
\begin{eqnarray}
\left( h_1 \, , \, e_1 \, , \, f_1 \right) & = & \left( \sqrt{2} L_0^{[0]} \, , \, L_+^{[0]} \, , \,  L_-^{[0]} \right)
\nonumber\\
\left( h_0 \, , \, e_0 \, , \, f_0 \right) & = & \left( -\sqrt{2} L_0^{[m]} \, , \, L_-^{[m]} \, , \,  L_+^{[m]}
\right) \quad ; \quad \mbox{for any choice of $0 \ne m\in \mathbb{Z}$}
\label{explicitCS}
\end{eqnarray}
This observation has  far reaching consequences in field--theory.
Suppose that we have two formally identical lagrangians $\mathcal{L}_1$, $\mathcal{L}_2 $ each with a
locally realized global symmetry $\mathrm{SL(2,\mathbb{R})}$, that we distinguish as $\mathrm{SL}(2,\mathbb{R})_{1}$ and $\mathrm{SL}(2,\mathbb{R})_2$, respectively. Let us
suppose furthermore that the local fields $\phi_{[1]}^i$ appearing
in the first Lagrangian have a non local (invertible) relation to the fields
$\phi_{[2]}^i$ of the second lagrangian.  Schematically we denote the
transformation from one set of fields to the others as the action of
a non local operator $ \mathcal{T}$:
\begin{eqnarray}
  {\mathcal{T}} & : & \phi_{[1]}^i(x) \, \mapsto \, \phi_{[2]}^i(x)
  \nonumber\\
 {\mathcal{T}^{-1}} & : & \phi_{[2]}^i(x) \, \mapsto \, \phi_{[1]}^i(x)
\label{Toperation}
\end{eqnarray}
Then, by use of $ \mathcal{T}$ and allowing also non local
transformations, we can define on the same set of fields, say $\phi_{[1]}^i(x)$,  two
sets of $\mathrm{SL(2,\mathbb{R})}$ transformations, the original
ones and also those associated with the second Lagrangian. Indeed if
$\delta_2 \phi_{[2]}(x)$ is a local $\mathrm{SL(2,\mathbb{R})_{2}}$ transformation of
the fields $\phi_{[2]}(x)$, we can define:
\begin{equation}
  \overline{\delta_2} \, \phi_{[1]}^i(x) \equiv {\mathcal{T}^{-1}} \, \delta_2
  \, {\mathcal{T}} \, \phi_{[1]}^i(x)
\label{nonlocal}
\end{equation}
which will be a non local transformation. In this way we can
introduce a set of six generators defined as follows:
\begin{eqnarray}
\left( h_1 \, , \, e_1 \, , \, f_1 \right) & = & \left( \sqrt{2} L_0 \, , \, L_+ \, , \,  L_- \right)
\nonumber\\
\left( h_0 \, , \, e_0 \, , \, f_0 \right) & = & \left( -\sqrt{2}  \, {\mathcal{T}^{-1}}   L_0   {\mathcal{T}}
\, , \, {\mathcal{T}^{-1}}  L_-   {\mathcal{T}} \, , \,  {\mathcal{T}^{-1}}  L_+   {\mathcal{T}}
\right)
\label{Csanstaz}
\end{eqnarray}
The names given to the generators anticipate that they might
constitute a Chevalley-Serre basis. In order for this to be true,
they should close the commutation relations (\ref{RelazieCS}).
This is not obvious a priori but it can be explicitly checked in
concrete field-theoretical models. The verification of
(\ref{RelazieCS}) is the necessary and sufficient condition to
prove that the system described by the two lagrangians admits the
full Ka\v c--Moody algebra as a symmetry. This is what Nicolai did
in the case of pure gravity \cite{NicolaiEHMM} by showing that the
two dimensional reduction schemes, Ehlers and Matzner--Misner,
lead to two $2D$--gravity coupled $\sigma$--models with target
manifold $\mathrm{SL(2,\mathbb{R})/O(2)}$, which exactly realize
the situation described above. He also advocated the same argument
to argue that in the case of maximal supersymmetry we have
$E_{8(8)}^\wedge \equiv E_{9(9)}$, but a systematic analysis of
the mechanism of generation of infinite symmetries was not
performed so far for generic supergravities. Our paper aims at
filling such a gap, clarifying also the relation with billiard
dynamics. As we are going to see, things change somewhat in the
case of generic supergravities since the two lagrangians
$\mathcal{L}^{(E)}$ and $\mathcal{L}^{(MM)}$, respectively
produced by the Ehlers and by the Matzner--Misner reduction, do
not have two identical copies of the same symmetry algebra. The Ehlers lagrangian has symmetry algebra $\mathbb{U}_{D=3}$ and it is extended with the 
$\mathrm{SL}(2\mathbb{R})_{MM}$ part of the Matzner--Misner symmetry (coming from the Einstein gravity), that gives the affine extension
 $\mathbb{U}_{D=3}^\wedge$.
\subsubsection{Systematics of the affine extension}\label{systematics}
Let us now consider in more detail the structure of the decomposition
(\ref{gendecompo}) which is the crucial ingredient for the affine
(and the hyperbolic) extension. The various cases corresponding to the
various values of $N_Q$ are listed in table \ref{tavolona} which
contains also more information, namely the completely split \textit{Tits
Satake subalgebra} of each duality algebra $\mathbb{U}_D$. This is
relevant for the discussion of the billiard phenomenon   as we are
going to discuss extensively in a forthcoming paper \cite{noiTSpape}
and just touch upon later on in the present one.
\par
As we recalled in eq.(\ref{simplerutte}), crucial for the affine
extension $ \mathbb{G}^\wedge $ of any simple Lie algebra $\mathbb{G}$ is the
highest root $\psi$ of this latter, since it is by means of $\psi$  that
we write the additional affine root $\alpha_0$ and correspondingly a
Chevalley-Serre basis. Indeed, in view, of eq.(\ref{simplerutte}) the
Chevalley-Serre basis of any $\mathbb{G}^\wedge$, which extends that
of $\mathbb{G}$ displayed in eq.(\ref{Csexemplo}) is the following
one:
\begin{eqnarray}
    \left( h_i \, , \, e_i \, , \, f_i \right) & = & \left( H_{\alpha_i} \, , \, E^{\alpha_i} \, , \,
    E^{-\alpha_i}
    \right) \quad ; \quad \left ( i=1, \dots , r \right)\nonumber\\
    \left( h_0 \, , \, e_0 \, , \, f_0 \right) & = & \left( H_{\alpha_0} \, , \, E^{\alpha_0} \, , \,
    E^{-\alpha_0}
    \right) \nonumber\\
  \null & = & \left(\ft 12 c_g \, k  -H_{\psi} \, , \, E^{-\psi}_1 \, , \,
    E^{\psi}_{-1}
    \right)
\label{Csexemplo2}
\end{eqnarray}
From the algebraic view-point a crucial property of the general
decomposition in eq.(\ref{gendecompo}) is encoded into the
following statements which are true for all the cases \footnote{An
apparent exception is given by the case of N=3 supergravity of
which we shall give a short separate discussion. The extra
complicacy, there, is that the duality algebra in $D=3$, namely
$\mathbb{U}_{D=3}$ has rank $r+2$, rather than $r+1$ with respect
to the naive algebra $\mathbb{U}_{D=4}$. By this latter we mean
the isometry algebra of the scalar manifold in $D=4$ supergravity.
Actually in this case there is an extra $\mathrm{U(1)_Z}$ factor
that is active on the vectors, but not on the scalars and which is
responsible for the additional complications. Indeed it happens in
this case that there are two vector roots, one for the complex
representation to which the vectors are assigned and one for its
conjugate and the phase group and they have opposite charges under
$\mathrm{U(1)_Z}$. }:
\begin{enumerate}
  \item The  $A_1$ root-system associated with the
  $\mathrm{SL(2,\mathbb{R})_E}$ algebra in the decomposition
  (\ref{gendecompo}) is made of $\pm \, \psi$ where $\psi$ is the
  highest root of $\mathbb{U}_{D=3}$.
  \item Out of the $r$ simple roots $\alpha_i$ of $\mathbb{U}_{D=3}$
  there are $r-1$ that have grading zero with respect to $\psi$ and
  just one $\alpha_{W}$ that has grading $1$:
\begin{eqnarray}
  \left ( \psi \, ,\, \alpha_i\right ) & = & 0 \quad \quad i \ne W
  \nonumber\\
\left ( \psi \, ,\, \alpha_W\right ) & = & 1
\label{bingopsi}
\end{eqnarray}
  \item The only simple root $\alpha_W$ that has non vanishing grading with
  respect $\psi$ is just the highest weight of the symplectic
  representation $\mathbf{W}$ of $\mathbb{U}_{D=4}$ to which the vector fields
  are assigned.
  \item The Dynkin diagram of $\mathbb{U}_{D=4}$ is obtained from that of
  $\mathbb{U}_{D=3}$ by removing the dot corresponding to the special
  root $\alpha_W$.
  \item Hence we can arrange a basis   for the simple roots
  of the rank $r$ algebra $\mathbb{U}_{D=3}$ such that:
\begin{equation}
  \begin{array}{rcl}
     \alpha_i & = & \left\{ \overline{\alpha}_i , 0 \right\}  \quad ; \quad i \ne W\\
      \alpha_W & = & \left\{ \overline{\mathbf{w}}_h , \frac{1}{\sqrt{2}} \right\}   \\
     \psi  & = & \left\{  \mathbf{0}  , {\sqrt{2}} \right\} \
  \end{array}
\label{lucullus}
\end{equation}
where $\overline{\alpha}_i$ are $(r-1)$--component vectors representing
a basis of simple roots for the Lie algebra $\mathbb{U}_{D=4}$,
$\overline{\mathbf{w}}_h$ is also an $(r-1)$--vector representing the \textit{highest
weight} of the representation $\mathbf{W}$.
\end{enumerate}
The above properties imply that the Dynkin diagram of
$\mathbb{U}^\wedge_{D=4}$ is just obtained by attaching $\alpha_0$
with a single line to $\alpha_W$. Furthermore, from the point of
view of the Chevalley-Serre presentation, any triplet $(h_0, e_0,
f_0)$ which, added to the generators of
$\mathrm{SL(2,\mathbb{R})_E}$ promotes this latter to its affine
extension, automatically promotes the entire $\mathbb{U}_{D=3}$ to
its own affine extension. This is so because the root of
$\mathrm{SL(2,R)_E}$ is the highest root of $\mathbb{U}_{D=3}$.
From the field-theory view point this is just what happens.
Indeed, as we prove in next sections, in the Matzner--Misner
reduction we obtain $\mathrm{SL(2,\mathbb{R})_{MM}}$ which yields
the affine extension of $\mathrm{SL(2,\mathbb{R})_E}$ and as a
consequence of the full $\mathbb{U}_{D=3}$.
\par
Before considering the field theoretic realization, let us conclude
this section by discussing the various instances of
eq.(\ref{gendecompo}) in some detail, by making reference to table
(\ref{tavolona}).
\vskip 0.2cm
\begin{table}
  \centering
  {\scriptsize \begin{tabular}{|c|c|c|c|c|c|}
\hline
 \# Q.s & & D=4 & D=3 & D=2 & D=1 \\
  \hline
 $\mathcal{N}=8$ & $\mathbb{U}$ & $\mathrm{E_{7(7)}}$ & $\mathrm{E_{8(8)}}$ &
 $\mathrm{E_9}$ & $\mathrm{E_{10}}$ \\
 & $\mathbb{H}$ & $\mathrm{SU}(8)$ & $\mathrm{SO}(16)$ & $\mathrm{KE}_9$ & $\mathrm{KE}_{10}$\\ \hline
 $\mathcal{N}=6$ & $\mathbb{U}$ & $\mathrm{SO^\star(12)}$ & $\mathrm{E_{7(-5)}}$ & $\mathrm{E_{7(-5)}}^\wedge$ &
 $\mathrm{E_{7(-5)}}^{\wedge\wedge}$ \\  & $\mathbb{H}$ & $\mathrm{SU(6) \times U(1)}$ & $\mathrm{SO(12) \times SO(3)}$ &
 $\mathrm{KE}_{7(-5)}^\wedge$ &  $\mathrm{KE}_{7(-5)}^{\wedge\wedge}$ \\ \cline{2-6} & $\mathbb{U}^{TS}$ &
 $\mathrm{Sp(6,\mathbb{R})}$ & $\mathrm{F_{4(4)}}$ & $\mathrm{F_{4(4)}}^\wedge$ & $\mathrm{F_{4(4)}}^{\wedge\wedge}$ \\
& $\mathbb{H}^{TS}$ & $ \mathrm{SU(3) \times U(1)}$ & $\mathrm{Usp(6) \times SU(2)}$ &
 $\mathrm{KF}_{4(4)}^\wedge$ &$\mathrm{KF}_{4(4)}^{\wedge\wedge}$ \\ \hline
 $\mathcal{N}=5$ &  $\mathbb{U}$ & $\mathrm{SU}(5,1)$ & $\mathrm{E}_{6(-14)}$ & $\mathrm{E}_{6(-14)}^\wedge$ &
 $\mathrm{E}_{6(-14)}^{\wedge\wedge}$ \\ & $\mathbb{H}$ & $\mathrm{SU}(5)\times \mathrm{U}(1)$ &
 $\mathrm{SO}(10)\times \mathrm{SO}(2)$ & $\mathrm{KE}_{6(-14)}^\wedge$ & $\mathrm{KE}_{6(-14)}^{\wedge\wedge}$ \\
 \cline{2-6} & $\mathbb{U}^{TS}$ & $\mathrm{SU(1,1)}$ & $bc_2$ & $\mathrm{A}_{4}^{(2)}$ & $\mathrm{A}_{4}^{(2)\wedge}$ \\
 & $\mathbb{H}^{TS}$ & $\mathrm{U(1)}$ & -- & $\mathrm{KA}_{4}^{(2)}$ & $\mathrm{KA}_4^{(2)\wedge}$ \\ \hline
 $\mathcal{N}=4$ & $\mathbb{U}$ & $\mathrm{SO}(6,n)\times\mathrm{SU}(1,1)$ & $\mathrm{SO}(8,n+2)$ &
 $\mathrm{SO}(8,n+2)^\wedge$ & $\mathrm{SO}(8,n+2)^{\wedge\wedge}$ \\ & $\mathbb{H}$ &
 $\mathrm{SO}(6)\times\mathrm{SO}(n)\times\mathrm{U}(1)$ & $\mathrm{SO}(8)\times\mathrm{SO}(n+2)$ &
 $\mathrm{KSO}(8,n+2)^\wedge$ & $\mathrm{KSO}(8,n+2)^{\wedge\wedge}$ \\ \cline{2-6} $n < 6$ &
 $\mathbb{U}^{TS}$ & $\mathrm{{SO(n,n)}\times SU(1,1)}$ & $\mathrm{{SO(n+2,n+2)}}$ & $\mathrm{SO(n+2,n+2)}^\wedge$ &  $\mathrm{SO(n+2,n+2)}^{\wedge\wedge}$ \\
  & $\mathbb{H}^{TS}$ & $\mathrm{SO(n)\times SO(n) \times U(1)}$ & $\mathrm{SO(n+2)\times SO(n+2) }$ &
  $\mathrm{KSO(n+2,n+2)}^\wedge$ & $\mathrm{SO(n+2,n+2)}^{\wedge\wedge}$ \\ \hline
 $\mathcal{N}=4$ & $\mathbb{U}$ & $\mathrm{SO}(6,n)\times\mathrm{SU}(1,1)$ & $\mathrm{SO}(8,8)$ &
 $\mathrm{SO}(8,8)^\wedge$ & $\mathrm{SO}(8,8)^{\wedge\wedge}$ \\ $n=6$ & $\mathbb{H}$ &
 $\mathrm{SO}(6)\times\mathrm{SO}(6)\times\mathrm{U}(1)$ & $\mathrm{SO}(8)\times\mathrm{SO}(8)$ &
 $\mathrm{KSO}(8,8)^\wedge$ & $\mathrm{KSO}(8,8)^{\wedge\wedge}$ \\ \hline
 $\mathcal{N}=4$ & $\mathbb{U}$ & $\mathrm{SO}(6,n)\times\mathrm{SU}(1,1)$ & $\mathrm{SO}(8,n+2)$ &
 $\mathrm{SO}(8,n+2)^\wedge$ & $\mathrm{SO}(8,n+2)^{\wedge\wedge}$ \\ & $\mathbb{H}$ &
 $\mathrm{SO}(6)\times\mathrm{SO}(n)\times\mathrm{U}(1)$ & $\mathrm{SO}(8)\times\mathrm{SO}(n+2)$ &
 $\mathrm{KSO}(8,n+2)^\wedge$ & $\mathrm{KSO}(8,n+2)^{\wedge\wedge}$ \\ \cline{2-6} $n > 6$ &
 $\mathbb{U}^{TS}$ & $\mathrm{SO(6,6) \times SU(1,1)}$ & $\mathrm{SO(8,8)}$ & $\mathrm{SO(8,8)}^\wedge$ &  $\mathrm{SO(8,8)}^{\wedge\wedge}$ \\
  & $\mathbb{H}^{TS}$ & $\mathrm{SO(6) \times SO(6) \times U(1)} $ & $\mathrm{SO(8) \times SO(8)} $ & $\mathrm{KSO(8,8)}^\wedge$ & $\mathrm{KSO(8,8)}^{\wedge\wedge}$  \\ \hline
  $\mathcal{N}=3$ &  $\mathbb{U}$ & $\mathrm{SU}(3,n)$ & $\mathrm{SU}(4,n+1)$ & $\mathrm{SU}(4,n+1)^\wedge$
  & $\mathrm{SU}(4,n+1)^{\wedge\wedge}$ \\ & $\mathbb{H}$ & $\mathrm{SU(3) \times SU(n) \times U(1)}$ & $\mathrm{SU(4) \times SU(n+1) \times U(1)}$
  & $\mathrm{KSU}(4,n+1)^\wedge$ &
  $\mathrm{KSU}(4,n+1)^{\wedge\wedge}$ \\
  \hline  $\mathcal{N}=2$ & geom. & $\mathcal{SK}$ & $\mathcal{Q}$ & $\mathcal{Q}^\wedge$ &
  $\mathcal{Q}^{\wedge\wedge}$ \\ \cline{2-6} & $\mathrm{TS}[\mbox{geom.}]$ & $\mathrm{TS}[\mathcal{SK}]$&
  $\mathrm{TS}[\mathcal{Q}]$ & $\mathrm{TS}[\mathcal{Q}^\wedge]$ & $\mathrm{TS}[\mathcal{Q}^{\wedge\wedge}]$ \\ \hline
\end{tabular}
}
  \caption{In this table we present the duality algebras $\mathbb{U}_D$ in $D=4,3,2,1$, for various
  values of the number of supersymmetry charges. We also mention the corresponding Tits Satake projected algebras
    (where they are well defined) that are relevant
  for the discussion of the cosmic billiard dynamics}\label{tavolona}
\end{table}
\vskip 0.2cm
\paragraph{N=8}
\vskip 0.2cm
This is the case of maximal supersymmetry and it is illustrated by
fig. \ref{except2}.
\par
In this case all the involved Lie algebras are maximally split and we
have
\begin{equation}
  \mbox{adj} \, \mathrm{E_{8(8)}} = \mbox{adj} \, \mathrm{E_{7(7)}} \oplus
  \mbox{adj} \, \mathrm{SL(2,\mathbb{R})_E} \oplus \left( \mathbf{2},\mathbf{56}\right)
\label{urgeE8}
\end{equation}
The highest root of $\mathrm{E_{8(8)}}$ is
\begin{equation}
  \psi = 3\alpha_1 +4\alpha_2 +5\alpha_3+6\alpha_4 +3\alpha_5+4\alpha_6
+2\alpha_7 +2\alpha_8
\label{highE8}
\end{equation}
and the unique simple root not orthogonal to $\psi$ is $\alpha_8 =
\alpha_W$, according to the labeling of roots as in fig.
\ref{except2}. This root is the highest weight of the fundamental
$\mathbf{56}$-representation  of $E_{7(7)}$. As a consequence of this
the affine extension of $E_{8(8)}$ has the same Dynkin diagram as it
would have $E_{9(9)}$ formally continuing the $E_{r(r)}$ series to
$r>8$.
\begin{figure}
\centering
\begin{picture}(100,70)
      \put (-70,35){$E_8$} \put (-20,35){\circle {10}} \put
(-23,20){$\alpha_7$} \put (-15,35){\line (1,0){20}} \put
(10,35){\circle {10}} \put (7,20){$\alpha_6$} \put (15,35){\line
(1,0){20}} \put (40,35){\circle {10}} \put (37,20){$\alpha_4$}
\put (40,65){\circle {10}} \put (48,62.8){$\alpha_5$} \put
(40,40){\line (0,1){20}} \put (45,35){\line (1,0){20}} \put
(70,35){\circle {10}} \put (67,20){$\alpha_{3}$} \put
(75,35){\line (1,0){20}} \put (100,35){\circle {10}} \put
(97,20){$\alpha_{2}$} \put (105,35){\line (1,0){20}} \put
(130,35){\circle {10}} \put (127,20){$\alpha_1$} \put
(135,35){\line (1,0){20}} \put (160,35){\circle{10}}\put (160,35){\circle{9}}
\put (160,35){\circle{8}}\put (160,35){\circle{7}}\put (160,35){\circle{6}}
\put (160,35){\circle{5}}\put (160,35){\circle{4}} \put (160,35){\circle{3}}
\put (160,35){\circle{2}}\put (160,35){\circle{1}}\put
(157,20){$\alpha_8$}
\end{picture}
$$\begin{array}{l}\psi = 3\alpha_1 +4\alpha_2 +5\alpha_3+6\alpha_4 +3\alpha_5+4\alpha_6
+2\alpha_7 +2\alpha_8 \\(\psi \, , \,\alpha_8) = 1 \quad; \quad  (\psi \, , \, \alpha_i ) = 0 \quad i \ne 8 \
\end{array}$$
\begin{picture}(100,70)
      \put (-70,35){$E_9$} \put (-20,35){\circle {10}} \put
(-23,20){$\alpha_7$} \put (-15,35){\line (1,0){20}} \put
(10,35){\circle {10}} \put (7,20){$\alpha_6$} \put (15,35){\line
(1,0){20}} \put (40,35){\circle {10}} \put (37,20){$\alpha_4$}
\put (40,65){\circle {10}} \put (48,62.8){$\alpha_5$} \put
(40,40){\line (0,1){20}} \put (45,35){\line (1,0){20}} \put
(70,35){\circle {10}} \put (67,20){$\alpha_{3}$} \put
(75,35){\line (1,0){20}} \put (100,35){\circle {10}} \put
(97,20){$\alpha_{2}$} \put (105,35){\line (1,0){20}} \put
(130,35){\circle {10}} \put (127,20){$\alpha_1$} \put
(135,35){\line (1,0){20}} \put (160,35){\circle{10}}\put (160,35){\circle{9}}\put (160,35){\circle{8}}
\put (160,35){\circle{7}}\put (160,35){\circle{6}}
\put (160,35){\circle{5}}\put (160,35){\circle{4}} \put (160,35){\circle{3}}\put (160,35){\circle{2}}
\put (160,35){\circle{1}}\put
(157,20){$\alpha_8$}\put
(165,35){\line (1,0){20}}\put (190,35){\circle {10}}\put
(187,20){$\alpha_0$}
\end{picture}
\vskip 1cm \caption{The Dynkin diagram of $E_{8(8)}$ and of its
affine Ka\v c --Moody extension $E_{8(8)}^\wedge =E_9$. The only
simple root which has grading one with respect to the highest root
$\psi$ is $\alpha_8$ (painted black). With respect to the algebra
$\mathbb{U}_{D=4}=E_{7(7)}$ whose Dynkin diagram is obtained by
removal of the black circle, $\alpha_8$ is the highest weight of
the symplectic representation of the vector fields, namely
$\mathbf{W}=\mathbf{56}$. The affine extension is originated by
attaching the extra root $\alpha_0$ to the root corresponding to
the vector fields.} \label{except2}
\end{figure}
\par
The well adapted basis of simple  $E_{8}$  roots is
constructed as follows:
\begin{equation}
\begin{array}{rclcl}
\alpha_1 & = & \{ 1,-1,0,0,0,0,0,0\} & = &
\{\overline{\alpha}_1,0\} \\
\alpha_2  & = & \{ 0,1,-1,0,0,0,0,0\}& = &
\{\overline{\alpha}_2,0\} \\
\alpha_3  & = & \{ 0,0,1,-1,0,0,0,0\} & = &
\{\overline{\alpha}_3,0\}\\
\alpha_4  & = & \{ 0,0,0,1,-1,0,0,0\} & = &
\{\overline{\alpha}_4,0\}\\
\alpha_5  & = & \{ 0,0,0,0,1,-1,0,0\}& = &
\{\overline{\alpha}_5,0\} \\
\alpha_6  & = & \{ 0,0,0,0,1,1,0,0\}& = &
\{\overline{\alpha}_6,0\} \\
\alpha_7  & = & \{ - \frac{1}{2}  ,
  - \frac{1}{2}  ,
  - \frac{1}{2}  ,
  - \frac{1}{2}  ,
  - \frac{1}{2}  ,
  - \frac{1}{2}  ,
  \frac{1}{{\sqrt{2}}},0\}& = &
\{\overline{\alpha}_7, 0\}\\
\alpha_8  & = & \{ - 1,0,0,
  0,0,0,- \frac{1}{{\sqrt{2}}} , \frac{1}{{\sqrt{2}}}   \}
  & = & \{ \mathbf{w}_h,\frac{1}{{\sqrt{2}}}\}\
  \end{array}
\label{alfeE8}
\end{equation}
In this basis we recognize that the seven $7$-vectors
$\bar{\alpha}_i$ constitute a simple root basis for the $E_7$ root
system, while:
\begin{equation}
 \mathbf{w}_h \, = \, \{ -1,0,0,
  0,0,0,- \frac{1}{{\sqrt{2}}} \}
\label{ierunda}
\end{equation}
is the highest weight of the fundamental $\mathbf{56}$ dimensional
representation. Finally in this basis the highest root $\psi$
defined by eq.(\ref{highE8}) takes the expected form:
\begin{equation}
  \psi = \{ 0,0,0,0,0,0,0,\sqrt{2}\}
\label{expepsi8}
\end{equation}
\paragraph{N=6}
\vskip 0.2cm
In this case the $D=4$ duality algebra is
$\mathbb{U}_{D=4}=\mathrm{SO^\star(12)}$, whose maximal compact subgroup is $\mathrm{H=SU(6) \times U(1)}$. The
scalar manifold:
\begin{equation}
  \mathcal{SK}_{N=6} \equiv \frac{\mathrm{SO^\star(12)}}{\mathrm{SU(6) \times U(1)}}
\label{SK6}
\end{equation}
is an instance of special K\"ahler manifold which can also be
utilized in an $\mathrm{N}=2$ supergravity context. The $D=3$ algebra
is just dictated by the $c$-map of homogeneous special K\"ahler
manifolds \cite{specHomgeo} which yields quaternionic manifolds.
Indeed in $D=3$ we obtain the quaternionic manifold:
\begin{equation}
  \mathcal{Q} = \frac{\mathrm{E_{7(-5)}}}{\mathrm{SO(12) \times SO(3)}}
\label{QN6}
\end{equation}
and we have $\mathbb{U}_{D=3}=E_{7(-5)}$. The $16$ vector fields
of $D=4,N=6$ supergravity with their electric and magnetic field
strengths fill the spinor representation $\mathbf{32}_s$ of
$\mathrm{SO^\star(12)}$, so that the decomposition
(\ref{gendecompo}), in this case becomes:
\begin{equation}
  \mbox{adj} \, \mathrm{E_{7(-5)}} = \mbox{adj} \, \mathrm{SO^\star(12)} \oplus
  \mbox{adj} \, \mathrm{SL(2,\mathbb{R})_E} \oplus \left( \mathbf{2},\mathbf{32}_s\right)
\label{decompoN6}
\end{equation}
\begin{figure}
\centering
\begin{picture}(100,70)
      \put (-180,35){$E_{7(-5)}$} \put (-20,35){\circle {10}} \put (-20,35){\circle {9}}
      \put (-20,35){\circle {8}}\put (-20,35){\circle {7}}
\put (-20,35){\circle {6}}\put (-20,35){\circle {5}}\put (-20,35){\circle {4}}\put (-20,35){\circle {3}}
\put (-20,35){\circle {2}}
\put (-20,35){\circle {1}}\put
(-23,20){$\alpha_7$} \put (-15,35){\line (1,0){20}} \put
(10,35){\circle {10}} \put (7,20){$\alpha_6$} \put (15,35){\line
(1,0){20}} \put (40,35){\circle {10}} \put (37,20){$\alpha_4$}
\put (40,65){\circle {10}} \put (48,62.8){$\alpha_5$} \put
(40,40){\line (0,1){20}} \put (45,35){\line (1,0){20}} \put
(70,35){\circle {10}} \put (67,20){$\alpha_{3}$} \put
(75,35){\line (1,0){20}} \put (100,35){\circle {10}} \put
(97,20){$\alpha_{2}$} \put (105,35){\line (1,0){20}} \put
(130,35){\circle {10}} \put (127,20){$\alpha_1$}
\end{picture}
$$\begin{array}{l}\psi = \alpha_1 +2\alpha_2 +3\alpha_3+4\alpha_4 +2\alpha_5+3\alpha_6
+2\alpha_7  \\(\psi \, , \,\alpha_7) = 1 \quad; \quad  (\psi \, , \, \alpha_i ) = 0 \quad i \ne 7 \
\end{array}$$
\begin{picture}(100,70)
      \put (-180,35){$E_{7(-5)}^\wedge$} \put (-53,20){$\alpha_0$}\put (-50,35){\circle {10}}
      \put(-45,35){\line(1,0){20}}\put (-20,35){\circle {10}} \put (-20,35){\circle {9}}\put (-20,35){\circle {8}}
      \put (-20,35){\circle {7}}
\put (-20,35){\circle {6}}\put (-20,35){\circle {5}}\put (-20,35){\circle {4}}\put (-20,35){\circle {3}}
\put (-20,35){\circle {2}}
\put (-20,35){\circle {1}}\put
(-23,20){$\alpha_7$} \put (-15,35){\line (1,0){20}} \put
(10,35){\circle {10}} \put (7,20){$\alpha_6$} \put (15,35){\line
(1,0){20}} \put (40,35){\circle {10}} \put (37,20){$\alpha_4$}
\put (40,65){\circle {10}} \put (48,62.8){$\alpha_5$} \put
(40,40){\line (0,1){20}} \put (45,35){\line (1,0){20}} \put
(70,35){\circle {10}} \put (67,20){$\alpha_{3}$} \put
(75,35){\line (1,0){20}} \put (100,35){\circle {10}} \put
(97,20){$\alpha_{2}$} \put (105,35){\line (1,0){20}} \put
(130,35){\circle {10}} \put (127,20){$\alpha_1$}
\end{picture}
\vskip 1cm \caption{The Dynkin diagram of $E_{7(-5)}$ and of its affine Ka\v c --Moody extension $E_{7(-5)}^\wedge$.
The only simple root which has grading one
with respect to the highest root $\psi$ is $\alpha_7$ (painted black).
With respect to the algebra $\mathbb{U}_{D=4}=\mathrm{SO^\star(12)}$ whose Dynkin diagram is
obtained by removal of the black circle, $\alpha_7$ is the highest weight of the symplectic representation of
the vector fields, namely the $\mathbf{W}=\mathbf{32}_s$. The affine
extension is originated by attaching the extra root $\alpha_0$ to the root corresponding to the vector fields.}
\label{except3}
\end{figure}
The simple root $\alpha_W$  is $\alpha_7$ and the highest root is:
\begin{equation}
\begin{array}{l}\psi = \alpha_1 +2\alpha_2 +3\alpha_3+4\alpha_4 +2\alpha_5+3\alpha_6
+2\alpha_7  \
\end{array}
\label{highestE7}
\end{equation}
Correspondingly the affine
extension is described by the Dynkin diagrams in fig.\ref{except3}.
\par
A well adapted basis of simple  $E_{7}$  roots can be
written as follows:
\begin{equation}
\begin{array}{rclcl}
\alpha_1 & = & \{ 1,-1,0,0,0,0,0\} & = &
\{\overline{\alpha}_1,0\} \\
\alpha_2  & = & \{ 0,1,-1,0,0,0,0\}& = &
\{\overline{\alpha}_2,0\} \\
\alpha_3  & = & \{ 0,0,1,-1,0,0,0\} & = &
\{\overline{\alpha}_3,0\}\\
\alpha_4  & = & \{ 0,0,0,1,-1,0,0\} & = &
\{\overline{\alpha}_4,0\}\\
\alpha_5  & = & \{ 0,0,0,0,1,-1,0\}& = &
\{\overline{\alpha}_5,0\} \\
\alpha_6  & = & \{ 0,0,0,0,1,1,0\}& = &
\{\overline{\alpha}_6,0\} \\
\alpha_7  & = & \{ - \frac{1}{2}  ,
  - \frac{1}{2}  ,
  - \frac{1}{2}  ,
  - \frac{1}{2}  ,
  - \frac{1}{2}  ,
  - \frac{1}{2}  ,
  \frac{1}{{\sqrt{2}}}\}& = &
\{\overline{\mathbf{w}}_h,\frac{1}{{\sqrt{2}}} \}\
  \end{array}
\label{alfeE7}
\end{equation}
In this basis we recognize that the six $6$-vectors
$\bar{\alpha}_i$ ($i=1,\dots,6$) constitute a simple root
basis for the $D_6\simeq \mathrm{SO^\star(12)}$ root
system, while:
\begin{equation}
 \mathbf{w}_h \, = \, \{ - \frac{1}{{{2}}}  ,-
 \frac{1}{{{2}}},- \frac{1}{{{2}}},- \frac{1}{{{2}}},
  - \frac{1}{{{2}}},- \frac{1}{{{2}}},- \frac{1}{{{2}}},
  - \frac{1}{{{2}}} \}
\label{ierunda2}
\end{equation}
is the highest weight of the spinor $\mathbf{32}$-dimensional
representation of $\mathrm{SO^\star(12)}$. Finally in this basis
the highest root $\psi$
defined by eq.(\ref{highestE7}) takes the expected form:
\begin{equation}
  \psi = \{ 0,0,0,0,0,0,\sqrt{2}\}
\label{expepsi7}
\end{equation}
\par
As we anticipated, in this case, as in most cases of lower
supersymmetry, neither the algebra $\mathbb{U}_{D=4}$ nor the
algebra $\mathbb{U}_{D=3}$ are \textbf{maximally split}. In short
this means that the non-compact rank $r_{nc} < r $ is less than
the rank of $\mathbb{U}$, namely not all the Cartan generators are
non-compact. Rigorously $r_{nc}$ is defined as follows:
\begin{equation}
  r_{nc}\, = \, \mbox{rank} \left( \mathrm{U/H}\right)  \, \equiv \, \mbox{dim} \,
  \mathcal{H}^{n.c.} \quad ; \quad \mathcal{H}^{n.c.} \, \equiv \,
  \mbox{CSA}_{\mathbb{U}(\mathbb{C})} \, \bigcap \, \mathbb{K}
\label{rncdefi}
\end{equation}
For instance in our case $r_{nc}=4$.  As we extensively discuss in
the forthcoming paper \cite{noiTSpape}, when this happens it means
that the billiard dynamics is effectively determined by a
\textit{maximally split subalgebra} $\mathbb{U}^{TS} \subset
\mathbb{U}$ named the \textit{Tits Satake} subalgebra of
$\mathbb{U}$, whose rank is equal to $r_{nc}$. Effectively determined
does not mean that the smooth billiard solutions of the big system
$\mathrm{E_{7(-5)}}/\mathrm{SO(12) \times SO(3)}$ coincide with those
of the smaller system $\mathrm{F_{4(4)}}/\mathrm{Usp(6) \times
SU(2)}$, rather it means that the former can be obtained from the
latter by means of rotations of a compact subgroup of the big algebra
$\mathrm{G_{paint} \subset U}$ which we name the \textit{paint
group}.
\begin{figure}
\centering
\begin{picture}(90,50)
      \put (-70,35){$F_4$}  \put
(10,35){\circle {10}} \put (7,20){$\varpi_4$} \put (15,35){\line
(1,0){20}} \put (40,35){\circle {10}} \put (37,20){$\varpi_3$}\put (45,38){\line (1,0){20}}
\put (55,35){\line (1,1){10}} \put (55,35){\line (1,-1){10}}\put (45,33){\line (1,0){20}} \put
(70,35){\circle {10}} \put (67,20){$\varpi_{2}$} \put
(75,35){\line (1,0){20}} \put (100,35){\circle {10}} \put (100,35){\circle {9}}\put (100,35){\circle {8}}\put (100,35){\circle {10}}\put (100,35){\circle {7}}
\put (100,35){\circle {6}}\put (100,35){\circle {5}}\put (100,35){\circle {4}}\put (100,35){\circle {3}}\put (100,35){\circle {2}}\put (100,35){\circle {1}}\put
(97,20){$\varpi_{1}$}
\end{picture}
$$ \begin{array}{l}\psi = 2\varpi_1 +3\varpi_2 +4\varpi_3+2\varpi_4   \\(\psi \, , \,\varpi_1) = 2
\quad; \quad  (\psi \, , \, \varpi_i ) = 0 \quad i \ne 1 \
\end{array}     $$
\begin{picture}(90,50)
      \put (-70,35){$F_4^\wedge$}  \put
(10,35){\circle {10}} \put (7,20){$\varpi_4$} \put (15,35){\line
(1,0){20}} \put (40,35){\circle {10}} \put (37,20){$\varpi_3$}\put (45,38){\line (1,0){20}}
\put (55,35){\line (1,1){10}} \put (55,35){\line (1,-1){10}}\put (45,33){\line (1,0){20}} \put
(70,35){\circle {10}} \put (67,20){$\varpi_{2}$} \put
(75,35){\line (1,0){20}} \put (100,35){\circle {10}} \put (100,35){\circle {9}}\put (100,35){\circle {8}}\put (100,35){\circle {10}}\put (100,35){\circle {7}}
\put (100,35){\circle {6}}\put (100,35){\circle {5}}\put (100,35){\circle {4}}\put (100,35){\circle {3}}\put (100,35){\circle {2}}\put (100,35){\circle {1}} \put
(97,20){$\varpi_{1}$}\put
(104,38){\line (1,0){20}}\put
(104,32){\line (1,0){20}}\put (130,35){\circle {10}}\put(127,20){$\varpi_{0}$}
\end{picture}
\vskip 1cm \caption{The Dynkin diagram of $F_{4(4)}$ and and of its affine Ka\v c --Moody extension $F_{4(4)}^\wedge$. The only root which is not orthogonal
to the highest root is $\varpi_V = \varpi_1$ and consequently the Dynkin diagram of $F_{4(4)}^\wedge$ is that displayed above. In the Tits Satake projection
$\Pi^{TS}$ the highest root $\psi$ of $F_{4(4)}$ is the image of the highest root of $E_{7(-5)}$ and the root $\varpi_V = \varpi_1 = \Pi^{TS} \left( \alpha_7\right)$
is the image of the root associated with the vector fields and yielding the Ka\v c--Moody extension of $E_{7(-5)}$. \label{F4dynk}}
\end{figure}
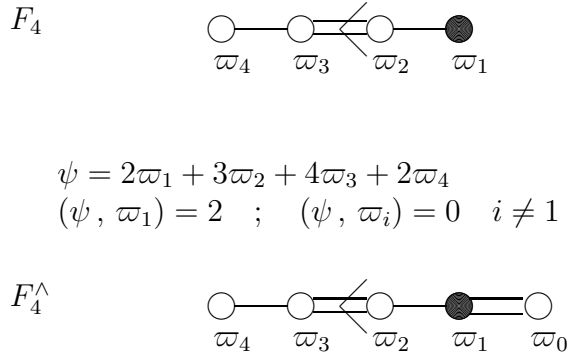
We refer for all details to the forthcoming paper \cite{noiTSpape}. Here we just emphasize a very important
fact, relevant for the affine extensions. To this effect we recall that the Tits Satake algebra is obtained from the
original algebra via a projection of the root system of $\mathbb{U}$ onto the smaller rank root system of $\mathbb{U}^{TS}$:
\begin{equation}
  \Pi^{TS} \quad ; \quad \Delta_\mathbb{U} \,\mapsto \,
  \overline{\Delta}_{\mathbb{U}^{TS}}
\label{Tsproj}
\end{equation}
In this projection the essential algebraic features of the affine
extension are preserved. In particular we have that the
decomposition (\ref{gendecompo}) commutes with the projection,
namely:
\begin{equation}
\begin{array}{rcl}
\mbox{adj}(\mathbb{U}_{D=3}) &=&
\mbox{adj}(\mathbb{U}_{D=4})\oplus\mbox{adj}(\mathrm{SL(2,\mathbb{R})_E})\oplus
W_{(2,W)} \\
\null &\Downarrow & \null\\
\mbox{adj}(\mathbb{U}^{TS}_{D=3}) &=&
\mbox{adj}(\mathbb{U}^{TS}_{D=4})\oplus\mbox{adj}(\mathrm{SL(2,\mathbb{R})_E})\oplus
W_{(2,W^{TS})} \\
\end{array}
\label{gendecompo2}
\end{equation}
In other words the projection leaves the $A_1$ Ehlers subalgebra untouched and has a non trivial effect only
on the duality algebra $\mathbb{U}_{D=4}$. Furthermore the image under the projection of the highest root
of $\mathbb{U}$ is the highest root of $\mathbb{U}^{TS}$:
\begin{equation}
  \Pi^{TS} \quad : \quad \psi \, \rightarrow \, \psi^{TS}
\label{Pionpsi}
\end{equation}
The explicit form of eq.(\ref{gendecompo2}) is the following one:
\begin{equation}
\begin{array}{rcl}
\mbox{adj}(\mathrm{E_{7(-5)}}) &=&
\mbox{adj}(\mathrm{SO^\star(12)})\oplus\mbox{adj}(\mathrm{SL(2,\mathbb{R})_E})\oplus
(\mathbf{2},\mathbf{32}_s)  \\
\null &\Downarrow & \null\\
\mbox{adj}(\mathrm{F_{4(4)}}) &=&
\mbox{adj}(\mathrm{Sp(6,\mathbb{R})}\oplus\mbox{adj}(\mathrm{SL(2,\mathbb{R})_E})\oplus
(\mathbf{2},\mathbf{14})  \
\end{array}
\label{Tsdecompo6}
\end{equation}
and the affine extension of the Tits Satake algebra $F_{4(4)}$ is described in
fig.\ref{F4dynk}. The representation $\mathbf{14}$ of
$\mathrm{Sp(6,\mathbb{R})}$ is that of an antisymmetric symplectic
traceless tensor:
\begin{equation}
\begin{array}{ccc}
 \mbox{dim}_{\mathrm{Sp(6,\mathbb{R})}} & \begin{array}{c}
  \vspace{-0.6cm} \\ \widetilde{\yng(1,1)}\
 \end{array} & \, = \, \mathbf{14}
\end{array}
\label{14repre}
\end{equation}
\begin{figure}
\centering
\begin{picture}(100,50)
      \put (-180,35){$E_{6(-14)}$} \put (-20,35){\circle {10}}
\put(-23,20){$\alpha_6$} \put (-15,35){\line (1,0){20}} \put
(10,35){\circle {10}} \put (7,20){$\alpha_5$} \put (15,35){\line
(1,0){20}} \put (40,35){\circle {10}} \put (37,20){$\alpha_3$}
\put (40,65){\circle {10}}\put (40,65){\circle {9}}\put (40,65){\circle {8}}\put (40,65){\circle {7}}\put (40,65)
{\circle {6}}\put (40,65){\circle {5}}\put (40,65){\circle {4}}\put (40,65){\circle {3}}\put (40,65){\circle {2}}
\put (40,65){\circle {1}}
\put (48,62.8){$\alpha_4$} \put
(40,40){\line (0,1){20}} \put (45,35){\line (1,0){20}} \put
(70,35){\circle {10}} \put (67,20){$\alpha_{2}$} \put
(75,35){\line (1,0){20}} \put (100,35){\circle {10}} \put
(97,20){$\alpha_{1}$}
\end{picture}
$$\begin{array}{l}\psi = \alpha_1 +2\alpha_2 +3\alpha_3+2\alpha_4 +2\alpha_5+\alpha_6
  \\(\psi \, , \,\alpha_4) = 1 \quad; \quad  (\psi \, , \, \alpha_i ) = 0 \quad i \ne 4 \
\end{array}$$
\begin{picture}(100,90)
      \put (-180,35){$E_{6(-14)}^\wedge$} \put (-20,35){\circle {10}}\put
(-23,20){$\alpha_6$} \put (-15,35){\line (1,0){20}} \put
(10,35){\circle {10}} \put (7,20){$\alpha_5$} \put (15,35){\line
(1,0){20}} \put (40,35){\circle {10}} \put (37,20){$\alpha_3$}
\put(40,95){\circle{10}}\put(40,70){\line (0,1){20}}\put
(48,90){$\alpha_0$}
\put (40,65){\circle {10}}\put (40,65){\circle {9}}\put (40,65){\circle {8}}\put (40,65){\circle {7}}\put (40,65)
{\circle {6}}\put (40,65){\circle {5}}\put (40,65){\circle {4}}\put (40,65){\circle {3}}\put (40,65){\circle {2}}
\put (40,65){\circle {1}} \put (48,62.8){$\alpha_4$} \put
(40,40){\line (0,1){20}} \put (45,35){\line (1,0){20}} \put
(70,35){\circle {10}} \put (67,20){$\alpha_{2}$} \put
(75,35){\line (1,0){20}} \put (100,35){\circle {10}} \put
(97,20){$\alpha_{1}$}
\end{picture}
\vskip 1cm \caption{The Dynkin diagram of $E_{6(-14)}$ and of its affine Ka\v c --Moody extension $E_{6(-14)}^\wedge$.
The only simple root which has grading one
with respect to the highest root $\psi$ is $\alpha_7$ (painted black).
With respect to the algebra $\mathbb{U}_{D=4}=\mathrm{SU(5,1))}$ whose Dynkin diagram is
obtained by removal of the black circle, $\alpha_4$ is the highest weight of the symplectic representation of
the vector fields, namely the $W=\mathbf{20}$. The affine
extension is originated by attaching the extra root $\alpha_0$ to the root corresponding to the vector fields.}
\label{except4}
\end{figure}
\vskip 0.2cm
\paragraph{N=5}
\vskip 0.2cm The case of $N=5$ supergravity is described by
fig.\ref{except4}. From the point of view of the Tits - Satake
projection this case has some extra complications since the
projected root system is not the root system of a simple Lie
algebra. This explains the entry $bc$-system appearing in table
\ref{tavolona} and because of that we postpone the discussion of
its Tits Satake projection to a next paper. Indeed, as we already
stressed, the focus of the present paper is in a different
direction.
\par
In the $N=5$ theory the scalar manifold is a complex coset of rank
$r=1$,
\begin{equation}
  \mathcal{{M}}_{N=5,D=4}=\frac{\mathrm{SU(1,5)}}{\mathrm{SU(5) \times U(1)}}
\label{MN5D4}
\end{equation}
and there are $\mathbf{10}$ vector fields whose electric and magnetic field strengths are assigned to the
$\mathbf{20}$-dimensional representation of $\mathrm{SU(1,5)}$, which
is that of an antisymmetric three-index tensor
\begin{equation}\begin{array}{ccc}
  \mbox{dim}_{\mathrm{SU(1,5)}} \, & \begin{array}{c}
     \vspace{-0.4cm} \\
     \yng(1,1,1) \
  \end{array} & \, = \, 20
\end{array}
\label{orpo}
\end{equation}
The decomposition (\ref{gendecompo}) takes the explicit form:
\begin{equation}
 \mbox{adj}(\mathrm{E_{6(-14)}}) =\mbox{adj}(\mathrm{SU(1,5})\oplus\mbox{adj}(\mathrm{SL(2,\mathbb{R})_E})\oplus
(\mathbf{2},\mathbf{20})
\label{decompe6}
\end{equation}
and we have that the highest root of $\mathrm{E_6}$, namely
\begin{equation}
  \psi =\alpha_1 +2\alpha_2 +3\alpha_3+2\alpha_4 +2\alpha_5+\alpha_6
\label{psie6}
\end{equation}
has non vanishing scalar product only with the root $\alpha_4$
originating the affine extension in the form depicted in fig.\ref{except4}.
\par
Writing a well adapted basis of $E_6$ roots is a little bit more laborious
but it can be done. We find:
\begin{equation}
\begin{array}{rclcl}
\alpha_1 & = & \left \{ 0,0,- \frac{{\sqrt{3}}}{2},
  \frac{1}{2\,{\sqrt{5}}},
  {\sqrt{\frac{6}{5}}},0\right \} & = &
\left \{\overline{\alpha}_1,0\right \} \\
\alpha_2  & = & \left \{ - \frac{1}
      {{\sqrt{2}}}  ,
  \frac{1}{{\sqrt{6}}},
  \frac{2}{{\sqrt{3}}},0,0,
  0\right \}& = &
\left \{\overline{\alpha}_2,0\right \} \\
\alpha_3  & = & \left \{ {\sqrt{2}},0,0,0,0,0\right \} & = &
\left \{\overline{\alpha}_3,0\right \}\\
\alpha_4  & = & \left \{ - \frac{1}
      {{\sqrt{2}}}  ,
  \frac{1}{{\sqrt{6}}},
  - \frac{1}
      {{\sqrt{3}}}  ,
  \frac{1}{{\sqrt{5}}},
  -{\sqrt{\frac{3}{10}}},
  \frac{1}{{\sqrt{2}}}\right \} & = &
\left \{\overline{\mathbf{w}}_h, \frac{1}{{\sqrt{2}}} \right \}\\
\alpha_5  & = & \left \{ - \frac{1}
      {{\sqrt{2}}}  ,
  -{\sqrt{\frac{3}{2}}},0,0,0,
  0\right \}& = &
\left \{\overline{\alpha}_4,0\right \} \\
\alpha_6  & = & \left \{ 0,{\sqrt{\frac{2}{3}}},
 - \frac{1}{2\,{\sqrt{3}}},
  -\frac{\sqrt{5}}{2},0,0\right \}& = &
\left \{\overline{\alpha}_5,0\right \} \\
  \end{array}
\label{alfeE6}
\end{equation}
In this basis we can check that the five $5$-vectors
$\bar{\alpha}_i$ ($i=1,\dots,5$) constitute a simple root
basis for the $A_5\simeq \mathrm{SU(1,5)}$ root
system, namely:
\begin{equation}
 \langle \bar{\alpha}_i \, , \, \bar{\alpha}_j \rangle = \left(\matrix{ 2 & -1 & 0 & 0 & 0
 \cr -1 & 2 & -1 & 0 & 0 \cr 0 & -1 & 2 &
     -1 & 0 \cr 0 & 0 & -1 & 2 & -1 \cr 0 & 0 & 0 & -1 & 2 \cr  }
     \right)\, = \, \mbox{Cartan matrix of $A_5$}
\label{a5carta}
\end{equation}
while:
\begin{equation}
 \mathbf{w}_h \, = \, \left \{ - \frac{1}
      {{\sqrt{2}}}  ,
  \frac{1}{{\sqrt{6}}},
  - \frac{1}
      {{\sqrt{3}}}  ,
  \frac{1}{{\sqrt{5}}},
  -{\sqrt{\frac{3}{10}}} \right \}
\label{ierundaE6}
\end{equation}
is the highest weight of the spinor $\mathbf{20}$-dimensional
representation of $\mathrm{SU(1,5)}$. Finally in this basis
the highest root $\psi$
defined by eq.(\ref{psie6}) takes the expected form:
\begin{equation}
  \psi = \{ 0,0,0,0,0,0,\sqrt{2}\}
\label{expepsi7bis}
\end{equation}
\par
\vskip 0.2cm
\paragraph{N=4}
\vskip 0.2cm
\begin{figure}
\centering
\begin{picture}(100,40)
      \put (-180,35){$D_{\ell=4+k+1}$}\put (-75,56){\line (1,-1){20}}
\put (-109,59){$\alpha_\ell$}
      \put
(-79,59){\circle {10}} \put (-75,14){\line (1,1){20}}
\put (-109,10){$\alpha_{\ell-1}$}
\put
(-79,10){\circle {10}} \put (-45,35){\line (1,0){20}} \put
(-50,35){\circle {10}} \put (-53,20){$\alpha_{\ell-2}$}\put (-20,35){\circle {10}}
\put(-23,20){$\alpha_{\ell-3}$} \put (-9,35){$\dots$} \put
(10,35){$\dots$}  \put (15,35){$\dots$} \put (40,35){\circle {10}} \put (37,20){$\alpha_3$}
 \put (45,35){\line (1,0){20}}
 \put (70,35){\circle {10}}
 \put (70,35){\circle {9}}
 \put (70,35){\circle {8}}
 \put (70,35){\circle {7}}
 \put (70,35){\circle {6}}
 \put (70,35){\circle {5}}
 \put (70,35){\circle {4}}
 \put (70,35){\circle {3}}
 \put (70,35){\circle {2}}
 \put (70,35){\circle {1}}
 \put (67,20){$\alpha_{2}$} \put
(75,35){\line (1,0){20}} \put (100,35){\circle {10}} \put
(97,20){$\alpha_{1}$}
\end{picture}
$$\begin{array}{l}\psi = \alpha_1 +2\alpha_2 +2\alpha_3+\dots
+2\alpha_{\ell-2}+\alpha_{\ell-1} +\alpha_{\ell}
  \\(\psi \, , \,\alpha_2) = 1 \quad; \quad  (\psi \, , \, \alpha_i ) = 0 \quad i \ne 2 \
\end{array}$$
\begin{picture}(100,70)
      \put (-180,35){$D_{\ell=4+k+1}^\wedge$}\put (-75,56){\line (1,-1){20}}
\put (-109,59){$\alpha_\ell$}
      \put
(-79,59){\circle {10}} \put (-75,14){\line (1,1){20}}
\put (-109,10){$\alpha_{\ell-1}$}
\put
(-79,10){\circle {10}} \put (-45,35){\line (1,0){20}} \put
(-50,35){\circle {10}} \put (-53,20){$\alpha_{\ell-2}$}\put (-20,35){\circle {10}}
\put(-23,20){$\alpha_{\ell-3}$} \put (-9,35){$\dots$} \put
(10,35){$\dots$}  \put (15,35){$\dots$} \put (40,35){\circle {10}} \put (37,20){$\alpha_3$}
 \put (45,35){\line (1,0){20}}
 \put (70,35){\circle {10}}
 \put (70,35){\circle {9}}
 \put (70,35){\circle {8}}
 \put (70,35){\circle {7}}
 \put (70,35){\circle {6}}
 \put (70,35){\circle {5}}
 \put (70,35){\circle {4}}
 \put (70,35){\circle {3}}
 \put (70,35){\circle {2}}
 \put (70,35){\circle {1}}
 \put (67,20){$\alpha_{2}$}
\put (70,40){\line(0,1) {20}}
\put (78,65){$\alpha_0$}
\put (70,65){\circle {10}}
 \put
(75,35){\line (1,0){20}} \put (100,35){\circle {10}} \put
(97,20){$\alpha_{1}$}
\end{picture}
\vskip 1cm \caption{The Dynkin diagram of $D_{4+k+1}$ and of its affine Ka\v c --Moody extension $D_{4+k+1}^\wedge$.
The algebra $D_{4+k+1}$ is that of the group $\mathrm{SO(8,2k+2)}$
corresponding to the Ehlers reduction of $N=4$ supergravity coupled to
$n_m=2k$ vector multiplets. The only simple root which has non
vanishing grading with respect to the highest one $\psi$ is $\alpha_2$.
Removing it (black circle) we are left with the algebra $D_{4+k-1}
\times A_1$ which is indeed the duality algebra in $D=4$, namely
$\mathrm{SO(6,2k)} \times \mathrm{SL(2,\mathbb{R})_0}$.
 The black root $\alpha_2$  is the highest weight of the symplectic representation of the vector fields,
 namely the $\mathbf{W}=(\mathbf{2_0},\mathbf{6+2k})$. The affine
extension is originated by attaching the extra root $\alpha_0$ to the root $\alpha_2$ which
corresponds to  vector fields.}
\label{except5}
\end{figure}
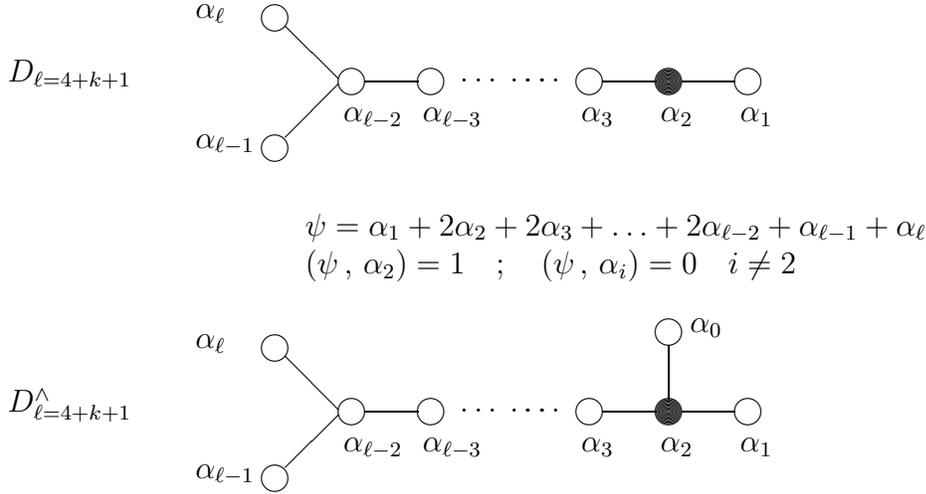
\vskip 0.2cm
\par
The case of $N=4$ supergravity is the first where the scalar manifold is
not completely fixed, since we can choose the number $n_m$ of vector multiplets
that we can couple to the graviton multiplet. In any case, once $n_m$ is
fixed the scalar manifold is also fixed and we have:
\begin{equation}
  \mathcal{M}_{N=4,D=4} = \frac{\mathrm{SL(2,\mathbb{R})_0}}{\mathrm{O(2)}} \, \otimes\,
  \frac{\mathrm{SO(6,n_m)}}{\mathrm{SO(6) \times SO(n_m)}}
\label{n6manif}
\end{equation}
The total number of vectors $n_V = 6+n_m$ is also fixed and the
symplectic representation $\mathbf{W}$ of the duality algebra
\begin{equation}
  \mathbb{U}_{D=4} =\mathrm{SL(2,\mathbb{R})_0} \, \times \,
  \mathrm{SO(6,n_m)}
\label{UD4}
\end{equation}
to which the vectors are assigned and which determines the embedding:
\begin{equation}
  \mathrm{SL(2,\mathbb{R})_0} \, \times \, \mathrm{SO(6) \times
  SO(n_m)}\, \mapsto \, \mathrm{Sp( 12+2\,n_m,\mathbb{R})}
\label{N4sympembed}
\end{equation}
is also fixed, namely $\mathbf{W} = \left( \mathbf{2_0,6+n_m} \right)$,
$\mathbf{2_0}$ being the fundamental representation of $\mathrm{SL(2,\mathbb{R})_0}$
and $\mathbf{6+n_m}$ the fundamental vector representation of
$\mathrm{SO(6,n_m)}$. The $D=3$ algebra that one obtains in the Ehlers dimensional
reduction is, for all number of vector multiplets given by
$\mathbb{U}_{D=3}=\mathrm{SO(8,n_m+2)}$, leading to the manifold:
\begin{equation}
  \mathcal{M}_{N=4,D=3} = \frac{\mathrm{SO(8,n_m+2)}}{\mathrm{SO(8) \times SO(n_m+2)}}
\label{MN8D3}
\end{equation}
Correspondingly the form taken by the general decomposition
(\ref{gendecompo}) is the following one, for all values of $n_m$:
\begin{equation}
 \mbox{adj}(\mathrm{SO(8,n_m+2)}) =\mbox{adj}(\mathrm{SL(2,\mathbb{R})_0} )\, \oplus \,
 \mbox{adj}(\mathrm{SO(6,n_m)})
 \oplus\mbox{adj}(\mathrm{SL(2,\mathbb{R})_E})\oplus
(\mathbf{2_E},\mathbf{2_0,6+n_m})
\label{decompN4}
\end{equation}
where $\mathbf{2_{E,0}}$ are the fundamental representations
respectively of $\mathrm{SL(2,\mathbb{R})_E}$ and of
$\mathrm{SL(2,\mathbb{R})_0}$.
\par
In order to give a Dynkin Weyl description of these algebras, we are
forced to distinguish the case of an odd and even number of vector
multiplets. In the first case both $\mathbb{U}_{D=3}$ and $\mathbb{U}_{D=4}$ are
non simply laced algebras of the $B$--type, while in the second case
they are both simply laced algebras of the $D$-type
\begin{equation}
  n_m = \cases{2k \quad \quad \quad\rightarrow \quad \mathbb{U}_{D=4} \simeq D_{k+3} \cr
  2k+1 \quad \, \rightarrow \quad \mathbb{U}_{D=4} \simeq B_{k+3} \cr}
\label{twocases}
\end{equation}
Just for simplicity and for shortness we choose to discuss only
the even case $n_m=2k$ which is described by fig.\ref{except5}.
\par
In this case we consider the $\mathbb{U}_{D=3} =\mathrm{SO(8,2k+2)}$
Lie algebra whose Dynkin diagram is that of $D_{5+k}$.
Naming $\epsilon_i$ the unit vectors in an Euclidean
$\ell$--dimensional space where $\ell=5+k$, a well adapted basis of
simple roots for the considered algebra is the following one:
\begin{eqnarray}
\alpha_1 & = & \sqrt{2} \, \epsilon _1 \nonumber\\
\alpha_2 & = & -\frac{1}{\sqrt{2}}\, \epsilon _1 -\epsilon _2 +\frac{1}{\sqrt{2}}\, \epsilon
_\ell\nonumber\\
\alpha_3 & = & \epsilon _2 -\epsilon _3 \nonumber\\
\alpha_4 & = & \epsilon _3 -\epsilon _4 \nonumber\\
\dots & = & \dots \nonumber\\
\alpha_{\ell-1} & = &\epsilon _{\ell-2} - \epsilon
_{\ell-1}\nonumber\\
\alpha_{\ell} & = &\epsilon _{\ell-2} + \epsilon
_{\ell-1}\nonumber\\
\label{welladapn4}
\end{eqnarray}
which is quite different from the usual presentation but yields the
correct Cartan matrix. In this basis the highest root of the algebra:
\begin{equation}
  \psi =  \alpha_1 +2\alpha_2 +2\alpha_3+\dots
+2\alpha_{\ell-2}+\alpha_{\ell-1} +\alpha_{\ell}
\label{highestn4}
\end{equation}
takes the desired form:
\begin{equation}
  \psi=\sqrt{2} \, \epsilon_\ell
\label{wellpsi}
\end{equation}
In the same basis the $\alpha_W =\alpha_2$ root has also the expect
form:
\begin{equation}
  \alpha_W = \left( \mathbf{w},\frac{1}{\sqrt{2}} \right)
\label{wellalphan4}
\end{equation}
where:
\begin{equation}
  \mathbf{w} =  -\frac{1}{\sqrt{2}}\, \epsilon _1 -\epsilon _2
\label{wellwn4}
\end{equation}
is the weight of the symplectic  representation $\mathbf{W}=\left( \mathbf{2_0} , \mathbf{6+2k}\right)
$. Indeed $-\frac{1}{\sqrt{2}}\, \epsilon _1$ is the fundamental
weight for the Lie algebra $\mathrm{SL(2,\mathbb{R})_0}$, whose root
is $\alpha_1 = \sqrt{2} \, \epsilon _1$, while $-\epsilon _2$ is the
highest weight for the vector representation of the algebra
$\mathrm{SO(6,2k)}$, whose roots are $\alpha_3, \alpha_4, \dots,
\alpha_\ell$.
\par
Next we briefly comment on the Tits Satake projection. The algebra
$\mathrm{SO(8,n_m+2)}$ is maximally split only for $n_m=6$ which,
from the superstring view point, corresponds to the case of
Neveu-Schwarz vector multiplets in a toroidal compactification.
For a different number of vector multiplets, in particular for
$n_m>6$ the study of billiard dynamics involves considering the
Tits Satake projection, which just yields the universal manifold:
\begin{equation}
  \mathcal{M}^{TS}_{N=4,D=3}=\frac{\mathrm{SO(8,8)}}{\mathrm{SO(8) \times SO(8)}}
\label{TSN8D3}
\end{equation}
The detailed study of these aspects is however postponed to future
publications as we have already stressed.
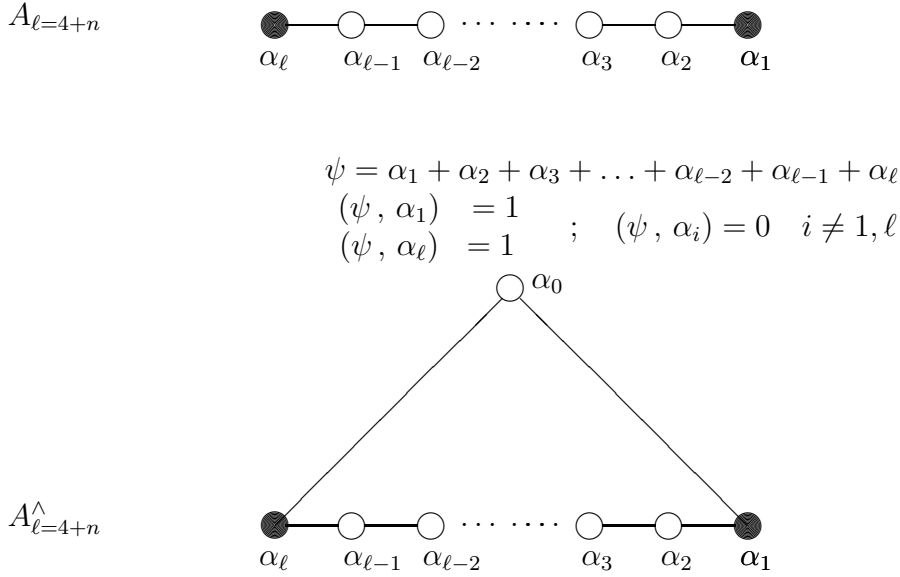
\begin{figure}
\centering
\vskip -1.768cm
\begin{picture}(100,70)
\put (-180,35){$A_{\ell=4+n}$}
\put (-75,35){\line (1,0){20}}
\put (-85,20){$\alpha_\ell$}
\put(-79,35){\circle {10}}
\put (-79,35){\circle {9}}
\put (-79,35){\circle {8}}
\put (-79,35){\circle {7}}
\put (-79,35){\circle {6}}
\put (-79,35){\circle {5}}
\put (-79,35){\circle {4}}
\put (-79,35){\circle {3}}
\put (-79,35){\circle {2}}
\put (-79,35){\circle {1}}
\put(97,20){$\alpha_{1}$}
\put (-45,35){\line (1,0){20}}
\put (-50,35){\circle {10}}
\put (-53,20){$\alpha_{\ell-1}$}
\put (-20,35){\circle {10}}
\put(-23,20){$\alpha_{\ell-2}$}
\put (-9,35){$\dots$}
\put (10,35){$\dots$}
\put (15,35){$\dots$}
\put (40,35){\circle {10}}
\put (37,20){$\alpha_3$}
 \put (45,35){\line (1,0){20}}
 \put (70,35){\circle {10}}
 \put (67,20){$\alpha_{2}$}
 \put (75,35){\line (1,0){20}}
 \put (100,35){\circle {10}}
 \put (100,35){\circle {9}}
 \put (100,35){\circle {8}}
 \put (100,35){\circle {7}}
 \put (100,35){\circle {6}}
 \put (100,35){\circle {5}}
 \put (100,35){\circle {4}}
 \put (100,35){\circle {3}}
 \put (100,35){\circle {2}}
 \put (100,35){\circle {1}}
 \put(97,20){$\alpha_{1}$}
\end{picture}
$$\begin{array}{l}\psi = \alpha_1 +\alpha_2 +\alpha_3+\dots
+\alpha_{\ell-2}+\alpha_{\ell-1} +\alpha_{\ell}
  \\
  \begin{array}{cc}
     (\psi \, , \,\alpha_1) & = 1 \\
      (\psi \, , \,\alpha_\ell) & = 1  \
  \end{array}  \quad; \quad  (\psi \, , \, \alpha_i ) = 0 \quad i \ne 1,\ell \
\end{array}$$
\begin{picture}(100,120)
\put (-180,35){$A_{\ell=4+n}^\wedge$}
\put (-75,35){\line (1,0){20}}
\put (-85,20){$\alpha_\ell$}
\put (-79,35){\line(1,1){86}}
\put (-79,35){\circle {10}}
\put (-79,35){\circle {9}}
 \put (-79,35){\circle {8}}
 \put (-79,35){\circle {7}}
 \put (-79,35){\circle {6}}
 \put (-79,35){\circle {5}}
 \put (-79,35){\circle {4}}
 \put (-79,35){\circle {3}}
 \put (-79,35){\circle {2}}
 \put (-79,35){\circle {1}}
 \put(97,20){$\alpha_{1}$}
\put (-45,35){\line (1,0){20}}
\put (-50,35){\circle {10}}
\put (-53,20){$\alpha_{\ell-1}$}
\put (-20,35){\circle {10}}
\put(-23,20){$\alpha_{\ell-2}$}
\put (-9,35){$\dots$}
\put (10,35){$\dots$}
\put (15,35){$\dots$}
\put (40,35){\circle {10}}
\put (37,20){$\alpha_3$}
 \put (45,35){\line (1,0){20}}
 \put (70,35){\circle {10}}
 \put (67,20){$\alpha_{2}$}
 \put (75,35){\line (1,0){20}}
 \put (100,35){\circle {10}}
 \put (100,35){\circle {9}}
 \put (100,35){\circle {8}}
 \put (100,35){\circle {7}}
 \put (100,35){\circle {6}}
 \put (100,35){\circle {5}}
 \put (100,35){\circle {4}}
 \put (100,35){\circle {3}}
 \put (100,35){\circle {2}}
 \put (100,35){\circle {1}}
 \put (100,35){\line(-1,1){86}}
 \put(97,20){$\alpha_{1}$}
\put (18,125){$\alpha_0$}
\put (10,125){\circle {10}}
\end{picture}
\caption{The Dynkin diagrams of $A_{4+n}$ and of its affine extension $A_{4+n}^{\wedge}$.
$A_{4+n}$ is the algebra of  $\mathrm{SU(4,n+1)}$
generated by the Ehlers reduction of $N=3$ supergravity with
$n_m=n$ vector multiplets. Two simple roots $\alpha_{1,\ell}$  have non
vanishing grading with respect to the highest root $\psi$.
Removing them (black circles) we are left with the algebra $A_{2+n}$
of $\mathrm{SU(3,n)}$. The black roots $\alpha_{1,\ell}$  are, respectively, the highest weight of the
 representation $(\mathbf{3+n})$ and of its conjugate
 $\overline{(\mathbf{3+n})}$ which together make up the
 symplectic representation of the vector fields, $\mathbf{W}$. The affine
extension is originated by attaching the extra root $\alpha_0$ to the
 \textit{vector roots.}\label{dynkN3}}
\end{figure}
\vskip 0.2cm
\paragraph{N=3}
The case of N=3 supergravity \cite{noiN3} is similar to that of N=4 since also
here the only free parameter is the number of vector multiplets $n_m$
leading to the following scalar manifold:
\begin{equation}
  \mathcal{M}_{N=3,D=4}=\frac{\mathrm{SU(3,n_m)}}{\mathrm{SU(3) \times SU(n_m)}}
\label{MscaN3}
\end{equation}
It might seem that the duality algebra in $D=4$ should be $\mathbb{U}_{D=4}=\mathrm{SU(3,n_m)}$, yet, in this case
there is a subtlety. The actual algebra is rather
\begin{equation}
\mathbb{U}_{D=4}=\mathrm{U(3,n_m)} = \mathrm{SU(3,n_m)} \times \mathrm{U(1)_Z}
\label{UD4N3}
\end{equation}
the overall phase group $\mathrm{U(1)_Z}$ having a vanishing action
on the scalars, but not on the vector fields. The symplectic representation $\mathbf{W}$ to which the vectors are
assigned is just made out of  the fundamental $\mathbf{3+n_m}$ of
$\mathrm{U(3,n_m)}$ plus its complex conjugate   $\overline{\mathbf{3+n_m}}$ leading to the embedding:
\begin{equation}
  \mathrm{U(3,n_m)}\, \mapsto \, \mathrm{Sp( 6+2\,n_m,\mathbb{R})}
\label{N3sympembed}
\end{equation}
which was explicitly described in \cite{noiN3}. In short it goes as follows.
Let $L$ be a  $(3+n) \times (3+n)$ matrix in the fundamental representation of
$\mathrm{SU(3,n)}$. We map it into a matrix belonging to
$\mathrm{Usp(3+n,3+n)}$ in the following way:
\begin{equation}
\mathrm{SU(3,n)} \, \ni \, L = \left( \begin{array}{c|c}
     X & Y \\
     \hline
     Z & W \
  \end{array}\right) \, \mapsto  \mathbb{L}_{Usp} = \left(
\begin{array}{c|c||c|c}
  X & 0 & 0 &Y \\
  \hline
  0 & W^\star &  Z^\star & 0 \\
  \hline
  \hline
  0 & Y^\star &  X^\star & 0 \\
  \hline
  Z & 0 &  0 & W\
\end{array}
  \right) \, \in \, \mathrm{Usp(3+n,3+n)}
\label{su11su3nusp}
\end{equation}
Then we use the Cayley isomorphism relating $\mathrm{Usp(n,n)}$ and
$\mathrm{Sp(2n,\mathbb{R})}$ to obtain a real symplectic matrix
$\mathbb{L}_{Sp}$ representing the original $\mathrm{SU(3,n)}$ element \cite{myparis}:
\begin{equation}
  \mathbb{L}_{Sp} = \frac{1}{\sqrt{2}} \, \left( \begin{array}{cc}
     \mathbf{1} & \mathbf{1} \\
     {\rm i} \, \mathbf{1} &  -{\rm i} \, \mathbf{1} \
  \end{array}\right) \,
  \mathbb{L}_{Usp} \,
  \frac{1}{\sqrt{2}} \left( \begin{array}{cc}
     \mathbf{1} & {\rm i} \, \mathbf{1} \\
     \mathbf{1} & -{\rm i} \, \mathbf{1} \
  \end{array}\right)
\label{ordigno}
\end{equation}
This matrix construction is the explicit definition of the
representation $\mathbf{W}$ of the vectors.
\par
The subtlety of the $N=3$ case is that the map (\ref{su11su3nusp})
extends to the full $\mathrm{U(3,n)}$ algebra, namely also to matrices
$ \left( \begin{array}{c|c}
     X & Y \\
     \hline
     Z & W \
  \end{array}\right) $ which are not traceless. In particular we see
  that the $\mathrm{U(1)}_Z$ group whose generator corresponds to $Y=Z=0$,
  $X=W= {\rm i}\, \mathbf{1}$ has a non trivial image in the
  symplectic group and hence a non trivial action on the vector
  fields, although its action is zero on the scalars of the coset
  manifold (\ref{MscaN3}). Hence it is correct to say that the true
duality algebra in $D=4$ is not just $\mathrm{SU(3,n)}$, rather it
is:
\begin{equation}
  \mathbb{U}_{D=4} = \mathrm{U(3,n)} \, \equiv \, \mathrm{SU(3,n)} \oplus
  \mathrm{U(1)}_Z
\label{eccoqui}
\end{equation}
With this proviso the general decomposition (\ref{gendecompo}) is
still true also in the $N=3$ case. Indeed
in $D=3$, via the Ehlers reduction, we obtain
$\mathbb{U}_{D=4}=\mathrm{SU(4,n_m+1)}$ and the form taken by the general decomposition (\ref{gendecompo}) is:
\begin{equation}
 \mbox{adj}\mathrm{SU(4,n_m+1)} =\mbox{adj}\mathrm{SU(3,n_m)}\, \oplus
 \,\mathrm{U(1)}_Z \,
 \oplus\mbox{adj}\mathrm{SL(2,\mathbb{R})_E}\oplus
(\mathbf{2_E},\mathbf{3+n_m}) \, \oplus  \, (\mathbf{2_E},\overline{\mathbf{3+n_m}})
\label{decompN3}
\end{equation}
The Weyl Dynkin description of this case is now easily provided and
it is summarized in fig.\ref{dynkN3}.
The simple roots of the Lie algebra $A_{\ell=4+n}$ can be written
according to a standard presentation as euclidean vectors in
$\ell+1 = 5+n$ dimensions and we can write:
\begin{eqnarray}
\alpha_1 & = & \epsilon_{5+n} -\epsilon_1 \nonumber\\
\alpha_{i+1} & = & \epsilon_i - \epsilon_{i+1} \quad (i=1,\dots,2+n)
\nonumber\\
\alpha_{\ell=4+n}&=&\epsilon_{3+n} - \epsilon_{4+n} \nonumber\\
\label{simpleruttenew}
\end{eqnarray}
The highest root $\psi=\sum_{i=1}^{\ell}\alpha_i$ takes the very simple form:
\begin{equation}
  \psi= - \epsilon_{4+n} + \epsilon_{5+n}
\label{psirotta}
\end{equation}
and has  grading one with respect to the roots $\alpha_1$ and
$\alpha_\ell$. The Dynkin diagram of $A_{2+n}$, namely of the Lie
algebra $\mathrm{SU(3,n)}$, is obtained by deleting
$\alpha_{1,\ell}$ which are indeed the weights of the fundamental
and anti-fundamental representation of $\mathrm{SU(3,n)}$, out of
which we cook up the symplectic $\mathbf{W}$--representation as we
have described above. Indeed calling $\lambda_{i}$ the fundamental
weights of $A_{2+n}$ described by the roots $\alpha_{i+1}$,
($i=1,\dots,2+n$), it follows by their own definition that we can
rewrite:
\begin{equation}
  \alpha_1 = \lambda_1 + \Lambda \quad ; \quad \alpha_{4+n} =
  \lambda_{n+2} + \Lambda
\label{agnizione2}
\end{equation}
where $\Lambda$ is the fundamental weight of $\mathrm{SL(2,R)_E}$,
\textit{i.e.} the vector orthogonal to all the other roots and such that $\psi\cdot\Lambda
=1$
This discussion shows that the N=3 case is actually no exception to the
general discussion we present in this paper. Yet since the $\mathbf{W}$
representation is made of one representation plus its conjugate there
are actually two conjugate weights and hence two roots participating
into the mechanism of affine extension.
\vskip 0.2cm
\paragraph{N=2} In N=2 supergravity we have geometries rather than
algebras since, from a general point of view, we just know that the
scalar manifold is of the special K\"ahler type. Yet many
considerations are still valid since the symplectic embedding is
just an integral part of the very definition of special geometry
and the so called c-map \cite{specHomgeo} is another name for the
Ehlers dimensional reduction. Hence the affine extension as well
as  the Tits Satake projection can be considered also at the level
of $N=2$ supergravity. We plan to devote a separate publication to
this extensions.
\par
Having concluded the algebraic discussion of the affine extensions,
we turn to its field theoretic realization, namely to the Ehlers and
Matzner--Misner reduction schemes.
\section{The Ehlers reduction}
\label{ehlersgrav}
The dimensional reduction \`a la Ehlers consists of three steps:
\begin{enumerate}
  \item First one dimensionally reduces the $D=4$ supergravity
  lagrangian (\ref{d4generlag}) to $D=3$ in the  standard way based
  on the triangular gauge for the vielbein and for the vector fields.
  \item Then one dualizes all the vector fields obtained in the $D=3$
  lagrangian to scalars: namely one dualizes both those vector fields that were already
  present in $D=4$ and the new ones generated by the Kaluza Klein
  mechanism. In this way one obtains $3D$--gravity coupled
  $\sigma$--model in $D=3$ which is based on a new coset manifold
  $\mathrm{U_{D=3}}/\mathrm{H}$ enlarging the original four-dimensional
  $\sigma$--model with the new scalars.
  \item Finally one further reduces the $D=3$ gravity coupled
  $\sigma$-model to $D=2$. In this step nothing new happens to the
  $\sigma$--model part of the lagrangian. The only novelty comes from
  the reduction of gravity which just produces the coupling to a
  dilaton field.
\end{enumerate}
According to the above plan we introduce the reduction ansatz for the
$D=4$ metric in the following form:
\begin{equation}
ds_{(4)}^2 = \Delta^{-1}ds_{(3)}^2 + \Delta(dx^3 + b_\mu dx^\mu)^2,
\end{equation}
where the index $\mu = 0,1,2$ corresponds to the $D=3$ space-time dimensions, $\Delta$ denotes the
Kaluza--Klein scalar and
$b_\mu$ denotes the Kaluza-Klein vector, for which we fix the following \textit{Coulomb} gauge
$b_0=0$.
In this frame the dimensional reduction of the Einstein term yields:
\begin{equation}
-2\sqrt{g^{(4)}}{R}[g^{(4)}] =
-2\sqrt{g}\left[{R}[g] - \frac{1}{4}(\partial\ln\Delta)^2 +
\Delta^2G_{\mu\nu}G^{\mu\nu}\right]
\label{einstreduce}
\end{equation}
where
\begin{equation}
  G_{\mu\nu} =\frac{1}{2}(\partial_\mu b_\nu -
\partial_\nu b_\mu)
\label{kkfielstren}
\end{equation}
is the field strength of the Kaluza Klein vector field.
\par
Let us now consider the dimensional reduction of the gauge fields.
These latter are redefined in the customary  way as follows:
\begin{equation}
A^\Lambda_{[D=4]} = \left( \bar{A}^{\Lambda}_\mu + \tau^\Lambda \, b_\mu \right ) dx^\mu +
\tau^\Lambda \, dx^3
\end{equation}
where $\bar{A}^{\Lambda}_\mu$ are the three-dimensional gauge fields, while
\begin{equation}
  \tau^\Lambda \equiv A_3^\Lambda
\label{taudefi}
\end{equation}
are the scalar fields  generated by the internal components of the $D=4$ gauge fields. Hence the field strengths
$\bar{F}^\Lambda = dA^\Lambda_{[D=3]}$ of
the $D=3$ vector fields are related to their higher dimensional
ancestors ${F}^\Lambda = dA^\Lambda_{[D=4]}$ by the following formula:
\begin{equation}
\bar{F}_{\mu\nu}^\Lambda = {F}_{\mu\nu}^{\Lambda} - \tau^\Lambda G_{\mu\nu}
- \frac{1}{2}[b_\nu\partial_\mu \tau^\Lambda - b_\mu\partial_\nu
\tau^\Lambda]
\label{FbarF}
\end{equation}
This being set, the  vector sector of the $D=4$ lagrangian
(\ref{d4generlag}) reduces as follows
\begin{eqnarray}
\ft 12 \, \mbox{Re}\mathcal{N}_{\Lambda\Sigma}
\, F^\Lambda_{\hat{\mu}\hat{\nu}} \, F^\Sigma_{\hat{\rho}\hat{\sigma}}
\epsilon^{\hat{\mu}\hat{\nu}\hat{\rho}\hat{\sigma}} &=&
\epsilon^{\mu\nu\rho}\mbox{Re}\, \mathcal{N}_{\Lambda\Sigma} \, (\bar{F}^{\Lambda}_{\mu\nu}
+ G_{\mu\nu}\tau^\Lambda)\partial_\rho \tau^\Sigma
\nonumber\\
\sqrt{-\mbox{det} \, g_{[4]}} \, \mbox{Im} \, \mathcal{N}_{\Lambda\Sigma} \, F^\Lambda_{\hat{\mu}\hat{\nu}} F^{\Sigma|\hat{\mu}\hat{\nu}}
 &=& \sqrt{-\mbox{det}g} \, \mbox{Im} \, \mathcal{N}_{\Lambda\Sigma}
 \,\left[\Delta \,
\left( \bar{F}^\Lambda_{\mu\nu} + \tau^\Lambda \, G_{\mu\nu} \right) \,
\left( \bar{F}^{\Sigma | {\mu\nu}} + \tau^\Sigma \, G^{\mu\nu} \right)
\right. \nonumber\\
&& \left. +
\ft 12 \frac{1}{\Delta} \, \partial_\mu \tau^\Lambda \, \partial^\nu \tau^\Sigma \right]
\label{vectreduce}
\end{eqnarray}
In this way we have completed step one of the Ehlers reduction
procedure. The second step is the dualization of all the vector
fields to scalars, which in three dimensions is always possible.
Explicitly we can replace the Kaluza Klein vector with an axion
$B$ and the vector fields $\bar{A}^\Lambda$ with as many axions
$\sigma_\Lambda$ by means of the following non--local relations:
\begin{eqnarray}
&& G_{\mu\nu} =
\frac{1}{4\Delta^2}\sqrt{g}\epsilon_{\mu\nu\sigma}g^{\sigma\rho}[\partial_\rho
B +
\ft 12 \sigma_\Lambda \stackrel{\leftrightarrow}{\partial}_\rho\tau^\Lambda] \\
&& \nonumber \bar{F}_{\mu\nu}^{\Sigma} =
-\frac{\sqrt{g}\epsilon_{\mu\nu\sigma}g^{\sigma\rho}}{2\Delta^2}\left[\Delta
\mbox{Im}\mathcal{N}^{-1\Sigma\Lambda}(\mbox{Re}\mathcal{N}_{\Lambda\Omega}\partial_\rho
\tau^{\Omega} +
\partial_\rho\sigma_\Lambda) + \ft 12 \left(\partial_\rho B +
\ft 12 \sigma_\Lambda
\stackrel{\leftrightarrow}{\partial}_\rho\tau^\Lambda \right)\tau^\Sigma\right]
\label{dualrula}
\end{eqnarray}
where we have introduced the short-hand notation: $a \stackrel{\leftrightarrow}{\partial} b = a\partial b -
b\partial a$.
\par
Upon use of the above construction blocks and collecting our results, the
lagrangian (\ref{d4generlag}) dimensionally reduced \`a la Ehlers takes the following
general form which holds true in all cases and is given by the sum of
three terms:
\begin{equation}
\mathcal{L}_E =-2\sqrt{g}\, {R}[g] +
\mathcal{L}_{\sigma-model} + \mathcal{L}_{vec+sl_2}
\label{treporcellini}
\end{equation}
the first term being the $D=3$ Einstein action, the second the
$\sigma$--model with target manifold
$\frac{\mathrm{U_{D=4}}}{\mathrm{H}}$, directly inherited from
higher dimension
\begin{equation}
\mathcal{L}_{\sigma-model} =-\frac{1}{6}\sqrt{g}\, h_{ab} \, \partial_\mu \phi^a \, \partial^\mu \phi^b
\label{d4sigmod}
\end{equation}
while the third term, coming from the reduction of gravity and
vectors, has the following structure:
\begin{eqnarray}
\mathcal{L}_{vec+sl_2} & = & -
\frac{\sqrt{g}}{4\Delta^2}\left((\partial\Delta)^2 + [\partial B +
\frac{1}{2}\sigma_\Lambda
\stackrel{\leftrightarrow}{\partial}\tau^\Lambda]^2\right) \nonumber\\
&& +
\frac{\sqrt{g}}{2\Delta}\left (\partial \tau \phantom{m}\partial
\sigma \right ) \, \mathbf{M}(\phi) \,\left(\matrix{\partial \tau \cr \partial\sigma}\right)
\label{genehlers}
\end{eqnarray}
The $2n \times 2n $ matrix $\mathbf{M}(\phi) $ has the following
form
\begin{equation}
\, \mathbf{M}(\phi) \, = \,
  \left(\matrix{\mbox{Im}\, \mathcal{N} + \mbox{Re}\, \mathcal{N}\mbox{Im}\, \mathcal{N}^{-1}\mbox{Re}\,
  \mathcal{N}  & \mbox{Re}\,
\mathcal{N}\mbox{Im}\, \mathcal{N}^{-1} \cr \mbox{Im}\, \mathcal{N}^{-1}\mbox{Re}\, \mathcal{N} &
\mbox{Im}\, \mathcal{N}^{-1}}\right)
\label{Qmatricia}
\end{equation}
and works as kinetic metric of the axionic scalar fields $\left\{
\tau^\Lambda \,, \, \sigma_\Sigma\right\} $: it depends only on
the original scalar fields $\phi$ of the $D=4$ lagrangian. This
particular matrix is not a newcomer. It already appeared in the
discussion of the geodesic potential for supersymmetric
black-holes (for a review see for instance \cite{ioericcabhole}).
This latter, which is $\mathrm{U}_{D=4}$ invariant is given in
terms of the black-hole electric $q_\Pi$ and magnetic charges
$p^\Lambda$ by:
\begin{equation}
  V^{geo}(\phi) = \left( p \, , \, q \right ) \, \mathbf{M}(\phi) \, \left(\matrix{p \cr q}\right)
\label{geopoto}
\end{equation}
The $\mathrm{U}_{D=4}$ invariance is guaranteed by the fact that
under a symplectic transformation $\Lambda_\xi$  generated by
an isometry $\xi$ of the $D=4$ $\sigma$-model (see eq.
(\ref{embeddusmatra})), the matrix (\ref{Qmatricia}) transforms as
follows:
\begin{equation}
  \mathbf{M}(\xi\phi) \, \mapsto \, \Omega \, \Lambda_\xi
  \,\mathbf{M}(\phi)\, \Lambda_\xi^T \, \Omega
\label{Qmtractransf}
\end{equation}
as a consequence of the linear fractional transformation
(\ref{Ntransfa}) of the vector kinetic matrix $ \mathcal{N} $.
\par
We can now group all the scalar fields into a single set:
\begin{eqnarray}
  \Phi^I &=& \left\{ \underbrace{\phi^a}_{m} ,\,  \underbrace{\Delta, \, B}_{2} , \, \underbrace{\tau^\Lambda , \,
  \sigma_\Pi}_{2n} \, \right\} \nonumber\\
    m &=& \mbox{dim}\frac{\mathrm{U_{D=4}}}{\mathrm{H_{D=4}}} \quad ;
  \quad 2 = \mbox{dim}\frac{\mathrm{SL(2,\mathbb{R})}}{\mathrm{O(2)}}
    \quad ; \quad n = \mbox{dim} \mathbf{W}
\label{origano}
\end{eqnarray}
and we can regard the sum  $\mathcal{L}_{\sigma-model} +
\mathcal{L}_{vec+sl_2}$ as the definition of  a new metric $\widetilde{h}_{IJ} \left (\Phi \right ) $
and a new $\sigma$--model
lagrangian on the set of $m +2 + 2n$ fields $\Phi^I$:
\begin{equation}
  \mathcal{L}_{\sigma-model} +
\mathcal{L}_{vec+sl_2} \equiv \, - \,  \sqrt{g} \, \widetilde{h}_{IJ} \left (\Phi \right ) \,
\partial_\mu \Phi^I \, \partial^\mu \, \Phi^J
\label{newmetrona}
\end{equation}
Eq. (\ref{origano}) is the field theoretical counterpart of the
general algebraic decomposition (\ref{gendecompo}) leading to the
presentation (\ref{genGD3}) of the Lie algebra $\mathrm{U}_{D=3}$.
This latter is the isometry algebra of the new metric
$\widetilde{h}_{IJ} \left (\Phi \right )$ defined in
eq.(\ref{newmetrona}). The invariance of $\widetilde{h}_{IJ} \left
(\Phi \right )$ under the original $D=4$ algebra
$\mathrm{U_{D=4}}$ is guaranteed by the symplectic embedding
(\ref{sympembed}) and by the transformation properties
(\ref{Qmtractransf}) of the matrix $\mathbf{M}(\phi)$. Indeed it
suffices to assign the scalar fields $\left\{ \tau^\Lambda,
\sigma_\Sigma\right\} $ to the $\mathbf{W}$ representation of
$\mathbb{U}_{D=4}$ realized by the matrices $\Lambda_\xi$ and the
invariance is proved. To this effect, note also that the term
$\tau^\Lambda \stackrel{\leftrightarrow}{\partial} \sigma_\Lambda$
can be more effectively written as:
\begin{equation}
  \tau^\Lambda \stackrel{\leftrightarrow}{\partial} \sigma_\Lambda = \left(
  \tau \, , \, \sigma \right) ^T \, \Omega \, \left( \begin{array}{c}
     \partial \tau \\
     \partial \sigma \
  \end{array}\right)
\label{tausigmadtau}
\end{equation}
which shows its symplectic invariance.
\par
Let us also note that, if we set $\tau^\Lambda = \sigma_\Sigma =0$,
namely if we suppress all the scalars coming from the vector fields
in $D=4$, the Ehlers lagrangian reduces to:
\begin{equation}
 \mathcal{L} = \sqrt{g}\,\left[ -2 {R}[g] -
\frac{1}{4\Delta^2}\left(\partial\Delta^2 + \partial
B^2\right)\right]
\label{reducedlarana}
\end{equation}
namely to an $\mathrm{SL(2,\mathbb{R})_E/O(2)}$ $\sigma$--model coupled
to $2D$--gravity. Indeed,
\begin{equation}
ds^2_{Poin} =  \frac{1}{\Delta^2}\left(d\Delta^2 + dB^2\right)
\label{poincaremetric}
\end{equation}
is the standard Poincar\'e metric on the upper complex plane
parametrized by the complex variable $z = B + {\rm i} \Delta$.
The Ehlers $\mathrm{SL(2,\mathbb{R})_E}$ generates isometries of the
reduced lagrangian (\ref{reducedlarana}) but also of the complete one
(\ref{newmetrona}). This is just the statement that the full isometry
algebra $\mathbb{U}_{D=3}$ always includes the
$\mathrm{SL(2,\mathbb{R})}$ subalgebra as asserted by
eq.(\ref{gendecompo}). Indeed the further invariance of the
metric $\widetilde{h}_{IJ} \left (\Phi \right
)$ under the transformations  $L^x$ and $W^{i\alpha}$ that close the
algebra (\ref{genGD3}) together with the generators $T^A$ of
$\mathrm{U_{D=4}}$ will be discussed in the next section
\ref{killini}.
\par
We can now conclude the Ehlers programme by performing the last step,
namely the further reduction from $D=3$ to $D=2$.
In the $\sigma$--model part of the lagrangian there is nothing to do
apart from restricting the dependence to the first two coordinates
$t,x_1$. The only novelty comes from the reduction of the Einstein
term. We choose the following reduction ansatz:
\begin{eqnarray}
ds_{(3)}^2 &=& ds^2_{(2)} + \rho^2dx_2^2
\nonumber\\
ds^2_{(2)} & = &-N^2dt^2 + \lambda^{2}dx_1^2
\label{reduceansatz}
\end{eqnarray}
which can be motivated as follows. In the first line there is no
off-diagonal term (internal-space-time), namely there is no
Kaluza-Klein vector. This is no restriction in the case of
reduction from $D=3$ to $D=2$ since in $D=2$ vector fields carry
no degrees of freedom. In the second line of
eq.(\ref{reduceansatz}) the diagonal form assigned to the $D=2$
metric is again an always available choice in two--dimensional
space--times, due to conformal invariance. With this choices we
obtain:
\begin{equation}
-2\sqrt{g^{[3]}}\, {R}[g^{[3]}] = -2\, \rho \, \sqrt{g^{[2]}}\, {R}[g^{[2]}]
\end{equation}
and the final form of the Ehlers lagrangian is:
\begin{equation}
  L^{D=2}_{Ehlers} \, = \, -2\, \rho \, \sqrt{g}\, {R}[g]
  - \rho \,  \sqrt{g} \, \widetilde{h}_{IJ} \left (\Phi \right ) \,
\partial_\mu \Phi^I \, \partial^\mu \, \Phi^J
\label{Ehlersfinal}
\end{equation}
which is just a standard $2D$ $\sigma$--model with target manifold the following coset
\begin{equation}
  \mathcal{M}^E_{target} = \mathrm{\frac{U_{D=3}}{H_{D=3}}}
\label{Ehlerstarget}
\end{equation}
and further  coupled to the dilaton $\rho$.
\subsection{Field theoretic realization of the $\mathbb{U}_{D=3}$ Lie algebra in the Ehlers reduction}
\label{killini} As just announced, in this section we write the
explicit form of the local transformations of the Ehlers fields
under the action of the duality algebra $\mathrm{U}_{D=3}$. As
explained above, the Ehlers lagrangian is that of  a standard
$\sigma$--model, with (\ref{Ehlerstarget}) as target manifold.
Hence the action of the Lie algebra generators on the fields is
obtained by standard techniques, once a parametrization of the
coset representative is given. Let us call $T_\mathcal{A}$ the
generators of $\mathrm{U}_{D=3}$, namely:
\begin{equation}
  T_\mathcal{A} =\left\{  T^A, L^x, W^{i\alpha}\right\}
\label{allgenera}
\end{equation}
and let $\mathbb{L}(\Phi) \, \in \, \mathrm{U}_{D=3}$ be the coset representative, depending on
the set of all $\sigma$-model fields. Following general prescriptions
we have:
\begin{equation}
 \xi^{\mathcal{A}} \,  T_\mathcal{A} \, \mathbb{L}(\Phi) = \xi^{\mathcal{A}} \, \left(  \mathbb{L}\left (
 \Phi +\delta_\mathcal{A}\Phi \right
  ) - \mathbb{L}(\Phi) \, W_\mathcal{A}(\Phi) \right)
\label{Tlambda}
\end{equation}
where $W_\mathcal{A}(\Phi) \, \in \,\mathbb{H}_{D=3}$ is a suitable
compact subalgebra compensator and $\xi^{\mathcal{A}}$ are generic parameters identifying an element of the Lie algebra
$\mathbb{U}_{D=4}$. With the definition (\ref{Tlambda})
the variations $\delta_\mathcal{A}\Phi$ fulfill the commutation relations
of the generators with identical structure constants:
\begin{eqnarray}
\left[ T_\mathcal{A} \, , \, T_\mathcal{B} \right]  & =  & f_{\mathcal{A} \mathcal{B}}^{\phantom{\mathcal{A}\mathcal{B}}
\mathcal{C}} \,
T_\mathcal{C} \nonumber\\
\delta_\mathcal{A} \, \delta_\mathcal{B} \, \Phi  - \delta_\mathcal{B} \, \delta_\mathcal{A} \, \Phi & = &
- f_{\mathcal{A} \mathcal{B}}^{\phantom{\mathcal{A}\mathcal{B}}\mathcal{C}}
\, \delta_\mathcal{C} \, \Phi
\label{deltadeltaphi}
\end{eqnarray}
and it is our programme to work them out explicitly.
\par
To this effect we consider the algebra (\ref{genGD3}) and we
introduce a new basis for the generators of the  $\mathrm{SL}(2,\mathbb{R})$ subalgebra,
and for those associated with the $\mathbf{W}$ representation. First
we recall that, by definition, the symplectic representation
$\mathbf{W}$ is even dimensional, namely $\mbox{dim} \, \mathbf{W} = 2\,
n_V$ where $n_V$ is the number of vector fields in $D=4$. Hence the
index $\alpha$ runs on a set of $2 n_V$ values that can be split into
two subsets of $n_V$ elements each, respectively corresponding to the
positive and negative weights of the representation, from the
algebraic view point, and to the electric and magnetic field strengths
from the physical view point. Hence we write:
\begin{equation}
  W^{i\alpha} \equiv \left\{ W^{i\Lambda} \, , \, W^{i}_{\phantom{i}\Sigma}\right\}
\label{Wrange}
\end{equation}
where the index $\Lambda$ is that which enumerates the vector field
strengths in the lagrangian (\ref{d4generlag}). Secondly we introduce
the further notations:
\begin{equation}
\begin{array}{lcrclcr}
 L_0 &=& {\sqrt{2}} \, \mathrm{L}^3   & ; &  L_{\pm}=\mathrm{L}^1\pm
\mathrm{L}^2 & \null & \null\\
  W^{\alpha} &\equiv& W^{1\alpha} &=& \left\{ W^{\Lambda} \, , \, W_{\Sigma}\right\} &\null & \null  \\
 \hat{W}^{\alpha} & \equiv &
 W^{2 \alpha}  & = &
 \left \{  \hat{W}^{\Lambda} \, , \,
\hat{W}_{\Sigma} \right\} & \null & \null \
\end{array}
\label{newnamesgener}
\end{equation}
In terms of these objects, the commutation relations (\ref{genGD3})
become:
\begin{eqnarray}
&&  \left [W^\Lambda,\hat{W}_\Sigma \right ] =
(\Lambda_A)^{\Lambda}_{\phantom{\Lambda}\Sigma}\, T^A \, + \,
\frac{1}{2\sqrt{2}} \, \delta^\Lambda_\Sigma \, L_0, \nonumber \\
&&  \left [W_\Lambda,\hat{W}^\Sigma \right ] =
(\Lambda_A)_{\Lambda}^{\phantom{\Lambda}\Sigma}\, T^A \, -\, \frac{1}{2\sqrt{2}} \, \delta^\Lambda_\Sigma \, L_0,
\nonumber \\
&&\left [L_+,W^\alpha \right] = \left [ L_-,\hat{W}^\alpha \right ] = 0, \nonumber\\
&& \nonumber \left [L_+,\hat{W}^\alpha \right ] = W^\alpha, \quad \left [L_-,W^\alpha \right] = \hat{W}^\alpha, \\
&& \nonumber \left [L_0,W^\alpha \right ] = \frac{1}{\sqrt{2}} \, W^\alpha, \quad
\left [L_0,\hat{W}^\alpha \right ] = -\frac{1}{\sqrt{2}} \,
\hat{W}^\alpha\nonumber\\
&&\left [L_0,L_\pm \right ] = \, \pm \sqrt{2} \, L_\pm \quad ; \quad \left [L_+,L_- \right
] \, = \, \sqrt{2} \, L_0
\label{betterGD3}
\end{eqnarray}
Let $T_a$ be a basis of generators for the solvable Lie algebra $Solv\left(\mathrm{U_{D=4}/H_{D=4}} \right)$ of
the scalar coset appearing in $D=4$ supergravity.
The full set of $D=3$ scalar fields and a basis for the solvable
Lie algebra $Solv\left(\mathrm{U_{D=3}/H_{D=3}} \right)$ can be paired in the
following way:
\begin{equation}
\begin{array}{cccl}
 \sqrt{2} \,  \log \Delta & \Leftrightarrow & L_0 & \null\\
  B & \Leftrightarrow & L_+ &\null\\
  \varpi^\alpha & \Leftrightarrow & W^\alpha &\alpha = 1,\dots, \, \ft 12 \, \mbox{dim} \,\mathbf{W} \, =  n_V  \\
  \phi^a & \Leftrightarrow & T_a & a=1,\dots, \mbox{dim} \, \left( {\mathrm{U_{D=4}}}/{\mathrm{H_{D=4}}}\right)
\end{array}
\label{fieldpairing}
\end{equation}
Correspondingly we write the coset representative as follows
\begin{equation}
\mathbb{L}(\Phi) = \exp\left(Solv\right) = \exp \left [{BL_+} \right ] \, \exp \left [{\varpi_\alpha
W^\alpha}\right] \, \mathbb{L}(\phi_a) \, \exp \left[\sqrt{2} \, \log \Delta \, L_0 \right ]
\end{equation}
where $\mathbb{L}(\phi_a) \in \mathrm{G}_{D=4}/\mathrm{H}_{D=4}$ is the coset representative for the scalar
$\sigma$--model of $D=4$ supergravity.
\par
Under a generic element  $\mathbb{U}_{D=3} \ni T \equiv \xi^{\mathcal{A}} \,
T_{\mathcal{A}}$ of the Lie algebra of parameters $\xi^{\mathcal{A}}$,
the Ehlers fields transform as follows:
\begin{itemize}
\item {\bf \textbf{generator} $\gamma =\xi^+ L_+$, that is the step--up operator of $\mathrm{SL(2,\mathbb{R})_E}$ }
\begin{equation}
\label{Lplus}
\delta_\gamma \varpi_\alpha = \delta_\gamma \Delta = \delta_\gamma \phi_a =
0  \quad ; \quad \delta B = \xi^+
\end{equation}
\item {\bf generator $\gamma \equiv \xi^-L_-$ that is the step--down operator of $\mathrm{SL(2,\mathbb{R})_E}$)}:
\begin{equation}
\delta_\gamma B = \left[\Delta^2 - B^2 -\frac{a}{2\cdot 4!}
\varpi_{\alpha_1}\dots
\varpi_{\alpha_4}\Lambda_A^{(\alpha_1\alpha_2}\Lambda^{A\alpha_3\alpha_4)}\right]\xi^-
\label{LminBvarpi}
\end{equation}
\begin{equation}
 \delta_\gamma \varpi_\alpha = \left[(\Delta - B)\varpi_\alpha -\frac{1}{3!}\varpi_{\alpha_1}\varpi_{\alpha_2}
\varpi_{\alpha_3}(\Lambda_A)_\alpha^{(\alpha_1}(\Lambda^A)^{\alpha_2\alpha_3)}\right]\xi^-
\label{LminBvarpi2}
\end{equation}
\begin{equation}
\delta_\gamma \Delta = - 2B\Delta\xi_-, \quad \delta_\gamma\phi^a =
- \frac{\varpi^\alpha
\varpi^\beta}{2}\Lambda^B_{\alpha\beta} \, k_B^a(\phi)\xi^-,
\label{LminDeltaphi}
\end{equation}
where $k_B^a(\phi)$ are the Killing vectors of $\mathbb{U}_{D=4}$.
\item {\bf \textbf{generator} $\gamma = \xi^0L_0$, that is the Cartan of $\mathrm{SL(2,\mathbb{R})_E}$}
\begin{equation}
\label{L0}
\delta_\gamma B = B \xi^0 \quad ; \quad \delta_\gamma \varpi_\alpha =
 \frac{1}{2}\varpi_\alpha\xi^0  \quad ; \quad  \delta \Delta = \xi^0\, \Delta
\quad ; \quad  \delta \phi^a = 0
\end{equation}
\item {\bf \textbf{generators}  $\gamma = \xi^A\, T_A$ of the $\mathbb{U}_{D=4}$ Lie algebra}
\begin{equation}
\label{Ta}
\delta_\gamma B = \delta_\gamma \Delta = 0  \quad ; \quad \delta_\gamma \varpi_\alpha
 =  \xi^A (\Lambda_A)_\alpha^{\phantom{\alpha}\beta}\, \varpi_\beta
 \quad ; \quad \delta_\gamma \phi^a  =  \xi^A \, k_A^a(\phi)
\end{equation}
\item {\bf \textbf{generators} $\xi_\alpha W^\alpha$ associated with weights of $\mathbf{W}$}
\begin{equation}
\label{W}
\delta_\gamma \varpi_\alpha = \xi_\alpha  \quad ; \quad
\delta_\gamma B = \frac{1}{4}\varpi_\alpha \, \Omega^{\alpha\beta} \, \xi_\beta
\quad ; \quad \delta_\gamma \Delta = \delta_\gamma \phi_a = 0
\end{equation}
\item {\bf \textbf{generators} $\gamma =\hat{\xi}_\alpha \hat{W}^\alpha$ associated with
the dual weights of $\mathbf{W}$}
\begin{equation}
\label{hatWB}
\delta_\gamma B = - \frac{1}{24}\, \hat{\xi}_\alpha \,
\varpi_{\beta_1}\varpi_{\beta_2}\varpi_{\beta_3}
(\Lambda_A)^{\beta_3\alpha}(\Lambda^A)^{\beta_1\beta_2} \, + \,
\frac{1}{2} \, \Omega^{\alpha\beta} \, \hat{\xi}_\alpha \, \varpi_{\beta} \, (B +
\Delta)
\end{equation}
\begin{equation}
\label{hatWvarpi}
\delta_\gamma \varpi_\beta =- \frac{1}{2}\left[\hat{\xi}_\alpha \,
\varpi_{\beta_1}\varpi_{\beta_2}
\Lambda_A^{\beta_2\alpha}(\Lambda^A)_\beta^{\beta_1} \, + \,
\frac{1}{4}\, \hat{\xi}_\alpha \, \varpi_\beta \varpi_\gamma
\Omega^{\gamma\alpha}\right]\, + \, (\Delta - B)\, \hat{\xi}_\beta
\end{equation}
\begin{equation}
\label{hatWDeltaphi}
\delta_\gamma \Delta= - \frac{1}{2}\, \hat{\xi}_\alpha \,
\varpi_\beta\, \Omega^{\beta\alpha}\, \Delta \quad ; \quad
 \delta_\gamma\phi^a = - \hat{\xi}_\alpha \,
 \varpi_\beta \, \left( \Lambda^A\right) ^{\beta\alpha} \,
 k_A^a(\phi)
\end{equation}
\end{itemize}
This concludes the analysis of the Ehlers lagrangian and of its
symmetries. In the next section we turn our attention to the Matzner
Misner dimensional reduction scheme.
\section{The   $\mathrm{Sp(2n,\mathbb{R})} \mapsto \mathrm{SO(2n,2n)}$ embedding and the
general Matzner--Misner lagrangian} The key point in the Maztner
Misner reduction of a general supergravity lagrangian from $D=4$
to $D=2$ is provided by the following pseudorthogonal embedding:
\begin{equation}
  \mathrm{Sp(2n,\mathbb{R})} \mapsto \mathrm{SO(2n,2n)}
\label{pseudorthogemb}
\end{equation}
which we presently discuss. In $D=4$ the duality group
$\mathbb{G}_{D=4}$ is simultaneously realized as an isometry group
of the scalar manifold metric $h_{ab} (\phi)$ and as a group of
electric-magnetic duality transformations on the vector field
strengths $\mathcal{F}^{ | \Lambda}_{\mu\nu}$ as we have already
emphasized. The general form of the lagrangian was given in
eq.(\ref{d4generlag}). Furthermore, as already stressed there is
always a symplectic embedding of the duality group
$\mathcal{G}_{D=4}$ as mentioned in eq.(\ref{sympembed}) where
$n=n_V$ is the total number of vector fields in the theory. Let us
now consider the general form of a lagrangian in $D=2$. Here we
have two type of scalars, namely the  scalars-scalars $\phi^a$ and
the \textit{twisted scalars} or \textit{scalar-forms}
$\pi^\alpha$. This distinction is important. The scalars-scalars
appear in the lagrangian under the form of a usual $\sigma$-model
while the \textbf{scalar-forms} appear  only covered by
derivatives and in two terms, one symmetric, one antisymmetric.
The coefficients of these two terms are matrices depending on the
scalars-scalars. Explicitly the lagrangian has the form (see
\cite{myparis} for a general review):
\begin{eqnarray}
L_{(D=2)} & = & \int \, d^2x \,\sqrt{-\mbox{det} g} \left \{-2 \, R[g]
 - \ft 1 6 \, h_{ab}(\phi) \partial_\mu \phi^a \partial _\mu \phi^b
\right. \nonumber\\
\null & \null & \left. +   \, \ft 12 \,\kappa \, \left [-
\partial_\mu \pi^\alpha \,  \gamma_{\alpha\beta}(\phi) \, \partial_\mu \pi^\alpha \,
+ \partial_\mu \pi^\alpha \, \theta_{\alpha\beta}(\phi)  \, \partial_\nu \pi^\beta \, \epsilon^{\mu\nu} \right]\right\}
\label{d2generlag}
\end{eqnarray}
where $\kappa$ is a normalization parameter that can always be
reabsorbed into the definition of the $0$-forms $\pi^\alpha$
and where, according to the general theory for the dimensions $D=4\nu + 2$
(see section 2.4 of \cite{myparis})
if $\mathcal{G}_{D=2}^{iso}$ is the isometry group of the
$\sigma$-model metric $h_{ab}(\phi)$, then there is a
pseudo-orthogonal embedding:
\begin{equation}
  \mathcal{G}_{D=2}^{iso} \mapsto \mathrm{SO(m,m) }
\label{ortoembed}
\end{equation}
where $m$ is the total number of the scalar-forms $\pi^\alpha$.
 Hence for each element $\xi \in \mathcal{G}_{D=2}^{iso}$ we have its
representation by means of a  suitable pseudorthogonal matrix:
\begin{equation}
  \xi \mapsto \Sigma_\xi \equiv \left( \begin{array}{cc}
     \mathcal{A}_\xi & \mathcal{B}_\xi \\
     \mathcal{C}_\xi & \mathcal{D}_\xi \
  \end{array} \right)
\label{embeddusmatra2}
\end{equation}
which satisfies the defining equation:
\begin{equation}
  \Sigma_\xi ^T \, \left( \begin{array}{cc}
     \mathbf{0}_{m \times m}  & { \mathbf{1}}_{m \times m} \\
     { \mathbf{1}}_{m \times m}  & \mathbf{0}_{m \times m}  \
  \end{array} \right) \, \Sigma_\xi = \left( \begin{array}{cc}
     \mathbf{0}_{m \times m}  & { \mathbf{1}}_{m \times m} \\
     { \mathbf{1}}_{m \times m}  & \mathbf{0}_{m \times m}  \
  \end{array} \right)
\label{definingorto}
\end{equation}
 implying the following relations on the $m \times m$ blocks:
\begin{eqnarray}
\mathcal{A}^T \, \mathcal{C} + \mathcal{C}^T \, \mathcal{A} & = & 0 \nonumber\\
\mathcal{A}^T \, \mathcal{D} + \mathcal{C}^T \, \mathcal{B} & = & \mathbf{1}\nonumber\\
\mathcal{B}^T \, \mathcal{C} + \mathcal{D}^T \, \mathcal{A}& = &  \mathbf{1}\nonumber\\
\mathcal{B}^T \, \mathcal{D} + \mathcal{D}^T \, \mathcal{B} & =  & 0
\label{pseudoerele}
\end{eqnarray}
Defining the $ m \times m$ matrix
\begin{equation}
  \mathcal{M} \equiv \theta + \gamma
\label{coulor}
\end{equation}
under the group $\mathcal{G}_{D=2}^{iso}$ it transforms as follows:
\begin{equation}
  \mathcal{M}^\prime = \left( \mathcal{ C} + \mathcal{D }\, \mathcal{M}\right) \,
    \left( \mathcal{A} + \mathcal{B} \,\mathcal{M}\right)
  ^{-1}
\label{Mtransfa2}
\end{equation}
\begin{equation}
 - {\mathcal{M}^T}^\prime = \left( \mathcal{ C} - \mathcal{D }\, \mathcal{M}^T\right) \,
    \left( \mathcal{A} -\mathcal{B} \,\mathcal{M}^T \right)
  ^{-1}
\label{Mtransfa2bis}
\end{equation}
We can now link the $D=4$ supergravity lagrangian (\ref{d4generlag}) to the $D=2$
lagrangian that will emerge from the Matzner-Misner
reduction, which is of the  type (\ref{d2generlag}).
\par
Consider the transformation rule of the matrix $\mathcal{N}$ and
multiply it by $-{\rm i}$, we obtain:
\begin{equation}
 \left(-{\rm i}  \mathcal{N}\right) ^\prime = \left[ -{\rm i}  C + D \, \left( -{\rm i} \mathcal{N}\right)
 \right] \, \left[ A + {\rm i} B \,\left(  -{\rm i}\mathcal{N}\right)
 \right]
  ^{-1}
\label{Ntransfabis2}
\end{equation}
Next let us represent the imaginary unit by a $2\times 2$ matrix
$\varepsilon$ such that
\begin{equation}
  \varepsilon^2 = -\mathbf{1}_{2 \times 2}
\label{quadrauno}
\end{equation}
In this way we can write:
\begin{equation}
  \begin{array}{ccccl}
     \mathcal{M }& \equiv & -{\rm i} \, \mathcal{N} & = & \mbox{Im} \mathcal{N} \otimes \mathbf{1}_{2\times 2}
     -\mbox{Re} \mathcal{N} \otimes \mathbf{\varepsilon}_{2\times 2} \\
     \mathcal{A} & \equiv & A & = & A \otimes \mathbf{1}_{2\times 2} \\
     \mathcal{B} & \equiv & {\rm i} B & = & B \otimes \mathbf{\varepsilon}_{2\times 2} \\
     \mathcal{C} & \equiv & - {\rm i}C & = &  -\, C \otimes \mathbf{\varepsilon}_{2\times 2} \\
     \mathcal{D} & \equiv & D & = & D \otimes \mathbf{1}_{2\times 2} \\
  \end{array}
\label{agnosi}
\end{equation}
and the transformation (\ref{Ntransfabis2}) becomes the
transformation (\ref{Mtransfa2}). Furthermore the $2n \times 2n$
blocks $ \mathcal{A},\mathcal{B},\mathcal{C},\mathcal{D}$, defined by
equation (\ref{agnosi}) satisfy the relations (\ref{pseudoerele}) as
a consequence of (\ref{symplerele}) and (\ref{quadrauno}). This
provides the required embedding (\ref{pseudorthogemb}) in the form:
\begin{equation}
\mathrm{Sp(2n,\mathbb{R})}  \, \ni \,  \left( \begin{array}{cc}
     A_\xi & B_\xi \\
     C_\xi & D_\xi \
  \end{array} \right) \, \mapsto \, \left(\begin{array}{cc||cc}
     A & 0 & 0 & B \\
     0 & A & -B & 0 \\
     \hline
     \hline
     0 & -C & D & 0 \\
     C & 0 & 0 & D \
  \end{array} \right) \, \in \,\mathrm{SO(2n,2n)}
\label{orpelli}
\end{equation}
The matrix $\mathcal{M}$ transforms correctly under the pseudorthogonal
embedded group $\mathcal{G}^{iso}_{D=2}$ as a consequence of the same
symplectic embedding of $\mathcal{G}^{iso}$. From its definition in
the first of equations (\ref{agnosi}), we derive the symmetric and
antisymmetric parts ($A,B=1,2$):
\begin{eqnarray}
  \gamma_{\alpha\beta} & = & \mbox{Im} \mathcal{N} \otimes \mathbf{1}_{2\times
  2}\, = \, \mbox{Im} \mathcal{N}_{\Lambda\Sigma} \, \delta_{AB}
  \nonumber\\
 \theta_{\alpha\beta} & = & - \mbox{Re} \mathcal{N} \otimes \mathbf{1}_{2\times
  2}\, = - \, \mbox{Re} \mathcal{N}_{\Lambda\Sigma} \, \varepsilon_{AB}
  \nonumber\\
\label{invasiocamp}
\end{eqnarray}
It follows that if the reduced lagrangian takes the $D=2$ form:
\begin{eqnarray}
L^{MM}_{(D=2)} & = & \int \, d^2x \,\sqrt{-\mbox{det} g} \left \{- 2 \, R[g]
 - \ft 1 6 \, h_{ab}(\phi) \partial_\mu \phi^a \partial _\mu \phi^b
\right. \nonumber\\
\null & \null & \left. +   \ft 12 \, \left [-
\nabla_\mu \pi^{\Lambda |A} \,  \mbox{Im} \mathcal{N}_{\Lambda\Sigma}(\phi) \, \delta_{AB}  \,
\nabla^\mu \pi^{\Sigma |B} \,
+ \nabla_\mu \pi^{\Lambda | A}  \, \mbox{Re} \mathcal{N}_{\Lambda\Sigma} \, \varepsilon_{AB}  \, \nabla_\nu
\pi^{\Sigma |B}
\, \epsilon^{\mu\nu} \right]\right.\nonumber\\
&&\left. \,  + \, \mbox{more} \right \} \label{d2generlagMM}
\end{eqnarray}
then the symmetry group $\mathbb{U}_{D=4}$ is realized in $D=2$
just as in $D=4$, namely as an isometry group on the
scalars-scalars and as as a group of generalized duality
transformations on the \textit{scalars-forms} or \textit{twisted
scalars}.  An enlargement of symmetries can arise, in the Matzner
Misner reduction only by the $\emph{more}$ that we mentioned in
eq.(\ref{d2generlagMM}). What is this more? It is just the
contribution of pure gravity, that in the Matzner--Misner
reduction yields another $\mathrm{SL(2,\mathbb{R})_{MM}/O(2)}$
sigma-model as we have already anticipated in the introduction.
Hence, taking into account also the constant shifts of the scalar forms $\pi^{\alpha}$, the symmetry of Matzner--Misner lagrangian will eventually
be:
\begin{equation}
 \mathbf{U}^{MM}_{D=2} = \mathbf{U}_{D=4} \times \mathrm{SL(2,\mathbb{R})_{MM}}\rtimes \mathbb{R}^m
\label{UMMD=2}
\end{equation}
opposed to the more extended $\mathbf{U}_{D=3}$ symmetry displayed by
the Ehlers lagrangian. 
\par
Let us now see the details of the Matzner--Misner reduction and
let us prove that the reduced lagrangian does indeed take the form
(\ref{d2generlagMM}).
\subsection{The Matzner--Misner reduction}
\label{MMgrav} The  reduction \`a la Matzner--Misner  is just the
straightforward dimensional reduction of the $D=4$ lagrangian
(\ref{d4generlag}) on a 2-dimensional torus $\mathrm{T^2}$,
without any dualization of the vector fields. To this effect we
split the space-time indices in the following way: $\mu ,\nu =
0,1$ and $i,j=2,3$.  Then the
 Matzner--Misner reduction consists of two steps:
\begin{enumerate}
\item dimensional reduction of the $D=4$ supergravity lagrangian
in eq. (\ref{d4generlag}) on a 2--torus. This gives a system that is a
mixture of an ordinary  $\sigma$--model and a twisted
$\sigma$--model, namely the coupling of $0$-forms with duality symmetries as described in the previous section.
Both sectors
are coupled to  $2$--dimensional dilaton
gravity and we have  a dilaton kinetic term.
 \item a rescaling of the 2--dimensional metric appropriate to
 cancel the kinetic term of the 2--dimensional dilaton, thus
 putting the lagrangian into a standard form.
 \end{enumerate}
The reduction ansatz for the $D=4$ metric is the following one:
\begin{equation}
ds_{(4)}^2 = ds_{(2)}^2 + g_{ij}dx^i dx^j,
\label{Mmmetansatz}
\end{equation}
where the indices $i,j = 2,3$ correspond to the dimensions of the internal torus.
More explicitly, we parametrize the internal metric
$g_{ij}$ as it follows:
\begin{equation}
g_{ij} = \rho\left(\matrix{\tilde{\Delta} +
\frac{\tilde{B}^2}{\tilde{\Delta}} &
\frac{\tilde{B}}{\tilde{\Delta}}\cr
\frac{\tilde{B}}{\tilde{\Delta}}&\frac{1}{\tilde{\Delta}} }\right)
\label{MMmetric}
\end{equation}
where $\tilde{\Delta}$ denotes the Kaluza--Klein scalar, $\rho =
\sqrt{-\mbox{det}(g_{ij})}$ is the two-dimensional dilaton and
$\tilde{B} = b_2 = g_{23}\tilde{\Delta}/{\rho}$ denotes the
internal component of the Kaluza-Klein vector. In this frame the
dimensional reduction of the Einstein term yields:
\begin{equation}
-2\sqrt{-\mbox{det}g^{(4)}}{R}[g^{(4)}]=
-2\sqrt{-\mbox{det}g_{(2)}}\rho\left[{R}[g_{(2)}] + \ft
18(g^{ik}g^{jl} - g^{il}g^{jk})\partial g_{ik}\partial
g_{jl}\right])
\end{equation}
Inserting the explicit form of the internal metric
(\ref{MMmetric}), we obtain the kinetic term for an
$\mbox{SL}(2,\mathbb{R})$ $\sigma$--model plus the kinetic term
for the 2--dimensional dilaton $\rho$
\begin{equation}
 \mathcal{L}_{grav} = -2\sqrt{-\mbox{det}g_{(2)}}\rho\left(\mathcal{R}_2[g_{(2)}] +
 \ft 14\frac{(\partial\rho)^2}{\rho^2} -
 \frac{1}{4\tilde{\Delta}^2}[(\partial\tilde{\Delta})^2 + (\partial\tilde{B})^2] \right)
 \end{equation}
Now we perform the second step, absorbing the dilaton kinetic term
in the gravity term by the rescaling $g^{(2)}_{\mu\nu} =
\bar{g}^{(2)}_{\mu\nu}\rho^{-\ft 12}$.\footnote{Any two dimensional
metric is conformally flat, so there will be no overall rescaling
of the gravity lagrangian, but there is a contribution from the
boundary term, that exactly cancels $ \ft
14\frac{(\partial\rho)^2}{\rho^2}$.}
\par
Eventually the $D=4$ gravity term reduced \`a la Matzner--Misner
gives a pure $\mbox{SL}(2,\mathbb{R})$ $\sigma$--model coupled to
 2D dilaton--gravity
\begin{equation}
\label{MMgravfinal}\mathcal{L}^{MM}_{grav} =
-2\sqrt{-\mbox{det}\bar{g}_{(2)}}\rho\left(\mathcal{R}_2[\bar{g}_{(2)}]
- \frac{1}{4\tilde{\Delta}^2}[(\partial\tilde{\Delta})^2 +
(\partial\tilde{B})^2] \right)
\end{equation}
\par
The lagrangian (\ref{MMgravfinal}) is formally identical to the
reduced lagrangian (\ref{reducedlarana}) of the Ehlers case and as
such admits an $\mathrm{SL(2,\mathbb{R})_{MM}}$ group of
isometries. Yet the fields $\Delta, B$ are different from the
fields ${\tilde \Delta}, {\tilde B}$ a non local relation existing
between the two. As noted years ago by Nicolai
(\cite{NicolaiEHMM}) the coexistence of the two formally identical
lagrangians (\ref{reducedlarana}) and (\ref{MMgravfinal}),
together with the non-local map between the fields $\Delta, B$ and
the fields ${\tilde \Delta}, {\tilde B}$  is the mechanism which
promotes the $\mathrm{SL(2,\mathbb{R})}$ symmetry to its affine
extension in pure gravity. Our present paper aims at generalizing
the same mechanism to all cases of supergravity. The further
complicacy in the general case is that two lagrangians, Ehlers and
Matzner--Misner are not formally identical and have different
isometry algebras. Yet they coincide in one sector and it is there
that the affine extension mechanism works.
\par
So let us continue by considering the dimensional reduction of the gauge fields.
To this effect we restrict the structure of the metric in $D=2$,
assuming that there is no off-diagonal term
\begin{equation}
ds_{(2)}^2 = -\tilde{N}^2dt^2 + \tilde{\lambda}^2dx_1^2
\end{equation}
Then, in the absence of Kaluza--Klein fields, there is no need to
redefine the $D=2$ vectors and they can be parametrized as
follows:
\begin{equation}
A^\Lambda_{[D=4]} = A_{(2)\mu}^{\Lambda} dx^\mu + A_i^\Lambda \,
dx^i
\end{equation}
where $A_{(2)\mu}^{\Lambda}$ are the two-dimensional gauge fields
that do not propagate. In what follows we set the gauge: $A_{(2)\mu}^{\Lambda}=0$.
\par
This being set, the  vector sector of the $D=4$ lagrangian
(\ref{d4generlag}) reduces as follows
\begin{eqnarray}
\ft 12 \, \mbox{Re}\mathcal{N}_{\Lambda\Sigma} \,
F^\Lambda_{\hat{\mu}\hat{\nu}} \,
F^\Sigma_{\hat{\rho}\hat{\sigma}}
\epsilon^{\hat{\mu}\hat{\nu}\hat{\rho}\hat{\sigma}} &=& 2 \,
\epsilon^{\mu\nu}\epsilon^{ij}\mbox{Re}\,
\mathcal{N}_{\Lambda\Sigma} \,\partial_\mu A_i^\Lambda
\partial_\nu A_j^\Sigma, \\ \sqrt{-\mbox{det} \, g_{[4]}} \, \mbox{Im} \,
\mathcal{N}_{\Lambda\Sigma} \, F^\Lambda_{\hat{\mu}\hat{\nu}}
F^{\Sigma|\hat{\mu}\hat{\nu}} &=& 2\sqrt{-\mbox{det}g_{(2)}}g^{ij}
\, \mbox{Im} \, \mathcal{N}_{\Lambda\Sigma}
 \,\partial_\mu A_i^\Lambda
\, \partial^\nu A_j^\Sigma  \label{vectreduceMM}
\end{eqnarray}
In order to make contact with the action (\ref{d2generlagMM}) described in the previous section, which implements
the $\mathrm{Sp(2n,\mathbb{R})} \mapsto \mathrm{SO(2n,2n)} $ embedding we have to introduce new notations.
We define the following doublets of $0$-forms
with flat internal indices
 \begin{equation}
 \pi^{\Lambda|A} =
e^{iA}A_i^\Lambda
 \end{equation}
where $e^{iA}$ is the internal vielbein defined as usual by:
\begin{equation}
  e^{iA} \, e^{jB} \, \delta_{AB} \, = \, g^{ij}
\label{turno}
\end{equation}
The $0$--forms $ \pi^{\Lambda|A}$ are by definition sections of the
$\mathrm{O(2)}$ vector bundle defined by the coset
$\mathrm{SL(2,\mathbb{R})_{MM}/O(2)}$. Introducing the coset
representative of this latter
\begin{eqnarray}
  \mathbb{L}_2(\tilde{\Delta},\tilde{B}) &=& \left(\begin{array}{cc}
     \sqrt{\tilde{\Delta}} & \frac{\tilde{B}}{\sqrt{\tilde{\Delta}}} \\
     0 & \frac{1}{\sqrt{\tilde{\Delta}}} \
  \end{array} \right) \equiv \exp \left [ \tilde{B} L_+ \right] \, \exp \left
  [ \sqrt{2} \,\log \tilde{\Delta} \, L_0\right] \nonumber \\
  L_+ & = & \left(\begin{array}{cc}
     0 & 1 \\
     0 & 0 \
  \end{array} \right) \quad ; \quad  L_0 \, = \,  \left(\begin{array}{cc}
     \frac{1}{\sqrt{2}} & 0 \\
     0 & - \frac{1}{\sqrt{2}} \
  \end{array} \right)
\label{cosetsl2}
\end{eqnarray}
we can identify $e^{iA}=\mathbb{L}^{-1}_2(\Delta,B)$
so that the  precise  relation between the vector
components $A_i^\Lambda$ and the fields $\pi^{\Lambda|A}$ reads as
follows:
\begin{eqnarray}
&& \nonumber A_3^\Lambda = \rho^{1/2}\tilde{\Delta}^{-1/2}\pi^{3|\Lambda}, \\
&& A_2^\Lambda = \rho^{1/2}\tilde{\Delta}^{1/2}\pi^{2|\Lambda} +
\rho^{1/2}\tilde{\Delta}^{-1/2}\pi^{3|\Lambda}.
\end{eqnarray}
Furthermore, separating the left invariant
one--form $\mathbb{L}_2^{-1} d\mathbb{L}_2$ into its vielbein and connection parts:
\begin{eqnarray}
  \mathbb{L}_2^{-1} d\mathbb{L}_2 &=&  \omega \,
  \mathbb{H} + \mbox{vielbein}\nonumber\\
  \mathbb{H} & \equiv & \left( \begin{array}{cc}
     0 & 1 \\
     -1 & 0 \
  \end{array}\right)
\label{decompsl2}
\end{eqnarray}
we can define the $\mathrm{O(2)}$--covariant derivatives of the
$0$--forms $\pi^{\Lambda|A}$ as follows:
\begin{equation}
  \nabla_\mu \, \pi^{\Lambda|A} \, = \, \partial_\mu \pi^{\Lambda|A}
  \, + \, \epsilon^{AB} \, \omega_\mu \,  \pi^{\Lambda|B}
\label{covderpi}
\end{equation}
In this way the fully reduced and redefined vector lagrangian in $D=2$
takes the following form:
\begin{eqnarray}
L^{MM}_{vec} &=&  2 \left [\sqrt{-\mbox{det} g_{(2)}}\,
\mbox{Im} \mathcal{N}_{\Lambda\Sigma}(\phi) \,\delta_{AB}  \,
\nabla_\mu \pi^{\Lambda |A} \, \partial_\mu \pi^{\Sigma |B} \,
\right. \nonumber\\
&& \left. +
\mbox{Re} \mathcal{N}_{\Lambda\Sigma} \,\varepsilon_{AB}  \,
\nabla_\mu \pi^{\Lambda | A}  \,  \nabla_\nu \pi^{\Sigma |B} \,
\epsilon^{\mu\nu} \right]
\label{MMvectoLag}
\end{eqnarray}
The reduction of the $D=4$ scalars is straightforward and gives
\begin{equation}
\mathcal{L}^{MM}_{\sigma}
=-\frac{\rho}{6}\sqrt{-\mbox{det}g_{(2)}} \, h_{ab}\, \partial^\mu\phi^a \, \partial_\mu\phi^{b}
\label{MmsigmaLag}
\end{equation}
Putting together these results  we have completed all the steps of
the Matzner--Misner reduction and we have shown that the complete
lagrangian
\begin{equation}
  \mathcal{L}^{MM}_{complete} = \mathcal{L}^{MM}_{grav} \, + \,
  \mathcal{L}^{MM}_{\sigma}\, + \, \mathcal{L}^{MM}_{vec}
\label{completeMM}
\end{equation}
is, as announced, of the form (\ref{d2generlagMM}) realizing the pseudorthogonal
embedding $\mathrm{Sp(2n,\mathbb{R})} \mapsto \mathrm{SO(2n,2n)}$ and
that the \textit{more} is the Matzner--Misner
$\mathrm{SL(2,\mathbb{R})_{MM}/O(2)}$ sigma model:
\begin{equation}
\mbox{more} \, = \, - \frac{1}{4\tilde{\Delta}^2}[(\partial\tilde{\Delta})^2 +
(\partial\tilde{B})^2]
\label{Mmsl2}
\end{equation}
\par
\subsection{Action of the $\mbox{SL}(2,\mathbb{R})_{MM}$ algebra on the Matzner--Misner fields}
We denote the generators of the Matzner--Misner
$\mbox{SL}(2,\mathbb{R})_{MM}$ group as $\left ( L^{mm}_0,
L^{mm}_+, L^{mm}_- \right )$, with just the same conventions as in
the case of the Ehlers $\mathrm{SL(2,\mathbb{R})_{E}}$ group. We
will then see that, using these three generators, we can cook up
the Chevalley-Serre triplet $\left( h_0,e_0,f_0\right) $ needed to
generate the affine extension of the ${\mathbb{U}}_{D=3}$ Lie
algebra. Yet the way $\left ( L^{mm}_0, L^{mm}_+, L^{mm}_- \right
)$ are associated with $\left( h_0,e_0,f_0\right) $ is different
in the case of \textit{pure gravity} and in the case of \textit{
extended supergravity}, namely when we have also the vector sector
of the theory. This is not surprising since the affine Cartan
matrix of pure gravity namely $A^\wedge_1$ is structurally
different from those obtained in the various  supergravity models:
in pure gravity we add a double line to the Dynkin diagram, while
in supergravity we always add a simple line. For this reason we do
not  immediately commit ourselves with the association:
\begin{equation}
  \left ( L^{mm}_0, L^{mm}_+, L^{mm}_- \right ) \, \Leftrightarrow \,
\left( h_0,e_0,f_0\right)
\label{association}
\end{equation}
and we just write the transformations of the $\mathrm{SL(2,\mathbb{R})_{MM}}$
generators on the Matzner--Misner fields $\tilde{\Delta}$, $\tilde{B}$ and
$\pi^{\Lambda|A}$. Applying the general formula (\ref{Tlambda}) to the
coset representative (\ref{cosetsl2}) of the Matzner--Misner model
we easily derive the local transformations of the $\mbox{SL}(2,\mathbb{R})_{MM}$ fields
$\tilde{\Delta}$, $\tilde{B}$ which is the standard one:
\begin{equation}
\begin{array}{lcl}
  \delta_{L^{mm}_0}\tilde{\Delta} = \sqrt{2}\tilde{\Delta} &; &  \delta_{L^{mm}_0}\tilde{B} =
 \sqrt{2}\tilde{B} \\
 \delta_{L_+^{mm}} \tilde{\Delta} = 0 & ; & \delta_{L_+^{mm}}\tilde{B} = 1, \\
  \delta_{L_-^{mm}}\tilde{\Delta} =
- 2\tilde{\Delta}\tilde{B},
 & ; & \delta_{L_-^{mm}}\tilde{B} =
\tilde{\Delta}^2 - \tilde{B}^2
\end{array}
\end{equation}
On the other hand, the scalars $\pi^{\Lambda|A}$ are, as we have emphasized in a linear doublet
representation of the $\mbox{O}(2)$--compensator so that:
\begin{eqnarray}
 && \nonumber \delta_{L^{mm}_0}\pi^{\Lambda|A} = \delta_{L_+^{mm}}\pi^{\Lambda|A} = 0,
 \\ && \delta_{L_-^{mm}}\pi^{\Lambda|2} =
 -\tilde{\Delta}\pi^{\Lambda|3}, \quad \delta_{L_-^{mm}}\pi^{\Lambda|3} =
 \tilde{\Delta}\pi^{\Lambda|2}
\end{eqnarray}
This concludes the discussion of the local symmetries of our two
lagrangian models. We can now go to the affine extension by
considering also non local symmetries. This involves consideration
of the non local map $\mathcal{T}$ which relates the two set of
fields the Ehlers and the Massner Misner ones which we can trace
back reconsidering the steps of our two reduction schemes.
\section{The $\mathcal{T}$-map and the affine extension}
The two lagrangian models  obtained via the two  dimensional
reductions schemes, respectively \`a la Ehlers and \`a la Matzner--Misner, are related to each other by
a  non--local  transformation which is the main token in combining
the symmetries of the Ehlers lagrangian with those of the
Matzner--Misner one. Let us describe this map in detail. We write
\begin{equation}
\mathcal{T}: \quad \begin{array}{ccc} \mbox{MM}  & \rightarrow &  \mbox{Ehlers} \\
\tilde{N}  & \rightarrow &  N
\\ \tilde{\lambda} & \rightarrow & \lambda \\ \tilde{\rho} & \rightarrow & \rho \\
\tilde{\Delta} & \rightarrow & \Delta \\
\tilde{B} & \rightarrow & B \\    \pi^{\Lambda|A}
 & \rightarrow &  \tau^\Lambda, \sigma_\Lambda \\
 \tilde{\phi}^a  & \rightarrow &  \phi^a
\end{array}
\end{equation}
which is a generalization of the so named
Kramer--Neugebauer transformation, firstly considered in the dimensional reduction  of  pure
$D=4$ Einstein gravity \cite{Kramer&Neug}.
\par
Tracing back the rescalings made for the two--dimensional metric
and the dilaton, we reconstruct immediately the following map:
\begin{eqnarray}
\nonumber\tilde{N} &=& N \rho^{1/4}\Delta^{-1/2} \\
\nonumber \tilde{\lambda} &=& \lambda \rho^{1/4}\Delta^{-1/2}\\
\nonumber\tilde{\rho} &=& \rho
\label{taumappa0}
\end{eqnarray}
The Kaluza--Klein scalars in Ehlers and Matzner--Misner models are
connected in the following way
\begin{equation}
\tilde{\Delta} = \frac{\rho}{\Delta}
\label{taumappa1}
\end{equation}
The scalars coming from the off--diagonal part of the 4--dimensional
metric and from the 4--dimensional vectors ($B, \tau^\Lambda,
\sigma_\Lambda$), which were partially dualized in the Ehlers
reduction (\ref{dualrula}), have a non--local relation to the
Matzner--Misner fields ($\tilde{B}, \pi^{\Lambda|A}$)
\begin{eqnarray}
\partial_\mu\tilde{B} &=& -\ft
12\frac{N\lambda\rho}{\Delta^2}g_{(2)}^{\rho'\rho}\epsilon_{\mu\rho}\left[\partial_\rho
B +
\ft 12\sigma_\Lambda\stackrel{\leftrightarrow}{\partial}_\rho \tau^\Lambda\right], \\
\nonumber  \pi^{\Lambda|3} &=& \tau^\Lambda \Delta^{-1/2}, \\
\nonumber \ft 12\partial_\mu(\rho\Delta^{-1/2}\pi^{\Lambda|2}) &=&
\frac{N\rho\lambda}{2\Delta}\mbox{Im}\mathcal{N}^{-1\Lambda\Sigma}\left[
\mbox{Re}\mathcal{N}_{\Sigma\Omega}\partial_\nu \tau^\Omega +
\partial_\nu\sigma_\Sigma\right]g^{\nu\nu}\epsilon_{\mu\nu}-\ft 12\tau^\Lambda\partial_\mu\tilde{B}
\label{taumappa2}
\end{eqnarray}
Finally, the $4$--dimensional scalars $\phi^a$ went through the dimensional reduction untouched and therefore are
the same in the Ehlers and in the Matzner--Misner models:
\begin{equation}
\tilde{\phi}^a = \phi^a
\label{taumappa3}
\end{equation}
This concludes the description of the map $\mathcal{T}$. By means of
this token we can trace the action of the Ehlers
symmetry algebra $\mathbb{U}_{D=3}$ on the Matzner--Misner fields and
vice versa, trace the action of the Matzner--Misner algebra
$\mathbb{U}_{D=4}\times \mathrm{SL}(2,\mathbb{R})_{MM}$ on the
Ehlers  fields. The found extension of the Kramer-Neugebauer
transformation $\mathcal{T}$ allows to merge these two algebras into
a larger one which, as claimed, turns out to be the affine extension
of the Ehlers algebra $\mathbb{U}_{D=3}$.
\par
We shall prove this by showing that by using the Matzner--Misner
generators ($ L^{mm}_0$, $L^{mm}_+$, $L^{mm}_-$) we can add
to a Chevalley--Serre presentation $\left( h_i,e_i,f_i\right) \,\, (i=1,\dots,r)$ of the Ehlers algebra
$\mathbb{U}_{D=3}$ a new Chevalley-Serre triplet $\left( h_0,e_0,f_0\right) $, which
has, with the generators $\left( h_i,e_i,f_i\right)$, the correct
commutation relations corresponding to the Cartan matrix
$\mathcal{C}^\wedge$, if $\mathcal{C}$ was the Cartan matrix of
$\mathbb{U}_{D=3}$.
\subsection{Field Theory identification of the affine Chevalley-Serre triplet}
As anticipated the identification of the affine Chevalley-Serre
triplet is different in the case of pure gravity and in that of
supergravity. Let us begin with the case of pure gravity.
\par
\paragraph{Pure Gravity}
\par
This is the case originally discussed by Nicolai in
\cite{NicolaiEHMM}. Here we just have two copies of the
$\mathrm{SL(2,\mathbb{R})}$ algebra, the Ehlers and the Matzner--Misner
realization. The transformations of each algebra on its own fields are:
\begin{eqnarray}
&&  \begin{array}{|c|c|}
  \hline
     \mathrm{SL(2,\mathbb{R})_E} & \mathrm{SL(2,\mathbb{R})_{MM}} \\
     \hline
     \null & \null \\
     \begin{array}{lcl}
  \delta_{L_0^{E}}{\Delta} = \sqrt{2}{\Delta} &; &  \delta_{L_0^{E}}{B} =
 \sqrt{2}{B} \\
 \delta_{L_+^{E}} {\Delta} = 0 & ; & \delta_{L_+^{E}}{B} = 1, \\
  \delta_{L_-^{E}}{\Delta} =
- 2{\Delta}{B},
 & ; & \delta_{L_-^{E}}{B} =
{\Delta}^2 - {B}^2
\end{array} & \begin{array}{lcl}
  \delta_{L^{mm}_0}\tilde{\Delta} = \sqrt{2}\tilde{\Delta} &; &  \delta_{L^{mm}_0}\tilde{B} =
 \sqrt{2}\tilde{B} \\
 \delta_{L_+^{mm}} \tilde{\Delta} = 0 & ; & \delta_{L_+^{mm}}\tilde{B} = 1, \\
  \delta_{L_-^{mm}}\tilde{\Delta} =
- 2\tilde{\Delta}\tilde{B},
 & ; & \delta_{L_-^{mm}}\tilde{B} =
\tilde{\Delta}^2 - \tilde{B}^2
\end{array} \\
     \hline
  \end{array}\nonumber\\
\label{EhlerMM}
\end{eqnarray}
and the relation between the two sets is obtained from the general
form of the $\mathcal{T}$-map (\ref{taumappa0}--\ref{taumappa3}) by
deleting all the scalars coming from vector fields $\varpi^\alpha,
\pi^{\Lambda|A}$. We have:
\begin{eqnarray}
\nonumber\tilde{\rho} = \rho & ; & \tilde{\Delta} =
\frac{\rho}{\Delta}\\
\nonumber \partial_0\tilde{B} =
-\frac{N\rho}{2\Delta^2\lambda}\, \partial_1B  & ; &
\partial_1\tilde{B} \, = \,
-\frac{\lambda\rho}{2N\Delta^2}\, \partial_0B  \\
\label{taumappaGrav}
\end{eqnarray}
Using eq.s(\ref{EhlerMM}) and (\ref{taumappaGrav}) we can combine the
two algebras. Inspired by the algebraic discussion following
eq.(\ref{explicitCS}) we set
\begin{eqnarray}
\left( h_1 \, , \, e_1 \, , \, f_1 \right) & = & \left( \sqrt{2} L_0^{E} \, , \, L_+^{E} \, , \,  L_-^{E} \right)
\nonumber\\
\left( h_0 \, , \, e_0 \, , \, f_0 \right) & = & \left( \sqrt{2} L_0^{mm} \, , \, L_+^{mm} \, , \,  L_-^{mm}
\right)
\label{puregravCS}
\end{eqnarray}
and we  calculate the commutators of one triplet with the other
triplet. This defines the Cartan matrix of the extended algebra. In
particular we evaluate the commutator $[h_0,e_1]$ that is nonvanishing only on the Ehlers
field $B$ since on the  other Ehlers fields $\delta_{e_1} = 0$.  The commutator
of the two transformations is calculated as
\begin{equation}
\label{commdef}
[\delta_{h_0},\delta_{e_1}]  =\delta_{h_0} \delta_{e_1} -
\delta_{e_1} \delta_{h_0} = - \delta_{[h_0,e_1]}
\end{equation}
and we find:
\begin{equation}
[\delta_{h_0},\delta_{e_1}]B  =  -
\delta_{e_1}\delta_{h_0}B =  2\delta_{e_1}\, B
\label{CserregravonB}
\end{equation}
The last step in eq.(\ref{CserregravonB}) is motivated by the
following calculation
\begin{eqnarray}
  \delta_{h_0}\, \partial_0 B &=& \delta_{h_0}\,\left(-\frac{2\, \tilde{N}}{\tilde{\lambda}} \,
  \frac{\tilde{\rho}}{\tilde{\Delta}^2} \, \partial_1 \tilde{B} \right)
  = 2 \, \left(-\frac{2\, \tilde{N}}{\tilde{\lambda}} \,
  \frac{\tilde{\rho}}{\tilde{\Delta}^2} \, \partial_1 \tilde{B} \right)
  \nonumber\\
  &=& - 2 \, \partial_0 B
\label{urcacalculo}
\end{eqnarray}
which is consistent only with
\begin{equation}
  \delta_{h_0}\, B = - 2 \, B
\label{h0onB}
\end{equation}
Eq.(\ref{CserregravonB}), on the other hand, implies
\begin{equation}
  [h_0,e_1] = -2\,e_1
\label{CserreGrave1}
\end{equation}
which is just one of the Serre relations for the affine extension of the
$A_1$ Lie algebra. Indeed this commutation relation verifies the entry $\langle \alpha_0 \,
,\, \alpha_1 \rangle$ of the Cartan matrix (\ref{A1WCarta}).
The other necessary relation is:
\begin{equation}
[h_0,f_1] = 2 \, f_1
\label{CserreGravF1}
\end{equation}
which can also be verified by evaluating the commutators:
\begin{eqnarray}
&& \delta_{[h_0,f_1]}\Delta = 2 \, \delta_{f_1}\Delta
 \label{h0f1onDelta1} \\
&& \delta_{[h_0,f_1]}B = 2\delta_{f_1}B  \label{h0f1onDelta2}
\end{eqnarray}
The results in eq.s (\ref{h0f1onDelta1}) and (\ref{h0f1onDelta2}) are
obtained with straightforward calculations similar to that in
eq.(\ref{urcacalculo}). This concludes the proof that the symmetry of
pure gravity reduced to $D=2$ is indeed $A_1^\wedge$ and that the
identification (\ref{puregravCS}) is the correct one for the
Chevalley Serre pair of triplets occurring in this case. Let us now
turn our attention to supergravity.
\paragraph{Supergravity}
\par
It is fairly simple, within the algebra (\ref{genGD3}), to identify the  Chevalley--Serre triple
$\{h_1,e_1,f_1\}$ which is relevant for the affine extension, once  we rewrite  the commutation relations in the form
(\ref{betterGD3}). From the algebraic view point we know that the Chevalley--Serre
triple $\{h_w,e_w,f_w\}$ generating the affine node is associated
with the unique \textbf{black root} $\alpha_W$, which is not orthogonal to the highest root
$\psi$. Namely we can write:
\begin{eqnarray}
h_w & = & \alpha_W \cdot \mathcal{H} \nonumber\\
e_w & = & E^{\alpha_W} \nonumber\\
f_w & = & E^{-\alpha_W}
\label{tripletta1}
\end{eqnarray}
having denoted by $\mathcal{H}$ the CSA of $\mathbb{U}_{D=3}$.
The problem, in order to perform field--theoretical calculations is to
identify $\alpha_W \cdot \mathcal{H}$ and $E^{\pm\alpha_W}$ within
the presentation (\ref{betterGD3}) of the Lie algebra.  To this
effect, we just have to recall that the universal subalgebra
$\mathrm{SL(2,\mathbb{R})_E}$ coming from the Ehlers reduction of
Einstein gravity is nothing else but the $A_1$ subalgebra associated
with the highest root $\psi$. In other words, with reference to
eq.s(\ref{betterGD3}), we have the identification:
\begin{equation}
  L_0 \, = \psi \cdot \mathcal{H} \quad ; \quad L_+ \, = \, E^\psi
  \quad ; \quad L_- \, = \, E^{-\psi}
\label{agnizio1}
\end{equation}
Next we recall that for the simple roots we can always choose a basis
of the form (\ref{lucullus}), namely as  Cartan generators we
can use:
\begin{equation}
  \underbrace{\mathcal{H}_I}_{I=1,\dots,r} =
  \left\{ \underbrace{\overline{\mathcal{H}}_i}_{i = 1,\dots,r-1} \, , \, L_0 \right \}
\label{longshortH}
\end{equation}
where $\overline{\mathcal{H}}_i$ is a Cartan basis for the $\mathbb{U}_{D=4}$
Lie algebra. With these considerations we conclude that:
\begin{equation}
  \alpha_W \cdot \mathcal{H} = \overline{\mathbf{w}}_h \cdot
  \overline{\mathcal{H}} \, + \, \frac{1}{\sqrt{2}} \, L_0
\label{cartone1}
\end{equation}
where the term $\overline{\mathbf{w}}_h \cdot
  \overline{\mathcal{H}}$ is necessarily a linear combination of the
  generators $T^A$ of $\mathbb{U}_{D=4}$; which one we still has to
  determine.
\par
From the commutation relations (\ref{betterGD3}) we see that the only
generators having non vanishing grading with respect to $L_0$ are the
$W^\Lambda$. Hence we learn that $W^\Lambda \propto E^{\alpha_V}$
where we have collectively denoted by $\alpha_V$ the $\mathbb{U}_{D=3}$
roots corresponding to weights of the symplectic representation
$\mathbf{W}$. For one particular value $\Lambda_h$ we retrieve the
highest weight $\mathbf{w}_h$, namely the black root $\alpha_W$. At
this point we have enough information to fix also the absolute
normalizations. Indeed comparing the first of eq.s (\ref{betterGD3})
with eq. (\ref{cartone1}) we conclude that:
\begin{eqnarray}
h_w & = & 2 \left(\Lambda_A \right)^{\Lambda_h}_{\phantom{\Lambda}\Lambda_h} \, T^A \, + \,
 \frac{1}{\sqrt{2}} \, L_0\nonumber\\
e_w & = & \sqrt{2} \, W^\Lambda \nonumber\\
f_w & = & \sqrt{2}\, \hat{W}_\Lambda
\label{Cstripletta1}
\end{eqnarray}
The relevance of this identification is that now via eq.s
(\ref{LminBvarpi}--\ref{Lplus}) we know the action of the
Chevalley-Serre generators on the Ehlers fields and via eq.s(\ref{taumappa0}-\ref{taumappa3})
also on the Matzner--Misner fields. Hence we can evaluate
commutators of the found triplet $(h_w, e_w,f_w)$ with the affine triplet that is the same as in eq.(\ref{puregravCS}).
We illustrate the resuot on the  for the commutator $[h_0,e_w]$ that is
nonvanishing only on the Ehlers fields $B$ and $\varpi_\alpha =
(\tau^\Lambda, \sigma_\Sigma)$.  For instance we calculate $\delta_{[h_0,e_w]}
\,\tau^\Lambda$. First from eq.(\ref{taumappa2}) we easily calculate the action of
$h_0$ on, for example, $\tau^\Lambda$. Indeed we have:
\begin{eqnarray}
  \delta_{h_0} \, \tau^\Lambda & = & \delta_{h_0} \left(
  \sqrt{\frac{\tilde{\rho}}{\tilde{\Delta}}} \, \pi^{\Lambda|3}
  \right)=\left(
  \sqrt{\frac{\tilde{\rho}}{\tilde{\Delta}}} \, \pi^{\Lambda|3}
  \right)\nonumber\\
  & = & - \tau^\Lambda
\label{cuccocalcolo}
\end{eqnarray}
Then, using eq.s(\ref{W}) and the identification (\ref{Cstripletta1}) we have
\begin{equation}
  [\delta_{h_0},\delta_{e_w}] \, \tau^\Lambda = \delta_{e_w} \, \tau^\Lambda
  \quad \Rightarrow \quad [h_0,e_w] = - e_w \label{newAffineCarta}
\end{equation}
Using eqns.(\ref{hatWB}--\ref{hatWDeltaphi}) it is straightforward to check that $[h_0,f_w] = f_w$.

In order to evaluate the commutators $[h_w,e_0]$ and $[h_w,f_0]$ we note that due th the fact that the generators $L^{mm}$ of the $\mathrm{SL}(2,\mathbb{R})_{MM}$ 
commute with the generators $T^A$ of $\mathbb{U}_{D=4}$, we have only to calculate the commutators $[L_0,e_0]$ and $[L_0,f_0]$.
Then, using (\ref{L0}) and tracing the action of the generators $e_0$ and $f_0$ on the Ehlers fields via the map $\mathcal{T}$, one finds that $[h_w,f_0]=f_0$ (the action of the generator $e_0$ on the Ehlers fields is trivial). 

This allows to recunstruct the Cartan matrix of the extended symmetry algebra that corresponds to the affine extension of $\mathbb{U}_{D=3}$. 
The final check of the Serre relations $\mbox{ad}[e_i]^{C_{ji}+1}e_j = 0$ and  $\mbox{ad}[f_i]^{C_{ji}+1}f_j = 0$ should be more involved, but 
relying on the algebraic arguments presented in the section (\ref{systematics}) it should follow.

So, differently from the case of pure gravity, in all supergravity
models the affine extended Dynkin diagram has a \textbf{new simple
line} and \textbf{not} a \textbf{double line}
as in the case of pure gravity. This is due to the replacement of the
eigenvalue $-2$ with the eigenvalue $-1$ shown in the above
calculation. This change is due to the fact that the affine node is
linked to the vector root $\alpha_W$ and not to the root of
$\mathrm{SL(2,\mathbb{R})_E}$ as in the pure gravity case. We stress again that the extension is possible due to the coexistence of
the Ehlers and Matzner Misner dimensional reduction schemes that
lead to well distinct results in the supergravity case, while they
lead to formally identical lagrangians in pure gravity.
\section{Conclusions}
As emphasized in the introduction, in this paper we have analyzed
the field theoretical mechanism that leads to the affine extension
of the duality algebra in $D=2$. We have shown that there is a
uniform pattern underlying such mechanism and that this is based
on the coexistence of two non locally related dimensional
reduction schemes, the Ehlers and the Matzner--Misner scheme,
respectively. In this way we have extended an original idea that
Nicolai had applied to pure gravity \cite{NicolaiEHMM} to the more
general setup of all $D=4$ supergravities. In particular we have
stressed the structural differences that arise in extended or
matter coupled supergravity with respect to pure gravity and which
are related to the presence of vector fields as well. This leads
to the general form of the Matzner--Misner lagrangian which is
different from the Ehlers one and which to our knowledge had not
been discussed so far in the literature.
\par
The main motivation for our study is provided by the issue of cosmic
billiards. Indeed we plan to use our results in order to discuss how
the compensator method to generate solutions of the first order
equations which we introduced in \cite{noiconsasha} and which we
recently rediscussed in the context of the Tits Satake projection for
non maximally split cosets \cite{noiTSpape} can be extended to the infinite
dimensional compact subalgebras of Ka\v c--Moody algebras.
\par
Alternatively the field theoretical understanding of the affine
extensions is relevant in studying wave--solutions of
supergravity, as already emphasized by Nicolai \cite{NicolaiEHMM}
and other authors. In this context special attention is to be
devoted to the $pp$-waves and to the Penrose limit. Indeed, as it
will be pointed out in a forthcoming paper \cite{ioconandrea}, the
Penrose limit in supergravity models can be thoroughly understood
within the framework of \textit{Lie algebra contractions} and the
relation between the isometry Lie algebra of a wave solution and
the duality (affine) Lie algebra that generates the solution
itself is a quite challenging conceptual issue, potentially very
important in the quest for a deeper understanding of string theory
and brane physics. It should also be stressed that while all
purely time dependent solutions (in particular cosmic billiards)
break all supersymmetries, wave--like solutions as the $pp$--waves
can preserve several $\mathbf{SUSY}$ charges (for a comprehensive
review of the vast recent literature on supergravity $pp$-wave
solutions, see for instance\cite{BlauAll}). So an obvious research
line streaming from our present results is the systematic
investigation of wave solutions dynamically generated by the
affine Ka\v c--Moody extension of the duality algebra, along the
lines already pioneered in pure gravity by \cite{Maison&Breiten}
and \cite{NicolaiEHMM}, and their classification according to
Killing spinors.
\par
Another direction which is  to be pursued is the systematic
analysis of the double, or hyperbolic, extension of the duality algebra
occurring in one--dimension.
\par
It is easy to anticipate that the field theoretical mechanism
underlying this is the coexistence of two dimensional
reduction schemes, the Ehlers, the analog of the Matzner--Misner
one, in which we
step directly down from $D=4$ to $D=1$, by compactifying on a
$T^3$--torus. In this way we obtain a rank two $\sigma$--model
$\mathrm{SL(3,\mathbb{R})/O(3)}$ which describes the degrees of
freedom of pure gravity. This is also an obvious research line
which we plan to  carry out in the immediate future.
\par
Let us finally mention that the analysis of cosmic billiards was
so far given only in the context of ungauged supergravities. The
extension to gauged supergravities and hence to flux
compactifications (see, for istance, \cite{fluxcompacta}) is
clearly overdue and is \textit{in agenda}. Here we know that the
crucial item governing the classification of gaugings is the so
named \textit{embedding matrix}, originally introduced in
\cite{gaugedsugrapot1} and later shown in \cite{Trigiasamleben} to
be algebraically described by suitable irreducible representations
of the duality algebras in $D=4$ and $D=3$. It goes without saying
that the affine and hyperbolic extension of such an analysis is
also a must.

\end{document}